\documentclass[%
 reprint,
 amsmath,amssymb,
 aps,
]{revtex4-2}

\usepackage{graphicx}
\usepackage{dcolumn}
\usepackage{bm}
\usepackage[colorlinks=true,citecolor=blue]{hyperref}
\usepackage[utf8]{inputenc}
\usepackage{soul}

\usepackage{aas_macros}

\newcommand{\vect}[1]{\boldsymbol{#1}}
\newcommand{\uvect}[1]{\vect{\hat{#1}}}
\newcommand{\lh}{\tilde{h}^{(l)}}

\usepackage[usenames, dvipsnames]{xcolor}

\newcommand{\citationneeded}[1][]{{\color{red} [citation!]}}

\begin{document}

\preprint{}

\title{Detecting gravitational lensing in hierarchical triples in galactic nuclei with space-borne gravitational-wave observatories}

\author{Hang Yu}
\email{hangyu@caltech.edu}
\affiliation{Department of Physics, California Institute of Technology, Pasadena, CA 91125, USA}
 
\author{Yijun Wang}
\affiliation{Department of Physics, California Institute of Technology, Pasadena, CA 91125, USA}
  
\author{Brian Seymour}
\affiliation{Department of Physics, California Institute of Technology, Pasadena, CA 91125, USA}

\author{Yanbei Chen}
\affiliation{Department of Physics, California Institute of Technology, Pasadena, CA 91125, USA}

\date{\today}

\begin{abstract}
Stellar-mass binary black holes (BBHs) may merge in the vicinity of a supermassive black hole (SMBH). It is suggested that the gravitational-wave (GW) emitted by a BBH has a high probability to be lensed by the SMBH if the BBH's orbit around the SMBH (i.e., the outer orbit) has a period of less than a year and is less than the duration of observation of the BBH by a space-borne GW observatory. 
For such a ``BBH + SMBH'' triple system, the de Sitter precession of the BBH's orbital plane is also significant. 
In this work, we thus study GW waveforms emitted by the BBH and then modulated by the SMBH due to effects including Doppler shift, de Sitter precession, and gravitational lensing. We show specifically that for an outer orbital period of 0.1 yr and an SMBH mass of $10^7\,M_\odot$, there is a $3\%-10\%$ chance for the standard, strong lensing signatures to be detectable by space-borne GW detectors such as LISA and/or TianGO. For more massive lenses ($\gtrsim 10^8\,M_\odot$) and more compact outer orbits with periods $\lesssim 0.1\,{\rm yr}$, retro-lensing of the SMBH (which is closely related to the glory-scattering) might also have a $1\%$-level chance of detection. Furthermore, by combining the lensing effects and the dynamics of the outer orbit, we find the mass of the central SMBH can be accurately determined with a fraction error of $\sim 10^{-4}$.
This is much better than the case of static lensing because the degeneracy between the lens' mass and the source's angular position is lifted by the outer orbital motion. Including lensing effects also allows the de Sitter precession to be detectable at a precession period 3 times longer than the case without lensing. Lastly, we demonstrate that one can check the consistency between the SMBH's mass determined from the orbital dynamics and the one inferred from gravitational lensing, which serves as a test on theories behind both phenomena. The statistical error on the deviation between two masses can be constrained to a $1\%$ level. 
\end{abstract}

\maketitle


\section{Introduction}
\label{sec:intro}





Since September 14, 2015~\cite{LSC:16}, ground-based gravitational-wave (GW) observatories including LIGO~\cite{LSC:15}, Virgo~\cite{Acernese:15}, and KAGRA~\cite{kagra:19} have achieved great success with tens of GW events detected so far~\cite{LSC:19, LSC:21}. A new window for human beings to observe the Universe using GW radiation has since been opened up. Looking towards future, more excitements await us as multiple space-borne GW observatories are proposed to be launched in the near future, including LISA~\cite{Amaro-Seoane:17}, TianQin~\cite{Luo:16}, Taiji~\cite{Hu:17}, B-DECIGO~\cite{Nakamura:16, Kawamura:20}, Decihertz  Observatories~\cite{Sedda:19}, and TianGO~\cite{Kuns:19}. Their sensitivity covers the 0.001 to 0.1 Hz band, allowing them to observe typical, stellar-mass binary black hole (BBH) systems for years prior to the final merger. It thus opens the possibility of using the slowly chirping GW raddiation from a stellar-mass BBH as a carrier signal to search for external modulations induced by environmental perturbations. 

One particularly interesting scenario is if the stellar-mass BBH is in a galactic nucleus where a supermassive black hole (SMBH) resides. Such a hierarchical triple system (``BBH + SMBH'') is expected because there are theories predicting that the environment in a galactic nucleus can facilitate the merger of stellar-mass BBH. One channel that has been studied extensively in the literature is due to gaseous effects~\cite{McKernan:12, Bartos:17, Stone:17, McKernan:18, Tagawa:19, Yang:19}. If the BBH lives in a gaseous disk of an active galactic nuclei (AGN), the gas can provide extra frictional force on the BBH in addition to the force induced by GW radiation and thus hardens its orbit (the inner orbit). In addition to shrinking the orbit of the BBH itself, gas can also help the center of mass of the BBH to migrate in the AGN disk and thus alter its orbit around the SMBH~\cite{Bellovary:16, Secunda:19}. Even without gas, dynamical interactions of a variety of flavors may also help the formation of compact BBHs~\cite{OLeary:09, Antonini:12, Antonini:16, VanLandingham:16, Petrovich:17, Leigh:18, Chen:18, Fragione:19, Han:19, Samsing:20}, providing yet another channel for the formation of ``BBH + SMBH'' triple systems. 

Similar to the formation channels, the environmental perturbation on the BBH's GW waveform in such a triple system can also be divided up into two main classes. One is still due to the gaseous friction, and its main effect is to make the BBH appear to be more massive than its true value~\cite{Chen:20, Chen:20b, Caputo:20, Toubiana:21}. This effect is most prominent when the GW decay timescale of the BBH is a few kilo-years (comparable to the hardening timescale due to the gas), but is subdominant for more compact (i.e., ``harder'') BBHs that will merge in a few years. 

The other type of modulation is directly related to the gravitational field of the SMBH and is the main focus of our discussion here. The leading-order effect arises from the orbital motion of the center of mass of the BBH orbiting around the SMBH (i.e., the outer orbit), leading to a Doppler phase shift in the GW waveform emitted by the BBH. It has been shown that this Doppler phase shift might be detectable for BBHs as far as 1\,pc away from the SMBH~\cite{Inayoshi:17}, and when the outer orbital period is less than a few years (set by the duration of the observation), the frequency of the Doppler phase shift can be measured, which further constrains the mass density enclosed by the outer orbit~\cite{Randall:19}. Beyond the leading-order effect, the Newtonian tidal effect (which typically manifested as the Lidov-Kozai effect) may also play a role for a triple system with an inner orbital separation of $\sim 0.1\,{\rm AU}$~\cite{Hoang:19, Deme:20, Chandramouli:21}. For more compact BBHs with separation of $\sim 10^{-3}\,{\rm AU}$, Ref.~\cite{Yu:20d} showed that the (post-Newtonian) de Sitter-like precession of the BBH's orbital plane~\cite{Will:18, Liu:19, Kuntz:21} is the more critical correction to the waveform and can be detectable by space-borne GW detectors out to a cosmological distance of $\sim 1\,{\rm Gpc}$. Combining the de Sitter-like precession and the Doppler shift, Ref.~\cite{Yu:20d} further demonstrated that the mass of the central SMBH can be determined, which complements the existing direct methods of measuring the mass of an SMBH~\cite{Peterson:14}. 

Recently, Ref.~\cite{DOrazio:20} further considered the strong gravitational lensing of the BBH's GW caused by the SMBH and showed that there is a high geometrical probability ($\sim 10\%$) for lensing to happen if the BBH is in an outer orbit with a period of less than a year (shorter than the duration of the observation). Ref.~\cite{DOrazio:20} referred to this as ``repeated lensing''. Indeed, if the outer period is long (much longer than the observation duration), then in order for the BBH to be lensed, it needs to have both the right azimuthal and polar angles such that the angular separation between the source and the lens is sufficiently small to be comparable with the Einstein ring. However, the short orbital period in the repeated-lensing regime allows the inner BBH to scan through the azimuthal angle and there will thus always an instant during the observation when the BBH is behind the SMBH. Therefore, the BBH only needs to have the right polar angle. Furthermore, because the outer orbital radius is smaller, it is also more likely for the BBH to be within the Einstein ring of the SMBH. 

Ref.~\cite{DOrazio:20} focused on the standard strong lensing, which treats the SMBH as a Newtonian point particle and lensing happens when the source is behind the lens (i.e., the source and the observer are on opposite sides of the lens). Meanwhile, the strong gravity field of the SMBH can further lead to relativistic lensing signatures~\cite{Virbhadra:00, Bozza:02, Bozza:07, Eiroa:11, Virbhadra:09}. One such example is retro-lensing (which is also closely related to the glory-scattering of an SMBH) ~\cite{Luminet:79, Holz:02, Eiroa:04, Bozza:10, Eiroa:12}. Retro-lensing happens when the BBH is in front of the SMBH: the GW emitted by the BBH towards the SMBH gets bent by the strong gravity potential of the SMBH by an angle of approximately $\pi$ and eventually reaches the observer. For the same reason that repeated strong lensing is likely in BBH + SMBH triples, repeated retro-lensing also has a relatively high probability to happen (as it corresponds to the same geometrical configuration as the standard lensing but just with the outer orbital phase shifted by $\pi$). Therefore, retro-lensing should also be incorporated in the waveform modeling. 

Furthermore, the parameter space where repeated lensing happens~\cite{DOrazio:20} largely overlaps with the parameter space where the de Sitter precession is detectable~\cite{Yu:20d}. It is thus critical to incorporate both effects in the waveform modeling. More importantly, we note that including the lensing effects does not introduce any new parameters compared to the one needed for modeling the orbital dynamics. As shown in Ref.~\cite{Takahashi:03}, the lensing effect can be parameterized in terms of the mass of the lens (i.e., the SMBH) and the sky projection of the source for point-source lenses. 
All of them can also be independently inferred from the combination of Doppler shift and de Sitter precession as illustrated in Ref.~\cite{Yu:20d}. Therefore, the two effects can be combined to enhance the overall parameter estimation (PE) accuracy, and checked against each other to test the consistency of theories behind each effect. 

Therefore, in this work our goal is to construct GW waveforms of a BBH in the vicinity of an SMBH, including effects of the SMBH on both the orbital dynamics (Doppler shift and de Sitter precession) and lensing (standard lensing and retro-lensing). We will further use the waveform to quantify the detectability of the lensing signatures. Moreover, we also assess the accuracy of PE of the triple system, in particular the mass of the central SMBH.

Throughout this study, we will refer to the stellar-mass BBH (consists of $M_1, M_2 \sim \text{a few}\times 10\,M_\odot$ ) as the inner binary and quantities associated with it will often be denoted with a subscript ``$i$''. The inner orbit decays via gravitational radiation; the GW it emits serves as the signal carrier in our study. The orbit of the inner binary's center of mass around the SMBH ($M_3\sim 10^5-10^{10}\,M_\odot$) is referred to as the outer orbit and is denoted with a subscript ``$o$''. The GW radiation of the outer binary can be safely ignored for systems of our interest. For simplicity, we ignore the spin of the SMBH and treat it as a Schwarzschild BH. We further restrict our discussion here to the simple case where both the inner and the outer orbits are circular. The general case that allows for orbital eccentricities is deferred to future studies. Moreover, all values in our study are measured in the detector frame. In other words, they are redshifted by the cosmological expansion ($z_{\rm cos}\sim 0.2-0.3$ for sources at a luminosity distance on the order of 1\,Gpc) and the gravity of the SMBH ($z_{\rm grav}\sim10^{-3}-0.01$ for typical outer orbits we consider). We use geometrical units $G=c=1$. 

The paper is organized as follows. In Sec.~\ref{sec:waveform} we describe our construction of the GW waveform including effects due to both standard strong lensing (Sec.~\ref{sec:std_lens}) and retro-lensing (Sec.~\ref{sec:retro_lens}). We then examine the detectability of lensing effects by space-based GW observatories like LISA~\cite{Amaro-Seoane:17} and TianGO~\cite{Kuns:19} in Sec.~\ref{sec:mismatch} by considering the mismatches between waveforms with and without lensing. The PE analysis including both lensing and orbital dynamics (Doppler + de Sitter) is presented in Sec.~\ref{sec:PE}. Specifically, we consider the enhancement in the PE accuracy of the 
SMBH's mass in Sec.~\ref{sec:PE_enhancement}, followed in Sec.~\ref{sec:PE_consistency} by a study on how well we can test the consistency between the SMBH's mass determined from the lensing signal and that from the orbital dynamics. Lastly, we conclude in Sec.~\ref{sec:conclusion} together with a discussion on effects to be further incorporated by future studies.

\section{GW waveforms including gravitational lensing}
\label{sec:waveform}
In this section, we describe our modeling of the GW waveform. We will start by briefly reviewing the waveform construction without lensing effects, which closely follows Ref.~\cite{Yu:20d}. This is followed by Sec.~\ref{sec:std_lens} in which we consider the standard lensing. In Sec.~\ref{sec:retro_lens}, we further incorporate the retro-lensing into the waveform. We conclude our waveform modeling in Sec.~\ref{sec:samp_waveforms} by examining a few representative waveforms including lensing effects and sketching out the parameter space for the lensing signatures to be potentially significant. 

Following Ref.~\cite{Takahashi:03}, we write the lensed waveform [denoted by a superscript ``$(l)$''] as
\begin{equation}
    \lh(f) = F(f)\tilde{h}(f),
    \label{eq:lensed_waveform}
\end{equation}
where $\tilde{h}(f)$ is the frequency-domain waveform without lensing and the quantity $F(f)$ is an amplification factor due to the gravitational lensing. 

To model $\tilde{h}(f)$, we follow Ref.~\cite{Yu:20d} and write (see also Refs~\cite{Apostolatos:94, Cutler:98})
\begin{align}
    \tilde{h}(f) &= \Lambda(t)\tilde{h}_{c}(f) \nonumber \\
    &= [A_+^2(t)F_+^2(t) + A_\times^2(t)F_\times^2(t)]^{1/2} \nonumber \\ 
    &\times \exp\left\{-i\left[\Phi_{p}(t)+ 2\Phi_{T}(t) + \Phi_{D}(t) \right] \right\} \tilde{h}_{c}(f),
    \label{eq:waveform_no_lens}
\end{align}
where $\tilde{h}_c(f)$ is the antenna-independent carrier waveform, which we further model using the quadrupole formula as
\begin{align}
    \tilde{h}_{c}(f) &= \left(\frac{5}{96}\right)^{1/2}\frac{\mathcal{M}^{5/6}}{\pi^{2/3}D}f^{-7/6} \nonumber \\
    & \times \exp\left\{i \left[2\pi f t_{\rm c} -\phi_{\rm c}-\frac{\pi}{4}+\frac{3}{4}(8\pi\mathcal{M}f)^{-5/3}\right]\right\},
    \label{eq:hc}
\end{align}
where $\mathcal{M}$, $D$, $t_c$, and $\phi_c$ are the chirp mass (in the detector frame), luminosity distance, time and phase of coalescence, respectively. 

The antenna response is incorporated under the leading-order stationary phase approximation (SPA), which first evaluates each quantity as a function of time $t$, and then express the time as a function of frequency, $t=t(f)$, following
\begin{equation}
    t(f) = t_{\rm c} - 5(8\pi f)^{-8/3}\mathcal{M}^{-5/3}.
    \label{eq:t_vs_f}
\end{equation}
Furthermore, in Eq.~(\ref{eq:waveform_no_lens}) we have defined $A_+{=}1 + \left(\uvect{L}_i\cdot \uvect{N}\right)^2$ and $A_\times {=} -2\left.\uvect{L}_i\cdot \uvect{N}\right.$, where $\uvect{L}_i$ is the orientation of the inner orbital angular momentum and $\uvect{N}$ is the line of sight.  The quantities $F_+$ and $F_\times$ are the ``detector beam-pattern'', $\Phi_p=\arctan \left[-A_\times F_\times/A_+F_+\right]$ is the polarization phase, and $\Phi_T$ is the Thomas precession phase. Their expressions can be found in, e.g., Refs.~\cite{Apostolatos:94, Cutler:98, Yu:20d}. Note that they are time-dependent because of motions of both the detector in the solar frame and the inner binary in the SMBH frame. Specifically, we assume the detector follows an orbit as described in Ref.~\cite{Dhurandhar:05} and the its explicit orientation can be found in e.g., Ref.~\cite{Yu:20d} for a $90^\circ$-detector like TianGO~\cite{Kuns:19}, and Ref.~\cite{Cutler:98} for a $60^\circ$-detector like LISA~\cite{Amaro-Seoane:17}. For the inner binary's orientation, we include the de Sitter-like precession~\cite{Will:18, Liu:19}, which can be expressed as~\cite{Yu:20d}
\begin{equation}
    \frac{d\uvect{L}_i}{dt} = \Omega_{\rm dS} \uvect{L}_o \times \uvect{L}_i = \frac{3}{2}\frac{M_3}{a_o}
    \Omega_o \uvect{L}_o \times \uvect{L}_i,\label{eq:dS_prec},
\end{equation}
for a circular outer orbit. Here $\Omega_o=2\pi/P_o = \sqrt{M_3/a_o^3}$ is the orbital period with $M_3$ the mass of the SMBH and $a_o$ the semi-major axis of the outer orbit. We denote the outer orbital angular momentum as $\vect{L}_o$ and the total angular momentum of the triple as $\vect{J}=\vect{L}_o + \vect{L}_i\simeq \vect{L}_o$. We will further define $\lambda_L \equiv \arccos \left(\uvect{L}_i \cdot \uvect{L}_o\right)$ as the opening angle between the inner and outer orbital angular momenta. 

Lastly, the center of mass motion of the inner binary around the SMBH and the detector around the Sun are included via a Doppler phase, $\Phi_D$, as~\cite{Cutler:98} 
\begin{align}
    \Phi_D = 2\pi f & \left[ r_{o,\parallel}\cos\left(\Omega_o t + \phi^{(0)}\right)\nonumber \right. \\
    + &\left.\   r_{\oplus, \parallel}\cos\left(\frac{2\pi t}{{\rm yr}} - \overline{\phi}_S\right)\right], 
\end{align}
where $r_{o,\parallel} {=} a_o\sin \iota_J$ and $r_{\oplus, \parallel}{=}{\rm AU}\sin \overline{\theta}_S$. Here $\iota_J {=} \arccos \left(\uvect{L}_o \cdot \uvect{N}\right)$ in the inclination of the outer orbit, $\phi^{(0)}$ is an initial phase, and $\overline{\theta}_S$ and $\overline{\phi}_S$ are the polar and azimuthal angles of $\uvect{N}$ in the solar frame (following the notations used in Ref.~\cite{Yu:20d}, we use a bar to denote angular coordinates in the solar frame). Consistent with the treatment in Ref.~\cite{Cutler:98}, we include only the phase term of the Doppler shift $\Phi_D$ but drop the amplitude boosts like $\sim (1 - \dot{r}_{o, \parallel})$ for simplicity. Indeed, when we consider each term's contribution to parameter estimation by computing $\left(\partial \tilde{h}/\partial r_{o, \parallel}\right)$, the magnitude of the phase term's contribution is $\propto 2\pi f$ while the amplitude term's contribution is $\propto \Omega_o \sim 10^{-5} \times 2\pi f$. For  similar reasons, we ignore the time shifts [due to the propagation of the wave from the inner binary to the SMBH $\sim r_{o, \parallel}\lesssim 0.5\,{\rm day}$ and from the Sun to the detector $\sim r_{\oplus, \parallel}\sim 500\,{\rm s}$, as well as the extra time delay induced by the SMBH $\sim M_3 \lesssim 500\,{\rm s}$] in other non-Doppler terms in $\Lambda(t)$ because their variation rate is much smaller compared to the GW frequency $f$. 

Before we proceed to discuss the lensing amplification factor $F(f)$ in Secs.~\ref{sec:std_lens} (for standard lensing) and \ref{sec:retro_lens} (for retro-lensing), we would like to emphasize that the various effects entering our waveform modeling are typically computed using the lowest-order approximations. 
This is because our goal is to examine the detectability of various lensing effects and to estimate their effects on the PE accuracy. The construction of sufficiently precise templates that can be used for, e.g., signal detection via matched-filtering, are deferred to future studies. Nonetheless, we derive in Appx.~\ref{appx:spa_fast_mod} a general expression that improves the accuracy of the waveform under the SPA when the antenna response has a fast temporal variation. The waveform in Appx.~\ref{appx:spa_fast_mod}, while not used in this work for simplicity, can be readily adopted by future studies when more accurate waveforms are desired.

\subsection{Standard lensing under weak-deflection limit}
\label{sec:std_lens}
\begin{figure}[bt]
  \centering
  \includegraphics[width=0.9\columnwidth, trim=150 0 180 0, clip]{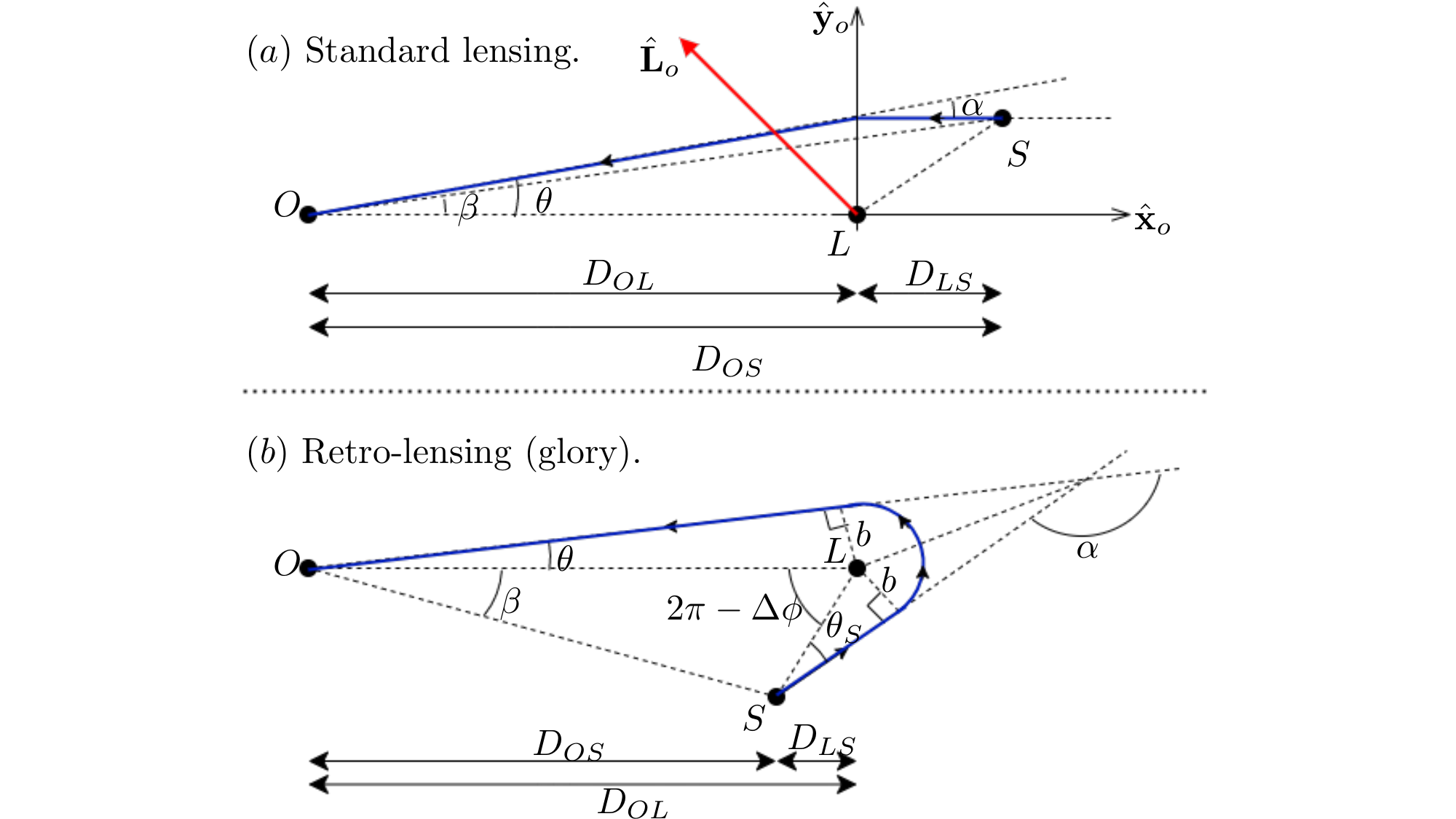}
  \caption{Cartoon illustrating the lensing geometry. The top part corresponds to the standard lensing scenario (i.e., the strong lensing) where the source is behind the lens and the deflection angle $\alpha \ll \pi$ (and here we specifically draw the instance when $z_o=\phi_o=0$). The bottom illustrates the geometry of retro-lensing (also known as the glory). Note in the repeated lensing scenario, we have $D_{OL}\simeq D_{OS}\simeq D\sim 1\,{\rm Gpc}$ and $D_{LS}\lesssim 100\,{\rm AU}\ll D_{OL}, D_{OS}$.  }
\label{fig:lens_geometry}
\end{figure}

We start the discussion of lensing effects by considering the standard lensing scenario, illustrated in the top part in Fig.~\ref{fig:lens_geometry}. This corresponds to the well-known strong lensing by the SMBH. In this scenario, the GW emitted by the source ($S$, which is the inner binary consisting of $M_1$ and $M_2$) is bent by the lens ($L$, which is the SMBH $M_3$) and then arrives at the observer $O$. We use $\beta$ and $\theta$ to indicate the angular location of the source and the image, respectively. The deflection angle is indicated by $\alpha$. Geometrically, we have
\begin{equation}
    \beta = \theta - \frac{D_{LS}}{D_{OS}} \alpha. 
\end{equation}
Note in this case, $\alpha \ll \pi$ and therefore the weak-deflection limit applies  (which is to be contrasted with the retro-lensing scenario in Sec.~\ref{sec:retro_lens}). 

For future convenience, we construct a reference frame $(x_o, y_o, z_o)$ centered on the SMBH $M_3$ and $\uvect{x}_o$ is aligned with the line of sight $\uvect{N}$ (i.e., along the line $OL$). The $x_o{-}y_o$ plane is the defined by the plane formed by the line of sight $\uvect{N}$ and the total angular momentum of the system $\uvect{J} \simeq \uvect{L}_o$ (as the spin of the SMBH is ignored). The inclination of the outer orbit is defined as $\iota_J$ with $\cos \iota_J = \uvect{N}\cdot \uvect{L}_o$. 

In this frame, we can write the source location as
\begin{equation}
\left\{
\begin{array}{l}
x_o(t) = a_o\sin \iota_J \cos \phi_o(t), \\
y_o(t) = a_o\cos \iota_J \cos \phi_o(t), \\
z_o(t) = -a_o \sin \phi_o(t),
\end{array}
\right.
\end{equation}
where $\phi_o(t)=\Omega_o t + \phi_o^{(0)}$ is the orbital phase of the outer orbit with $\Omega_o=\sqrt{M_3/a_o^3}$. Using these coordinates, we further have
\begin{align}
    D_{LS}(t) = x_o(t),\\
    \beta(t) = \frac{\sqrt{y_o^2(t) + z_o^2(t)}}{D_{OL}}.
\end{align}
Approximating the lens as a point mass (as we typically have $a_o\sim 100-1,000\,M_3\gg M_3$ for systems of interests), we then have a time-dependent $F(t; f)$ as~\cite{Takahashi:03} (see also Refs.~\cite{DOrazio:18, DOrazio:20}; note this is applied only when $x_o>0$ or the source is behind the lens) 
\begin{align}
    F(t; f) =& \exp\left\{\frac{\pi w}{4} + i \frac{w}{2}
    \left[\ln\left(\frac{w}{2}\right)-2\phi_m(\eta)\right]\right\}\nonumber \\
    &\times \Gamma(1-i\frac{w}{2})_1F_1(i\frac{w}{2}, 1; i\frac{w\eta^2}{2}),
    \label{eq:F_std_lens_full}
\end{align}
where $w(f)=8\pi M_3f$, $\phi_m\left[\eta(t)\right] = \left[\eta_m(t) - \eta(t)\right]^2/2 - \ln \eta(t)$, and $\eta_m(t) = \left\{\eta(t)+\left[\eta^2(t)+4\right]^{1/2}\right\}/2$. The quantity $\eta(t)$ is the angular location of the source normalized by the Einstein radius, 
$\eta(t) = \beta(t)/\theta_{\rm Ein}(t)$ with $\theta_{\rm Ein}(t) = \sqrt{4M_3D_{LS}(t)/\left(D_{OS}D_{OL}\right)}$ the Einstein radius. Under the SPA, the time $t$ can be further treated as a function of $f$ via Eq.~(\ref{eq:t_vs_f}). Note that both $\beta(t)$ and $D_{LS}(t)$ vary with the outer orbital phase. When the outer orbital period is shorter than the duration of the observation $P_{\rm obs}\sim 5\,{\rm yr}$, we can see multiple lensing-induced peaks (as we will see shortly in, e.g., Fig.~\ref{fig:samp_lensing_waveform}), and therefore the system is repeatedly lensed~\cite{DOrazio:20}. 

When $w \gg 1$ or $f\gg 1\, {\rm mHz}\, (M_3/10^7\,M_\odot)^{-1}$ (, which is a condition typically well-satisfied for systems we are interested in),  the full expression Eq.~(\ref{eq:F_std_lens_full}) reduces to the geometrical limit as a sum over images $j$~\cite{Nakamura:99, Takahashi:03}
\begin{equation}
    F(t; f) = \sum_j|\mu_j|^{1/2}\exp(2\pi i f t_{l, j} - i\pi n_j),
    \label{eq:F_geo_limit_def}
\end{equation}
where $\mu_j$ is the magnification of the $j$'th image, $t_{l, j}$ is the time delay of each image (we use the subscript ``$l$'' to indicate it is a quantity associated with lensing effects), and $n_j=0, 1/2, \text{ or }1$ when the image's traveling time is a minimum, saddle point, or maximum. For the standard-lensing configuration and treating $M_3$ as a point mass, two images form and 
\begin{equation}
    F(t; f) = |\mu_1(t)|^{1/2} -i |\mu_2(t)|^{1/2}e^{2\pi i f\Delta t_l(t)},
    \label{eq:F_std_lens_geo}
\end{equation}
where
\begin{align}
&\mu_{1,2}(t) = \frac{1}{2}\pm\frac{\eta^2(t)+2}{2\eta(t)\left[\eta^2(t)+4\right]^{1/2}}, \label{eq:std_lens_mag}\\
&\Delta t_l(t) = 4 M_3 \left(\frac{\eta(t)\left[\eta^2(t)+4\right]^{1/2}}{2}
\nonumber \right. \\
&\hspace{2cm}\left.+\ln \left\{\frac{[\eta^2(t)+4]^{1/2}+\eta(t)}{[\eta^2(t)+4]^{1/2}-\eta(t)}\right\}
\right).
\label{eq:std_lens_t_delay}
\end{align}

\subsection{Retro-lensing}
\label{sec:retro_lens}
\begin{figure}[bt]
  \centering
  \includegraphics[width=0.9\columnwidth]{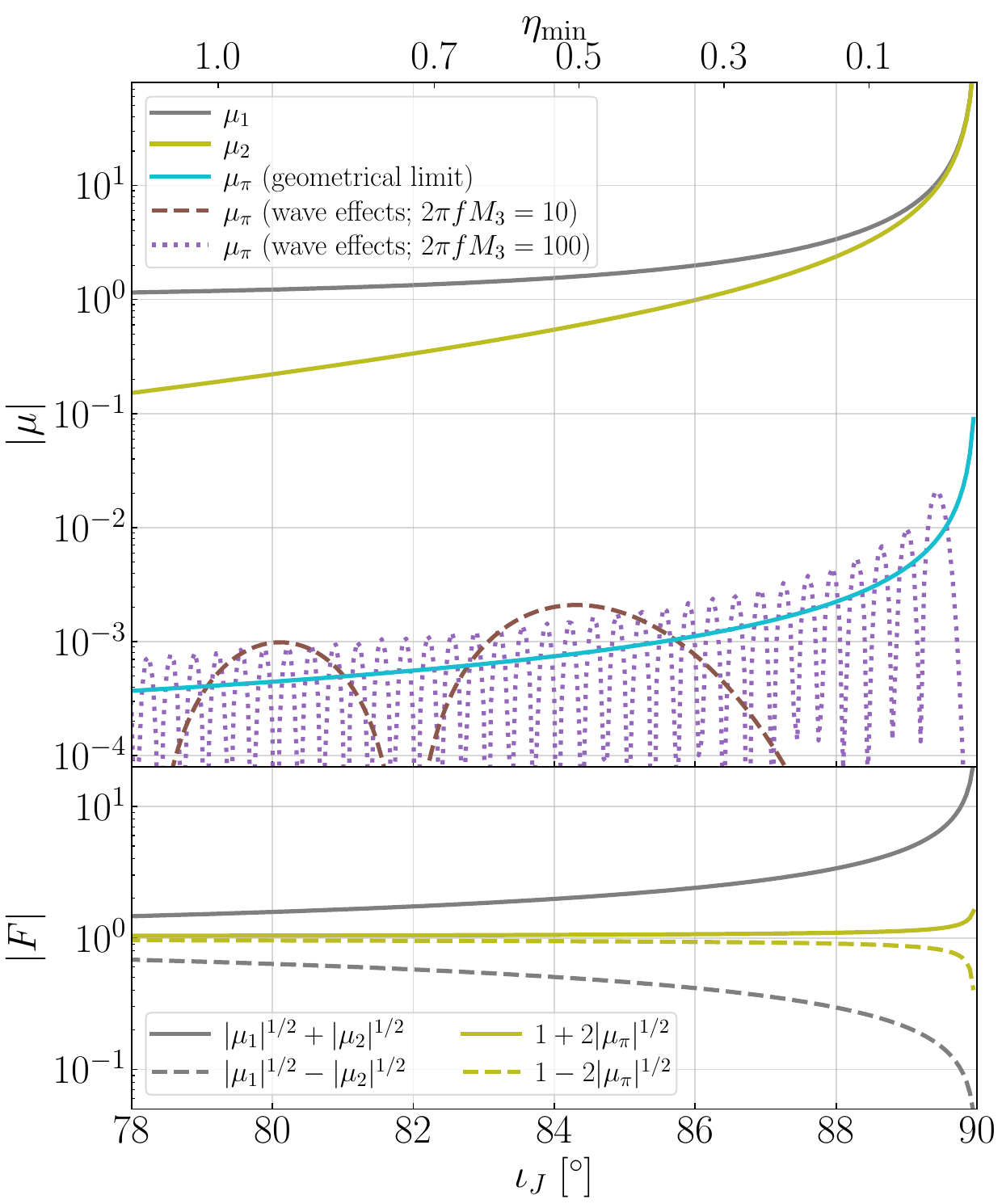}
  \caption{Top panel: magnification $|\mu|$ as a function of the inclination angle $\iota_J$ of the outer orbit.
  Bottom panel: upper and lower envelopes of the amplification factor. 
  Here we assume $M_3=10^8\,M_\odot$ and $a_o=100\,{\rm AU}$. The outer orbital phase is either $\phi_o=0$ (for standard lensing) or $\phi_o=\pi$ (for retro-lensing). For such $M_3=10^8\,M_\odot$, $2\pi f M_3=100$ correspond to a GW frequency of $f\simeq 0.03\,{\rm Hz}$.
  }
\label{fig:lens_mag_vs_iota}
\end{figure}

When the BBH is in front of the SMBH, its GW can still experience a retro-lensing with the wave bent by the SMBH by an angle of $\simeq \pi$. This is also known as the ``glory'' and is caused by the short-range attractive force of the SMBH. The geometry is illustrated in the bottom of Fig.~\ref{fig:lens_geometry}. 

We define $\Delta \phi = \phi_O - \phi_S + 2\pi$ such that $\Delta \phi \in [0, 2\pi)$, with $\phi_{O(S)}$ the azimuthal angle of the observer (source) in the projected plot shown in the bottom of Fig.~\ref{fig:lens_geometry}.  As we prove in Appx.~\ref{appx:geo_derivation}, we have the following geometrical relations when the observer, source, and lens are nearly aligned,
\begin{align}
    &\pi - \Delta \phi \simeq \frac{D_{OL}}{D_{LS}} \theta - \alpha, \label{eq:geo_relation_retro_1}\\
    &\tan \beta = \tan \theta - \frac{D_{LS}}{D_{OS}} \tan \alpha. 
    \label{eq:geo_relation_retro_2}
\end{align}

For a Schwarzschild lens, the closest impact a photon can make
is $b_{\rm ps}=3\sqrt{3}M_3$ for it to not be absorbed by the lens (see, e.g., Ref.~\cite{Luminet:79}). Light rays having an impact parameter $b$ slightly greater than $b_{\rm ps}$ may eventually reach the observer after making one or more turns around the SMBH. Infinitely many images thus form at~\cite{Eiroa:12}
\begin{equation}
    \theta_m = \theta_m^0\mp \zeta_m \Delta \alpha, 
\end{equation}
with
\begin{align}
    &\alpha = \pm \Delta \alpha \pm m\pi,\\
    &\theta_m^0 = \theta_{\rm ps} \left[1+e^{(c_2-m\pi)/c_1}\right],\\
    &\zeta_m = \frac{\theta_{\rm ps}}{c_1}e^{(c_2-m\pi)/c_1},
\end{align}
where $\theta_{\rm ps}=b_{\rm ps}/D_{OL}$, $m\in \mathbb{N}^+$, and $c_1$ and $c_2$ are constants determined by the metric. For the Schwarzschild metric, $c_1=1$ and $c_2=\ln\left[216 (7-4\sqrt{3})\right]-\pi\simeq-0.40$, leading to $\theta_{\rm ps}=5.35 M/D_{OL}$ and $\zeta_m=0.15 M/D_{OL} \exp\left[-(m-1)\pi\right]$~\cite{Eiroa:12}. 
For high alignment, the amplification of each image is given by~\cite{Bozza:10, Eiroa:12} 
\begin{align}
    \mu_{m\pi}(t) &= \left(\frac{D_{OS}}{D_{LS}}\right)^2\frac{\theta_m^0\zeta_m}{\sin\Delta \phi(t)} , \nonumber \\
    &\simeq\frac{0.80}{\sin \Delta \phi(t)} \left(\frac{M_3}{D_{LS}}\right)^2e^{-(m-1)\pi}, \nonumber 
    \\
    &\simeq \frac{0.80}{\beta(t)} \left(\frac{M_3^2}{D_{LS}D_{OL}}\right)e^{-(m-1)\pi}, 
    \label{eq:retro_mag_geo}
\end{align}
where in the second line we have plugged in the values for a Schwarzschild BH and used $D_{OS}\simeq D_{OL}$ in our case; in the third line we have expressed $\sin \Delta \phi$ in terms of $\beta$ using the geometrical relations Eqs.~(\ref{eq:geo_relation_retro_1}) and (\ref{eq:geo_relation_retro_2}). Although there are infinitely many images, the amplification decreases exponentially for large $m$. Thus, for the rest of the work we will focus only on the pair of $m=1$ images whose magnifications are denoted by $\mu_\pi$. 

While we derived retrolensing magnification in the geometric approximation, we can find the same result in wave scattering picture. 
Note that per unit time, the energy received by a detector with area $A_{\rm det}$ from a source with an isotropic luminosity $dE/dt$ is $A_{\rm det}/\left(4\pi D_{OS}^2\right) \left(dE/dt\right)$. At the same time, the source's emission may first reach the SMBH (lens) and then scatter towards the observer. The detector receives energy at a rate $1/\left(4\pi r_S^2\right)\left(d\sigma/d\Omega\right)(A_{\rm det}/D_{OL}^2)\left(dE/dt\right)$, where $r_S$ is the distance from the source to the lens with $r_{S}\simeq D_{LS}$ when the relative alignment is high, and $d\sigma/d\Omega$ is the cross section of the SMBH and it is further a function of the $\gamma(t)$, the angle of the outgoing rays with respect to the incoming ones. Geometrically, $\sin \gamma\simeq  (D_{OS}/D_{LS})\beta$. The magnification is thus
\begin{equation}
    \mu(t) = \left(\frac{D_{OS}}{D_{OL}D_{LS}}\right)^2 \left\{\frac{d\sigma}{d\Omega}\left[\gamma(t)\right]\right\}. 
    \label{eq:retro_mag_scat_def}
\end{equation}

The classical cross-section of a Schwarzschild SMBH is~\cite{Matzner:85} 
\begin{equation}
    \left\{\frac{d\sigma}{d\Omega}\left[\gamma(t)\right]\right\}_{\rm geo} =  \frac{b\left[\gamma(t)\right]\left[{db}/{d\gamma}(t)\right] }{\sin \gamma(t)}. 
    \label{eq:cross_sect_cl}
\end{equation}
where 
\begin{equation}
    b(\gamma)/M_3 \simeq 3\sqrt{3} + 3.48 \exp\left(-\gamma\right), \text{ when $\gamma\simeq\pi$}.
\end{equation}
We immediately see that magnification calculated using the classical cross section reduces to the one calculated in Eq.~(\ref{eq:retro_mag_geo}) under the geometrical limit. 

However, the classical/geometrical cross section assumes the scattering of a flow of particles and does not include effects of wave interference nor the spin of the wave. Ref.~\cite{Matzner:85} incorporated the wave effects (interference and the polarization) under a semi-classical approach and found that near glory, the cross section for each ray can be written as 
\begin{align}
    &\left\{\frac{d\sigma}{d\Omega}\left[\gamma(t)\right]\right\}_{\rm wave} = \frac{2\pi^2}{\lambda}b_g^2\left(\frac{db}{d\gamma}\right)
    J_{2s}^2\left[\frac{2\pi}{\lambda}b_g\sin \gamma(t)\right], \nonumber 
    \\
    &=84.65 M_3^2\left(fM_3\right) J_{2s}^2\left[33.62 f M_3\sin\gamma(t)\right]
    \label{eq:cross_sect_qu}
\end{align}
where $\lambda = 1/f=2\pi/\omega$ is the GW wavelength, $s$ is the spin of the scattered wave ($s=2$ for GW), $b_g$ is the impact parameter at the glory point ($b_g=3\sqrt{3}M$ for a Schwarzschild BH), and $J_{2s}$ is the Bessel function of order $2s$. In the second line, we have plugged in numerical values for a Schwarzschild BH. 
Note further that for sufficiently small $x$ and $s=2$, the Bessel function can be expanded as $J_{4}(x) \simeq x^4/384$. This means that the glory will be dark for a polarized wave while it has an infinite magnification under the geometrical limit. On the other hand, the location of the first peak of $J_4(x)$ is at $x\simeq 5.32$, which corresponds to 
\begin{equation}
    \sin \gamma \simeq \frac{D_{OS}}{D_{LS}}\beta\simeq 9.94 \times 10^{-3}\left(\frac{2\pi f M_3}{100}\right)^{-1}. 
    \label{eq:bessel_4_first_pk}
\end{equation}
Therefore, for GW at higher frequencies and/or for more massive lenses, the glory is dark for smaller range of alignments. In other words, the wave result [Eq.~(\ref{eq:cross_sect_qu})] approach better to the geometrical limit [Eq.~(\ref{eq:cross_sect_cl})] at greater values of $(2\pi f M_3)$ (i.e., shorter wavelength; see also Fig.~\ref{fig:lens_mag_vs_iota} which we will discuss shortly).

Once we have the magnification of each retro-lensed image $\mu_\pi$, we compute the magnification factor of the GW waveform $F(t; f)$ as [cf. Eq.~(\ref{eq:F_geo_limit_def})]
\begin{equation}
    F(t; f) = 1 + 2|\mu_\pi(t)|^{1/2}\exp\left[2\pi i f t_\pi(t)\right],
\end{equation}
where we simply approximate the time delay as $t_\pi(t)\simeq 2D_{LS}(t) + \pi b_g$~\cite{Bozza:04}. We do not further refine the solution because it does not affect the detectability of the glory. Because during the derivation process, we have made the assumption of high-alignment in various places, which would require $\Delta \alpha \simeq |\gamma-\pi|\simeq (D_{OS}/D_{LS})\beta \ll 1$ (see also, sec. 2.3 in Ref.~\cite{Eiroa:12}), we adopt here an \emph{ad hoc} cut and only apply the retro-lensing amplification when $(D_{OS}/D_{LS})\beta < \pi/12$. A more rigorous treatment of the magnification that is valid at arbitrary angles is deferred to future studies. Higher-order images are also ignored because their magnification drops exponentially with respect to the winding number of the SMBH. 

\subsection{Sample waveforms}
\label{sec:samp_waveforms}

In the top panel of Fig.~\ref{fig:lens_mag_vs_iota} we compare the magnifications of various images. Here $\mu_{1,2}$ are the magnifications corresponding to the primary and secondary images formed by the standard strong lensing under the weak-deflection limit [Eq.~(\ref{eq:std_lens_mag})]. The cyan curve, denoted by $\mu_\pi$, is the magnification of the retro-lensing under the geometrical limit [Eq.~(\ref{eq:retro_mag_geo}); it is also equivalent to the combination of  Eqs.~(\ref{eq:retro_mag_scat_def}) and (\ref{eq:cross_sect_cl})]. The glory magnification including wave effects [Eq.~(\ref{eq:cross_sect_qu})] are shown in the dashed-brown and dotted-purple traces for two different values of $2\pi f M_3=10$ and $100$, respectively. 
To generate the plot, we have assumed a lens with mass $M_3=10^8\,M_\odot$ and inclination $\iota_J=87^\circ$. We further set $\phi_o=0$ for the standard lensing under the weak-deflection limit and $\phi_o=\pi$ for the retro-lensing. Whereas the geometrical glory has an infinite magnification as $\iota_J\to 90^\circ$, the wave calculations lead to a vanishing glory at perfect alignment. On the other hand, at larger values of $2\pi f M_3$ (or effectively, shorter wavelengths), the wave result follows more closely the classical/geometrical value and the central dark spot has a smaller angular size [Eq.~(\ref{eq:bessel_4_first_pk})]. As a result, such systems will be more favorable for the detection of retro-lensing signatures. 

We also present the upper and lower envelopes of the amplification factor $|F|$ in the bottom panel in Fig.~\ref{fig:lens_mag_vs_iota}. Because we have $2\pi f\Delta t_l\sim w=8\pi f M_3 \gg 1$, the phase between different images changes rapidly, causing $|F|$ oscillates between the envelopes as the inner binary's frequency $f$ evolves. 



\begin{figure}[bt]
  \centering
  \includegraphics[width=0.9\columnwidth]{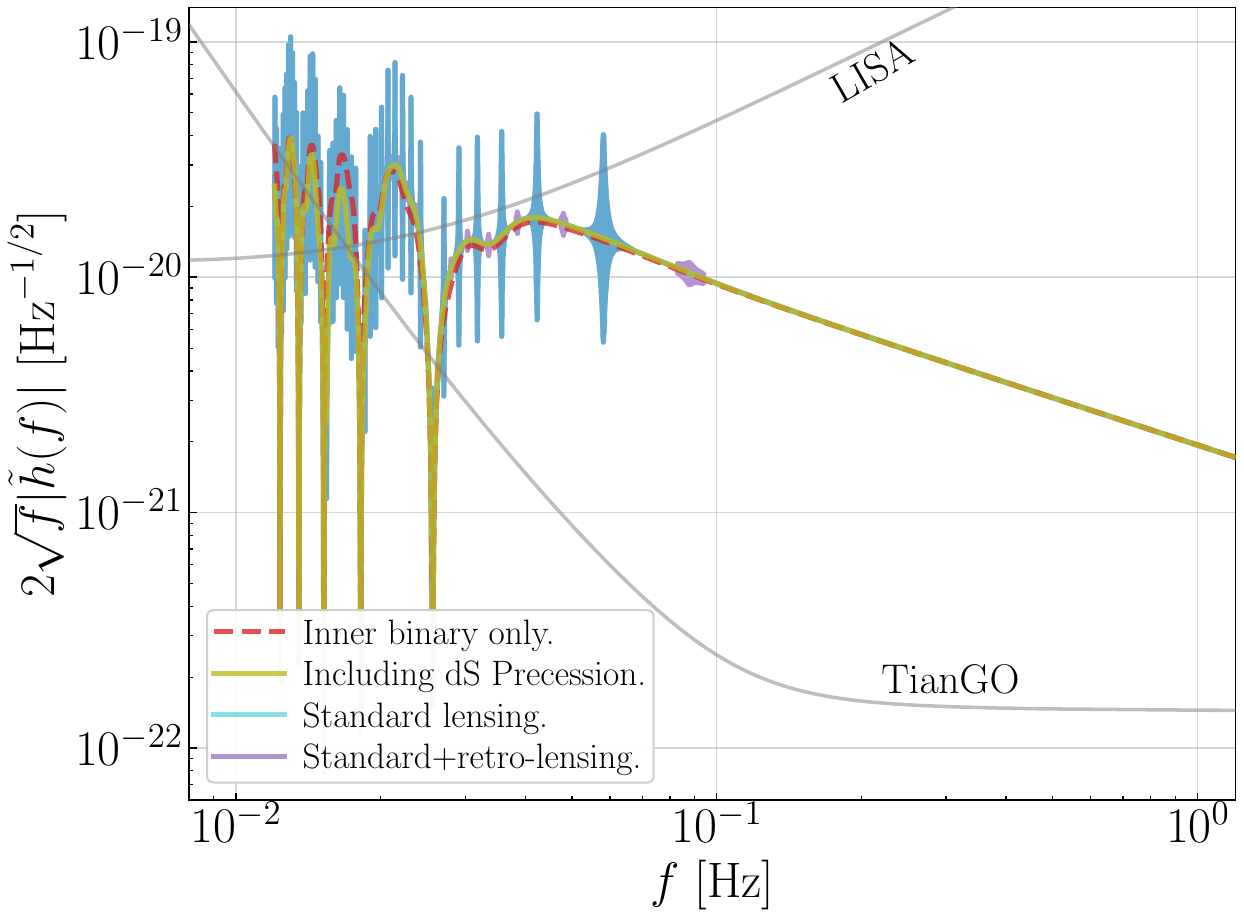}
  \caption{Sample waveforms including the lensing effects. Also shown in the grey traces are the proposed instrumental sensitivty of LISA~\cite{Robson:19} and TianGO~\cite{Kuns:19}. Here we assume $M_3=10^8\,M_\odot$, $a_o=100\,{\rm AU}\simeq 101 M_3$, and $\iota_J=87^\circ$. The outer orbit has a period of $P_o=0.10\,{\rm yr}$ and the inner orbit precesses with a period $P_{\rm dS}=6.8\,{\rm yr}$. Each time $\phi_o[t(f)]\simeq 0$ (source behind the lens), the standard lensing happens and is characterized by a sharp a cyan peak in the waveform. When $\phi_o[t(f)]\simeq\pi$ (source in front of the lens), we then have retro-lensing (glory), which is calculated including wave interference and polarization effects [Eq.~(\ref{eq:cross_sect_qu})]. Note that the starting frequency of each waveform is chosen so that the inner binary will merge in 5\,yr, which is also the fiducial duration of observation assumed in this study.}
\label{fig:samp_lensing_waveform}
\end{figure}

Putting different ingredients together, we show a sample waveform including various effects together in Fig.~\ref{fig:samp_lensing_waveform}. 
Here the dashed-red trace is the waveform for an isolated binary. When the SMBH is present, it induces a de Sitter-like precession of the inner binary's orbital plane~\cite{Yu:20d} and modifies its antenna response as shown in the solid-olive trace. Each time when the inner binary is behind the SMBH ($\phi_o\simeq 0$), it further experiences the standard strong lensing by the SMBH [Eq.~(\ref{eq:F_std_lens_full})], leading to the cyan peaks in the plot. The separation between two adjacent cyan peaks corresponds to the period of the outer orbit $P_o$ and 
the duration of each lensing event is approximately given by $0.5\mathcal{P}_{l}P_o$~\cite{DOrazio:20}, where 
\begin{align}
    \mathcal{P}_{l} &\simeq \frac{2}{\pi}\arcsin \left[\frac{D_{OL}\theta_{\rm Ein}(\phi_o=0)}{a_o}\right] \nonumber \\
    & \simeq 0.13 \left(\frac{a_o/M_3}{100}\right)^{-1/2},
    \label{eq:prob_std_lens}
\end{align}
is the geometrical probability of the inner binary to be significantly lensed by the SMBH (i.e.~the geometrical probability of $\eta=\beta/\theta_{\rm Ein}\leq 1$). 
When the inner binary is in front of the SMBH ($\phi_o\simeq \pi$), it is then retro-lensed by the SMBH, leading to the purple peaks in Fig.~\ref{fig:samp_lensing_waveform}. We have used Eq.~(\ref{eq:cross_sect_qu}) for the retro-lensing calculation to incorporate wave effects. 

To generate Fig.~\ref{fig:samp_lensing_waveform}, we assume $M_3=10^8\,M_\odot$, $a_o=100\,{\rm AU}$, leading to 
an outer orbit period of $P_o=0.10\,{\rm yr}$ and a de Sitter precession period of $P_{\rm dS}=6.8\,{\rm yr}$. We randomly choose $(\overline{\theta}_S, \overline{\phi}_S)=(33^\circ, 147^\circ)$ for the line of sight  in the solar frame. The orientation of the outer orbit $\uvect{L}_o$ in the solar frame (following the same notation as in  Ref.~\cite{Yu:20d}) is then set to $(\overline{\theta}_J, \overline{\phi}_J)=(120^\circ, 147.5^\circ)$ so that the outer orbit has an inclination $\iota_J=87^\circ$. We further set the phase of the outer orbit to be $\phi_o=-\pi/2$ at the merger of the inner binary as a conservative demonstration of the lensing effect, which essentially zeros the lensing signatures when the inner binary reaches the more sensitive ground-based GW detectors. The inner binary (signal carrier) has $M_1=M_2=50\,M_\odot$ and the initial frequency is set such that the binary merges in 5\,yr. The opening angle between $\uvect{L}_i$ and $\uvect{L}_o$ is $\lambda_L = \pi/4$.

\begin{figure}[bt]
  \centering
  \includegraphics[width=0.9\columnwidth]{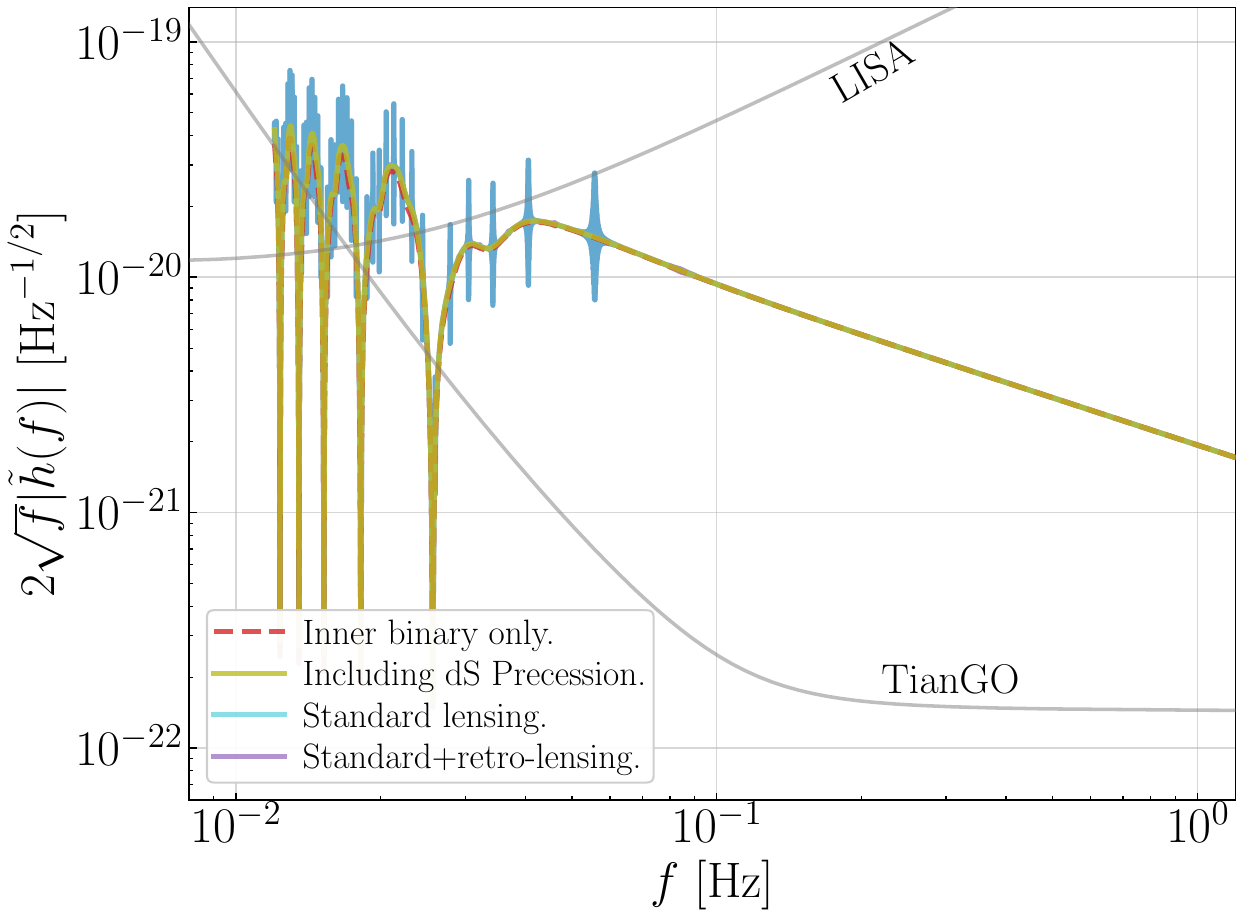}
  \caption{Similar to Fig.~\ref{fig:samp_lensing_waveform} but for $M_3=10^7\,M_\odot$, $a_o=50\,{\rm AU}\simeq 507 M_3$, corresponding to an outer orbital period of $P_o=0.11\,{\rm yr}$ and a dS precession period of $P_{\rm dS}=38\,{\rm yr}$. In this scenario only the standard lensing has a significant effect on the waveform.  }
\label{fig:samp_lensing_waveform_M3_1e7}
\end{figure}

As a comparison, we also consider the case where a less massive SMBH acts as the lens in Fig.~\ref{fig:samp_lensing_waveform_M3_1e7}. This time we set $M_3=10^7\,M_\odot$ and $a_o=50\,{\rm AU}$, leading to a similar outer orbital period of $P_o=0.11\,{\rm yr}$. The rest parameters are the same as in Fig.~\ref{fig:samp_lensing_waveform}. The standard lensing (cyan) still have easily visible features in the waveform, yet the retro-lensing can hardly be detectable. Indeed, for high alignment and at a fixed value of $P_o$, we have $\mu_{1,2}\propto M_3^{2/3}$ while $\mu_\pi\propto M_3^{5/3}$ [Eqs.~(\ref{eq:std_lens_mag}) and (\ref{eq:retro_mag_geo})], and therefore the detection of retro-lensing favors more massive SMBHs. 

\subsection{Parameter space for significant lensing}

\begin{figure}[bt]
  \centering
  \includegraphics[width=0.9\columnwidth]{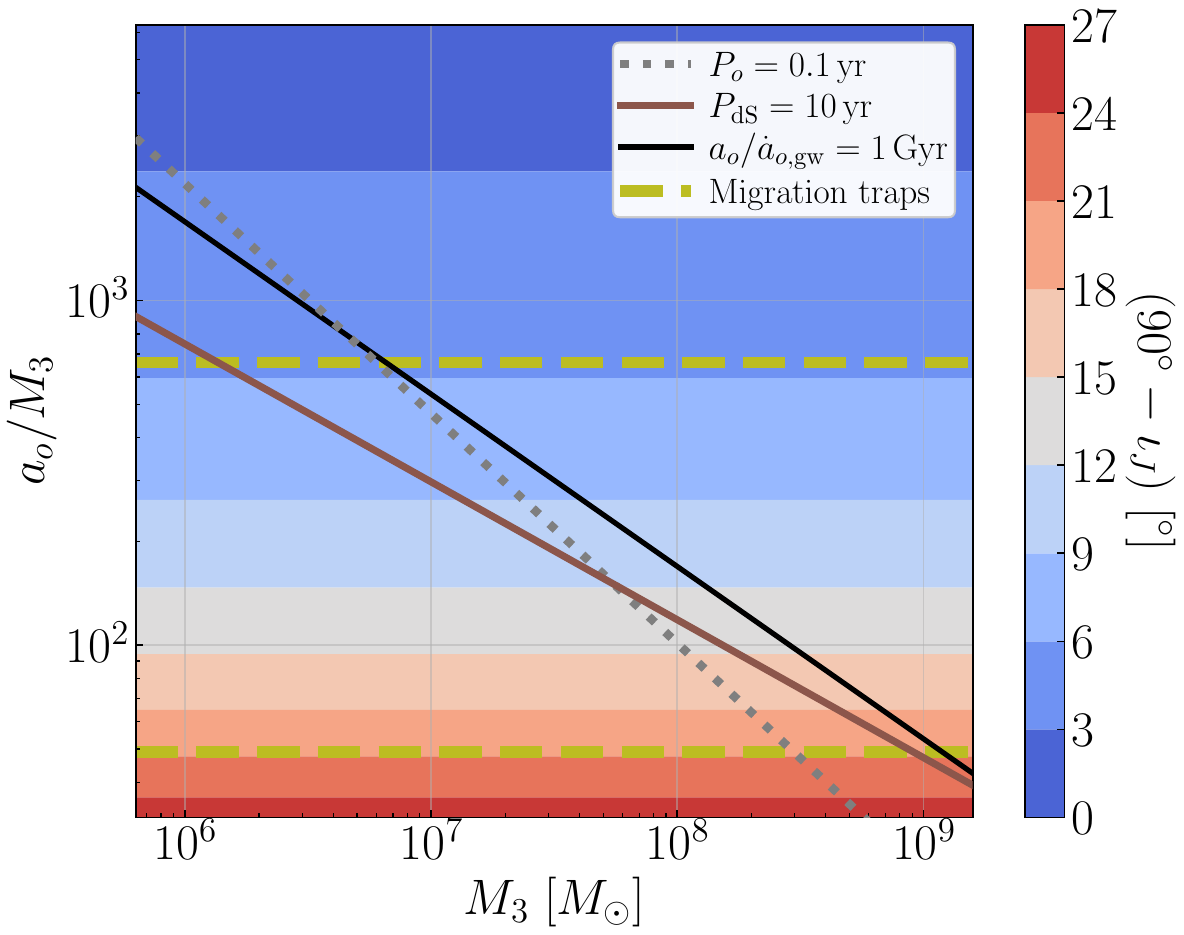}
  \caption{Threshold value of $(90^\circ-\iota_J)$ such that the secondary image in the standard-lensing case has a magnification of $|\mu_2|\geq 0.1$ [which corresponds to $\eta\leq 1.27$; Eq.~(\ref{eq:std_lens_mag})]. Also shown in the grey-dashed and brown-solid traces are lines corresponding to an outer orbital period of $P_o=0.1\,{\rm yr}$ and a dS precession period of $P_{\rm dS}=10\,{\rm yr}$. }
\label{fig:par_space_mu_m_0p1}
\end{figure}

\begin{figure}[bt]
  \centering
  \includegraphics[width=0.9\columnwidth]{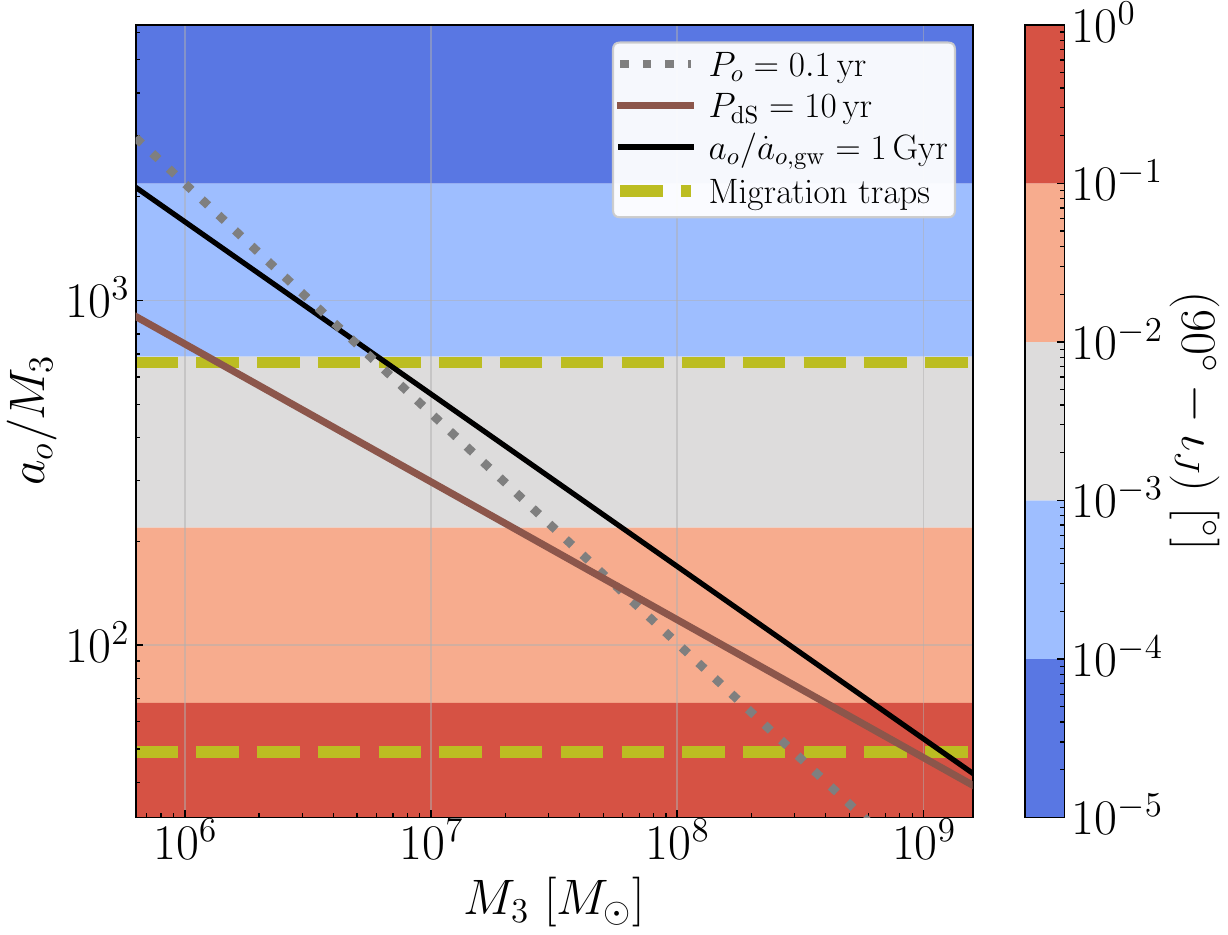}
  \caption{Threshold value of $(90^\circ-\iota_J)$ such that the magnification of the first retro-lensing image has $|\mu_\pi|\geq 0.1$. Note that in this case the color bar has a logarithm scale. }
\label{fig:par_space_mu_rel_cl_0p1}
\end{figure}

We can systematically examine the parameter space over which lensing is likely to be significant, as demonstrated in Figs.~\ref{fig:par_space_mu_m_0p1} and \ref{fig:par_space_mu_rel_cl_0p1}. In Fig.~\ref{fig:par_space_mu_m_0p1}, we consider the threshold $\iota_J$ required to make the magnification of the secondary image in the standard lensing scenario be $|\mu_2|\geq 0.1$ [Eq.~(\ref{eq:std_lens_mag})]. 
The upper envelope of the amplification factor $|F|\simeq 1+|\mu_2|^{1/2}\simeq 1.3$. 
Note this corresponds to $\eta\leq 1.27$, and thus the angle $90^\circ-\iota_J$ indicated by the color bar over $90^\circ$ is broadly consistent with the (repeated) lensing probability $\mathcal{P}_l$ defined in Eq.~(\ref{eq:prob_std_lens}). Also shown in the dashed-grey (or the solid-brown) trace is the line corresponding to the outer orbital period being $P_{o}=0.1\,{\rm yr}$ (or the period of the de Sitter-like precession of the inner orbital plane being $P_{\rm dS}=10\,{\rm yr}$). We see that along the line of $P_o=0.1\,{\rm yr}$, we might expect to see significant strong lensing over a range of $\sim 10^\circ$ for the outer inclination angle, meaning that the geometrical probability for the significant strong lensing to happen (repeatedly) could be $\sim 10\%$. 

To put this parameter space under astrophysical contexts, we also show in the solid-black line locations where the GW decay timescale of the outer orbit to be $a_o/\dot{a}_{o,{\rm gw}}=1\,{\rm Gyr}$, with $\dot{a}_{o,{\rm gw}}$ the decay rate of the outer orbit due to GW radiation and we have used $M_1+M_2=100\,M_\odot$. Along the line of $P_o= 0.1\, {\rm yr}$, the GW decay timescale of the outer orbit is typically between 10\,Myr to 1\,Gyr. This means that after the formation of  inner binary, it needs to be able to merge within $10\,{\rm Myr}$ in order for us to catch such a triple system. We will discuss this point more in Sec.~\ref{sec:conclusion}.
Furthermore, the locations of migration traps in AGN disks~\cite{Bellovary:16} are shown in the dotted-olive traces as a potential mechanism to form the inner binaries near the SMBH (see also the discussion in Ref.~\cite{DOrazio:20}). 

Similarly, we show in Fig.~\ref{fig:par_space_mu_rel_cl_0p1} the threshold inclination for the magnification of the retro-lensing (in the geometrical/classical limit) to be $|\mu_\pi|\geq0.1$ [Eq.~(\ref{eq:retro_mag_geo})]. For a massive lens with $M_3\gtrsim 10^8$, there is a $1\%$-level chance for the retro-lensing to be significant (under the geometrical limit). While rare, such effects are produced by the strong gravity field near the light ring of the SMBH and thus serve as valuable probes of gravity at a different regime than that probed by the standard (strong) lensing (see, e.g., \cite{Claudel:01}).

As a brief summary, we note a waveform including the Doppler shift, the de Sitter-like precession, and the gravitational lensing can be fully constructed with 13 parameters, $(\mathcal{M}, D, t_c, \phi_c, \overline{\theta}_S, \overline{\phi}_S, \overline{\theta}_J, \overline{\phi_J}, M_3, a_o, \lambda_L, \phi^{(0)}, \alpha_0)$. This is the same set of the parameters as used in Ref.~\cite{Yu:20d} because lensing does not introduce new unknown parameters (see the discussion below). Specifically, $(\mathcal{M}, D, t_c, \phi_c)$ are used to calculate the carrier signal [Eq.~(\ref{eq:hc})] under the quadrupole formula. One can also include corrections at higher post-Newtonian orders but they are critical near the final merger and are thus beyond the interest of this work focusing on the external modulation in the early inspiral stage. The line of sight direction $\uvect{N}$ is specified by the solar-frame coordinates $(\overline{\theta}_S, \overline{\phi}_S)$ and the orientation of the outer orbital angular momentum $\uvect{L}_o$ is given by $(\overline{\theta}_J, \overline{\phi_J})$. To determine the location of the inner binary in the outer orbit, we use $(M_3, a_o, \phi^{(0)})$ with $\phi^{(0)}$ a reference phase at $t=0$~\footnote{While using $(M_3, a_o)$ is conceptually simple, we nonetheless use $(M_3, \omega_o)$ when calculating the Fisher matrices in Sec.~\ref{sec:PE} as it is more numerically accurate.}. The orientation of the inner orbit can be further determined with an opening angle $\lambda_L = \arccos \left(\uvect{L}_i\cdot \uvect{L}_o\right)$ and a reference precession angle $\alpha_0$ at $t=0$. 

Before we proceed to following sections, we note that it is particularly interesting to combine gravitational lensing with the de Sitter-like precession of the inner orbital plane induced by the SMBH. Not only do the two effects share similar parameter space as shown in Fig.~\ref{fig:par_space_mu_m_0p1} (see also Refs.~\cite{DOrazio:20, Yu:20d}), but more importantly, combining the precession with the Doppler shift also determines \emph{all} the parameters entering the lensing calculation. Indeed, by measuring the frequency of the Doppler shift $\Omega_o = \sqrt{M_3/a_o^3}$ and the frequency of the de Sitter precession $\Omega_{\rm dS} = (3/2)(M_3/a_o)\Omega_o$~\cite{Will:18, Liu:19}\footnote{This expression assumes a circular outer orbit and $M_3\gg M_{1,(2)}$. When the outer orbit is elliptical, the eccentricity can be constrained from the Doppler shift~\cite{Yu:20d}. In this case, the instantaneous precession rate (see, e.g., Ref.~\cite{Barker:75}) should be used. }, we can determine the mass of the lens $M_3$ and the lens-source distance $a_o$~\cite{Yu:20d}. The outer orbital phase $\phi_o(t)$ can be measured from the Doppler phase shift. Lastly, as the inner orbital plane precesses around the outer orbit, we can further infer the orientation of the outer orbit and hence $\iota_J$ from the time evolution of the inner orbit's orientation. Consequently, lensing is a new effect to be incorporated to the study presented in Ref.~\cite{Yu:20d} without introducing new unknown parameters. It can thus be used to both enhance the PE uncertainty of the outer orbital parameters and test our understanding of the strong field gravity.

\section{Detectability of lensing signatures}
\label{sec:mismatch}

In this Section we examine the detectability of the lensing signatures by considering mismatches (to be defined below) of waveforms with and without the lensing effects. 

For this purpose, we first define the fitting factor (FF) between two waveforms as~\cite{Lindblom:08, Fang:19}, 
\begin{equation}
    {\rm FF}(h_1, h_2) = \frac{\langle h_1|h_2\rangle}{\sqrt{\langle h_1|h_1\rangle\langle h_2|h_2\rangle}}
\end{equation}
where 
\begin{equation}
    \langle h_1|h_2\rangle \equiv 2\int
    \frac{\tilde{h}^\ast_1(f)\tilde{h}_2(f) + \tilde{h}_1(f)\tilde{h}^\ast_2(f)}{S_n(f)} 
    df.
\end{equation}
We can then compute the mismatch $\epsilon$ 
\begin{equation}
    \epsilon = 1 - {\rm FF}. 
\end{equation}
There is a threshold mismatch, $\epsilon_{\rm th}$, given by
\begin{equation}
    \epsilon_{\rm th} = \frac{1}{\langle h_1|h_1\rangle + \langle h_2|h_2\rangle}
    \simeq \frac{1}{2\rho^2},
\end{equation}
where the second equality applies when $h_1\simeq h_2$ and $\rho$ is the signal-to-noise ratio (SNR) of the GW event. It is \emph{necessary} to have $\epsilon {>} \epsilon_{\rm th}\simeq 1/2\rho^2$ in order for the two waveforms to be distinguishable~\cite{Lindblom:08, Fang:19}. Note that the condition $\epsilon > \epsilon_{\rm th}$ is also equivalent to $\langle h_1-h_2|h_1-h_2\rangle {>} 1$. 

\begin{figure}[bt]
  \centering
  \includegraphics[width=0.9\columnwidth]{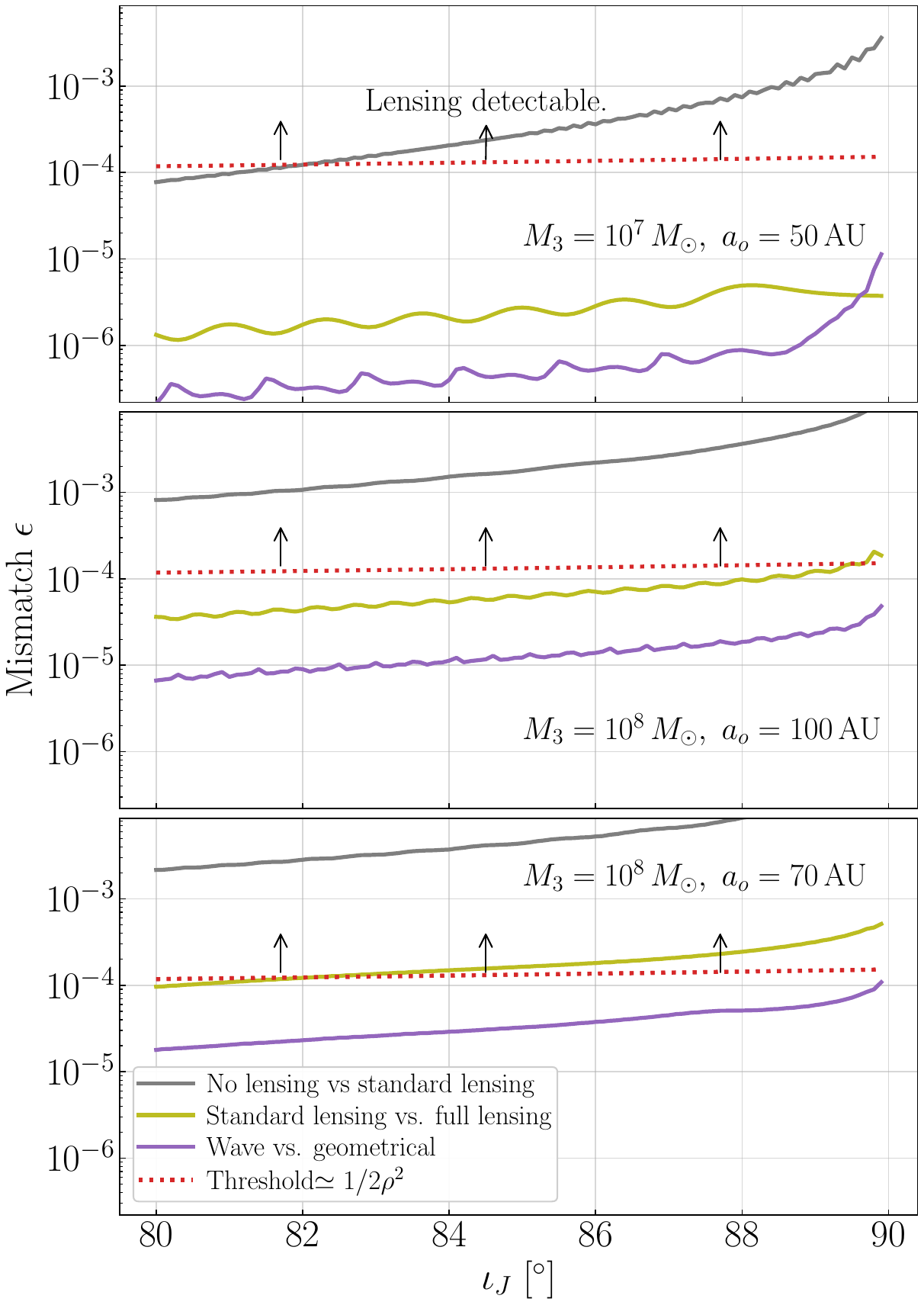}
  \caption{Mismatch $\epsilon$ of the waveform as a function of the inclination $\iota_J$ for three different outer orbital configurations, assuming a TianGO-like decihertz detector. In each panel, we use the grey trace to indicate the mismatch between a waveform without any lensing signatures with the one including the standard lensing (source behind the lens). The mismatches between  waveforms with standard lensing only and those further including retro-lensing effects are illustrated in the olive traces. The purple traces shows the mismatches between the lensing signatures calculated under the classical/geometrical limit (for both standard and retro lensing) and those incorporating wave effects.   
  The top and middle panels both have out orbital periods of $P_{o}\simeq 0.1\,{\rm yr}$ and the bottom panel has $P_{o}=0.06\,{\rm yr}$. For a relatively light lens with $M_3\lesssim 10^7\,M_\odot$, there is a decent chance of detecting the standard lensing. For more massive lenses $M_3\simeq 10^8\,M_\odot$, we might further detect the glory (retro-lensing). 
 Recall that we have $w=8\pi M_3 f = 12 \left(M_3/10^7\,M_\odot\right)\left(f/0.01\,{\rm Hz}\right)\gg 1$ and therefore the geometrical limit is typically a good approximation as indicated by the purple traces. 
  }
\label{fig:lens_ff}
\end{figure}

In Fig.~\ref{fig:lens_ff}, we compute the mismatch $\epsilon$ between different waveforms. The grey traces correspond to the mismatch between a waveform without any lensing signature and one including only the standard (strong) lensing [i.e., the source is only lensed when it is behind the lens or the SMBH and the deflection angle $\alpha\ll \pi$; Eq.~(\ref{eq:F_std_lens_full})]. The olive traces compare the waveforms with and without retro-lensing; wave effects are incorporated by computing the scattering cross-section using Eq.~(\ref{eq:cross_sect_qu}). They exhibit oscillatory features because the cross-section of retro-lensing is oscillatory (Fig.~\ref{fig:lens_mag_vs_iota}), yet the varying $\phi_o$ effectively allows different values of the scattering angle to be probed, which smooths out the oscillation. Lastly, in the purple traces we compare the mismatch between waveforms calculated under the geometrical limit and those include wave effects. More specifically, the geometrical waveforms are calculated using Eq.~(\ref{eq:F_std_lens_geo}) for the standard-lensing part and Eq.~(\ref{eq:cross_sect_cl}) for the retro-lensing part. The waveforms including wave effects are instead calculated using Eqs.~(\ref{eq:F_std_lens_full}) and (\ref{eq:cross_sect_qu}). We note further that the purple traces are in fact dominated by the contribution from retro-lensing [Eq.~(\ref{eq:cross_sect_cl}) vs. Eq.~(\ref{eq:cross_sect_qu}); see also Fig.~\ref{fig:lens_mag_vs_iota}]. The geometrical limit of the standard strong lensing typically provides a very good approximation to the full expression including wave effects as $2\pi f M_3\gg 1$ in our case.  

To generate the plot, we have assumed detection of the source with TianGO~\cite{Kuns:19}. The parameters we consider here are similar to the ones used in Fig.~\ref{fig:samp_lensing_waveform} except that we vary the orientation of $\uvect{L}_o$ by changing $\overline{\theta}_J$, which further varies the inclination $\iota_J$ between the line of sight and the outer orbit. 
Also shown in the plot as a comparison is the red-dotted line corresponding to $1/2\rho^2$ (assuming a waveform without lensing effects; though the SNR $\rho$ is generally similar with and without gravitational lensing).

In the three panels of Fig.~\ref{fig:lens_ff}, we consider three different combinations of $(M_3, a_o)$. From top to bottom, we have $(M_3/M_\odot, a_o/{\rm AU}){=}(10^7, 50),\ (10^8, 100), \ (10^8, 70)$, corresponding to $(P_o, P_{\rm dS}){=} (0.11, 38),\ (0.10, 6.8),\ (0.06, 2.8)\,{\rm yr}$. For the standard lensing, it might be detectable for a lens with $M_3=10^7\,M_\odot$ if $|\iota_J - 90^\circ|\lesssim 7.5^\circ$, which is nicely consistent with Fig.~\ref{fig:par_space_mu_m_0p1}. Along the line of fixed $P_{o}$, the lensing signature becomes more prominent as the mass of the lens $M_3$ increases. This is also shown in Fig.~\ref{fig:lens_ff} if we compare the middle panel with the top one. This indicates that for BBHs near massive SMBH with $M_3\sim 10^8\,M_\odot$, repeated strong lensing is indeed a critical component to be included in the waveform modeling. 


As for the retro-lensing, there is a small chance for it to be potentially detectable if the lens is sufficiently massive $M_3\gtrsim 10^8\,M_\odot$ and the outer orbit is compact with $P_o\lesssim 0.1\,{\rm yr}$. While a dark glory is expected for a polarized wave like GW at high alignment, for $M_3$ this massive, it is only dark for a very small range of angles [Eq.~(\ref{eq:bessel_4_first_pk})] and is further washed out by varying $\phi_o$. Indeed, for $2\pi f M_3\gtrsim 100$, the wave cross-section approaches the geometrical/classical value well (Fig.~\ref{fig:lens_mag_vs_iota}) and thus high alignment favors the detectability as shown in Fig.~\ref{fig:lens_ff}. On the other hand, distinguishing the GW diffraction signature from the classical one [Eq.~(\ref{eq:cross_sect_qu}) vs. Eq.~(\ref{eq:cross_sect_cl})] would be challenging given the sensitivity of TianGO, and it would rely on more sensitive detectors such as DECIGO~\cite{Kawamura:20} and/or the Big Bang Observer~\cite{Harry:06}.

Note specifically that we have kept the outer orbital phase to be $\phi_o=-\pi/2$ at the merger of the inner binary when we compute the mismatches. The values presented in Fig.~\ref{fig:lens_ff} are thus \emph{conservative} estimates (i.e. small mismatches) because in the frequency band of $f\gtrsim 0.1\,{\rm Hz}$ where TianGO is most sensitive (and this band includes the sensitivity band of ground-based GW observatories), the inner binary is far away from being affected by both the standard lensing and the retro-lensing. While the detectability could be enhanced if the inner binary happens to be lensed when $f\gtrsim0.1\,{\rm Hz}$, the inner binary spends only a small amount of time at this frequency band ($<1\,{\rm week}$ for the BBH we consider), therefore the probability is low [lower than the probability of repeated lensing by another factor of $\mathcal{P}_l$; Eq.~(\ref{eq:prob_std_lens})]. Consequently, we do not focus on the more optimistic case here. 

\begin{figure}[bt]
  \centering
  \includegraphics[width=0.9\columnwidth]{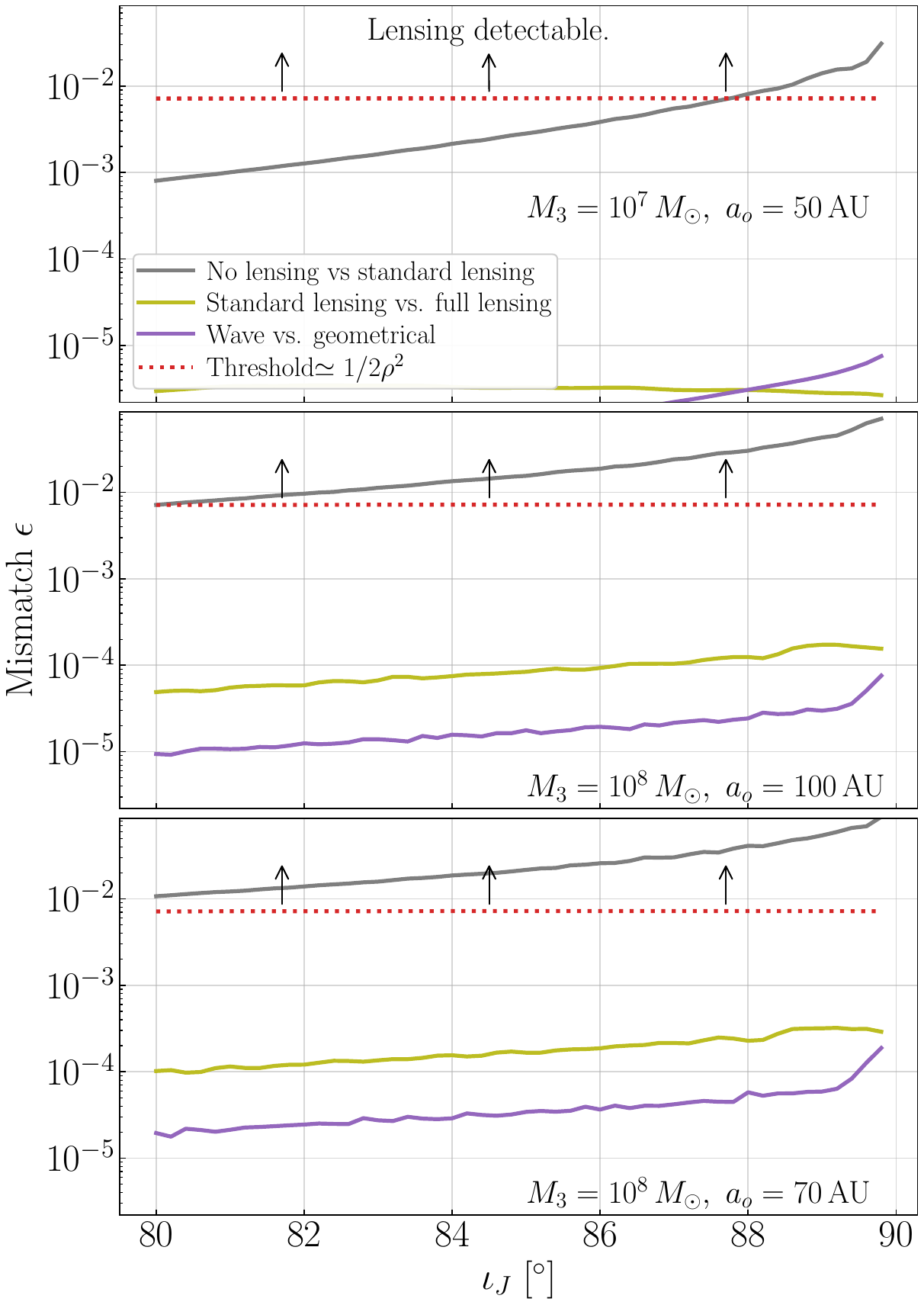}
  \caption{Similar to Fig.~\ref{fig:lens_ff} but for LISA. We also moved the source's luminosity distance from $1\,{\rm Gpc}$ to $200\,{\rm Mpc}$ so that the source's the SNR in LISA is greater than 8. For LISA, the standard lensing could have a decent detectability (a few to ten percent) while the retro-lensing are typically too weak to be detectable. }
\label{fig:lens_ff_LISA}
\end{figure}

The lensing effects could also be detected by detectors like LISA~\cite{Amaro-Seoane:17} (whose sensitivity is given by Ref.~\cite{Robson:19}) that are more sensitive in the millihertz band. The result is shown in Fig.~\ref{fig:lens_ff_LISA}. The parameters we assume are the same as in Fig.~\ref{fig:lens_ff} except for that we move the source's luminosity distance to $D=200\,{\rm Mpc}$ so that the SNR is greater than 8. The retro-lensing is too weak to be detectable with the sensitivity of LISA, yet the standard lensing may still have a decent chance to be detectable. For an outer orbital period of 0.1\,yr and a lens mass of $10^{7}\,M_\odot$ ($10^{8}\,M_\odot$), the probability for the standard lensing signature to be measurable is about $3\%$ ($10\%$). 

\section{PE accuracy including lensing effects} 
\label{sec:PE}

\subsection{Enhancing the PE accuracy of the SMBH properties}
\label{sec:PE_enhancement}
As we mentioned briefly at the end of Sec.~\ref{sec:samp_waveforms}, including the lensing effects does not introduce new free parameters to the waveform. Therefore, we naturally expect that including the lensing effects would enhance the PE accuracy of the SMBH properties compared to the results obtained in Ref.~\cite{Yu:20d} using orbital dynamics (Doppler shift and de Sitter precession) alone, as extra constraints are placed on the waveform. We examine this point quantitatively in this Section using the Fisher matrix approach. In particular, we write the waveform including lensing effects $\lh$ [Eq.~(\ref{eq:lensed_waveform})] in terms of 13 parameters as described in Sec.~\ref{sec:samp_waveforms}. We construct the Fisher matrix $\boldsymbol{\Gamma}$ whose elements are given by 
\begin{equation}
    \Gamma_{ij} = \left\langle \frac{\partial \lh}{\partial \theta_i} \Big{|} \frac{\partial \lh}{\partial \theta_j} \right\rangle,
\end{equation}
where $\theta_i$ is one of the 13 parameters. The PE error can then be obtained by inverting the Fisher matrix,
\begin{equation}
    \boldsymbol{\Sigma} = \left(\boldsymbol{\Gamma}\right)^{-1}.
\end{equation}
The diagonal element $\Sigma_{ii}$ corresponds to the statistical variance of parameter $\theta_i$, and the off-diagonal element $\Sigma_{ij}$ corresponds to the covariance between $\theta_i$ and $\theta_j$.

\begin{figure}[bt]
  \centering
  \includegraphics[width=0.9\columnwidth]{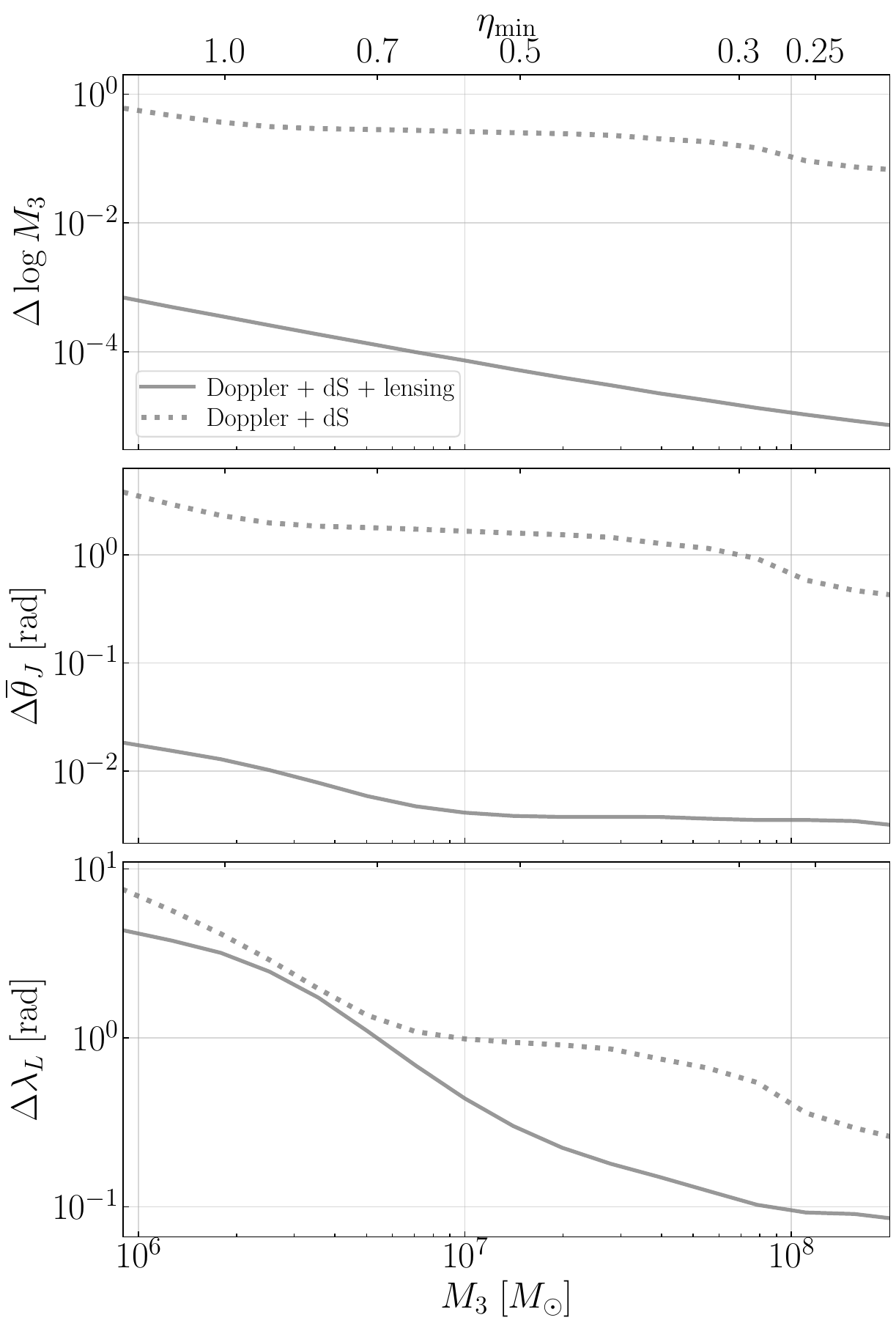}
  \caption{PE accuracy for systems with (solid traces) and without (dotted traces) the (standard) gravitational lensing, assuming detection by a single, TianGO-like detector. Here we fix $\iota_J=87^\circ$ and vary the mass of the central SMBH. The semi-major axis is chosen such that the outer orbital period is $P_o=0.1\,{\rm yr}$. For this $\iota_J$, we have $\epsilon > 1/2\rho^2$ when $M_3\gtrsim 1.7\times10^6$ or $\eta_{\rm min}\lesssim 1$. Both $\log M_3$ and $\overline{\theta}_J$ can be determined better by orders of magnitude. The determination of the opening angle $\Delta \lambda_L$ is also improved, especially when $P_{\rm dS}$ in the range of 10-45\,years. } 
\label{fig:PE_lens_vs_no_lens}
\end{figure}

In Fig.~\ref{fig:PE_lens_vs_no_lens}, we show the PE uncertainties of various parameters with (solid traces) and without (dotted traces) the lensing effects. Because retro-lensing is weak and the mismatch it induces is only marginally detectable (Sec.~\ref{sec:mismatch}), we thus ignore all the retro-lensing effects in the analysis here (and in Sec.~\ref{sec:PE_consistency}). The source is assumed to be detected by a TianGO-like decihertz observatory~\cite{Kuns:19}. We vary the mass of the SMBH $M_3$ and choose the outer orbit's semi-major axis $a_o$ such that the outer orbital period is fixed at $P_o=0.1\,{\rm yr}$. Other parameters are the same as in Fig.~\ref{fig:samp_lensing_waveform}, leading to $\iota_J=87^\circ$. Moreover, the outer orbital phase is fixed at $\phi_o=-\pi/2$ at the merger so that the BBH's signal is \emph{not} lensed in TianGO's most sensitive band, $f\gtrsim 0.1\,{\rm Hz}$ (corresponding to the last week of the inner BBH's inspiral). In the top x-axis, we also show the minimum value of $\eta_{\rm min}$ reached during the 5-year observation period. Note that we have restrict our discussion here to systems with $\eta_{\rm min}<1$, and the mismatches between waveforms with and without lensing satisfy $\epsilon > \epsilon_{\rm th}$. 

The top panel in Fig.~\ref{fig:PE_lens_vs_no_lens} shows the fractional error in mass of the SMBH. Without lensing (dotted trace), it can be constrained to $\Delta \log M_3=\mathcal{O}(10\%)$ from the periods of the outer orbit and the de Sitter precession, $P_o$ and $P_{\rm dS}$. If the inner BBH also experiences significant strong lensing by the SMBH (solid trace), then the PE uncertainty can be reduced by almost 3 orders of magnitude to $\Delta \log M_3=\mathcal{O}(10^{-4})$. 

It is worth to note that the error in $\Delta \log M_3$ is much smaller than the value obtained in Ref.~\cite{Takahashi:03} for \emph{static} lensing. The reason is detailed in Appx.~\ref{appx:PE_static_lens}. In brief, this is because the time delay between the primary and secondary images, $\Delta t_l\sim 8M_3\eta$ for $\eta\ll 2$ [or $2M_3\eta^2$ for $\eta\gg 2$; Eq.~(\ref{eq:std_lens_t_delay})], is the best measured quantity when lensing is static (i.e., $\eta$ stays as a constant during the observation period). As a result, $M_3$ and $\eta$ are highly correlated. Nonetheless, as $\eta$ varies due to the outer orbital motion, the waveform effectively samples different values of $\Delta t_l$. This thus breaks the degeneracy between $M_3$ and $\eta$ and allows $M_3$ to be determined to a much better accuracy than in the case of static lensing studied in Ref.~\cite{Takahashi:03}. 

Similarly, because most constraints are from combining information at different values of $\Delta t_l$ instead of from a single instance (e.g., at $\phi_o=0$), the results shown in Fig.~\ref{fig:PE_lens_vs_no_lens} does not depend sensitively on the value of $\iota_J$ as long as the lensing is detectable (Fig.~\ref{fig:lens_ff}) so that the formalism of Fisher matrix applies.

Once we have the uncertainty in $\log M_3$, the outer orbit's semi-major axis typically has an error $\Delta \log a_o\simeq \Delta \log M_3/3$, as the outer orbital period can be accurately determined by the Doppler shift~\cite{Yu:20d}. Thus, the lensing signatures can also help constraining $\iota_J$ from the instantaneous values of $\eta$, which further leads to a better determination of the orientation of the outer orbit's angular momentum $\uvect{L}_o$ (as $\uvect{N}$ can be measured from the motion of the detector around the Sun; see Refs.~\cite{Cutler:98, Kuns:19}). This point is illustrated in the middle panel in Fig.~\ref{fig:PE_lens_vs_no_lens} where the error in $\overline{\theta}_J$ is shown. 

As $\uvect{L}_o$ is the axis around which the inner orbital plane (i.e., $\uvect{L}_i$) precesses, a better determined $\uvect{L}_o$ also enhances the detectability of the precession signature. We demonstrate this in the bottom panel of Fig.~\ref{fig:PE_lens_vs_no_lens}. One may argue that the result of the Fisher matrix is self-contained only if $\Delta \lambda_L < \lambda_L$~\cite{Yu:20d}. Without lensing, this condition is not satisfied until $M_3 \gtrsim 3\times 10^7\,M_\odot$ or $P_{\rm dS}\lesssim 15\,{\rm yr}$ for the sources we consider here. On the other hand, this condition can be satisfied for less massive SMBHs with $M_3\gtrsim 6\times10^6\,M_\odot$ (corresponding to $P_{\rm dS}\lesssim 44\,{\rm yr}$) if the inner binary is also lensed by the SMBH. In other words, lensing effects help enhancing the detectability of the de Sitter precession of the inner binary and allow it to be measurable at a $P_{\rm dS}$ about 3 times greater than the one without lensing.

\subsection{Consistency tests}
\label{sec:PE_consistency}
In principle, the lensing does not introduce any new free parameter. We can nonetheless introduce an ad hoc parameter $\kappa$, defined via
\begin{equation}
    M_l = \kappa M_3,
    \label{eq:kappa_def}
\end{equation}
where $M_3$ is the mass of the SMBH for evaluating the orbital dynamics (Doppler phase shift and dS precession) and $M_l$ is the SMBH mass determining the lensing. In other words, the parameter $\kappa$ serves as an indicator of the consistency between the two effects. 
Nominally $\kappa=1$ and the mass of the SMBH creating the lensing is the same as that affecting the orbital dynamics. On the other hand, deviation may exist due to theoretical approximations made when constructing the waveform. After more careful waveform modeling, $\kappa$ can be further used to test the general theory of relativity, as the orbital dynamics and the lensing effects are induced by gravity at different regions around the SMBH. 
This is similar to how one may constrain deviations from general relativity using the Shapiro time delay~\cite{Will:03, Will:14}.
It is thus interesting to ask the question of how well we can measure the deviation of $\kappa$ from unity. 

\begin{figure}[bt]
  \centering
  \includegraphics[width=0.9\columnwidth]{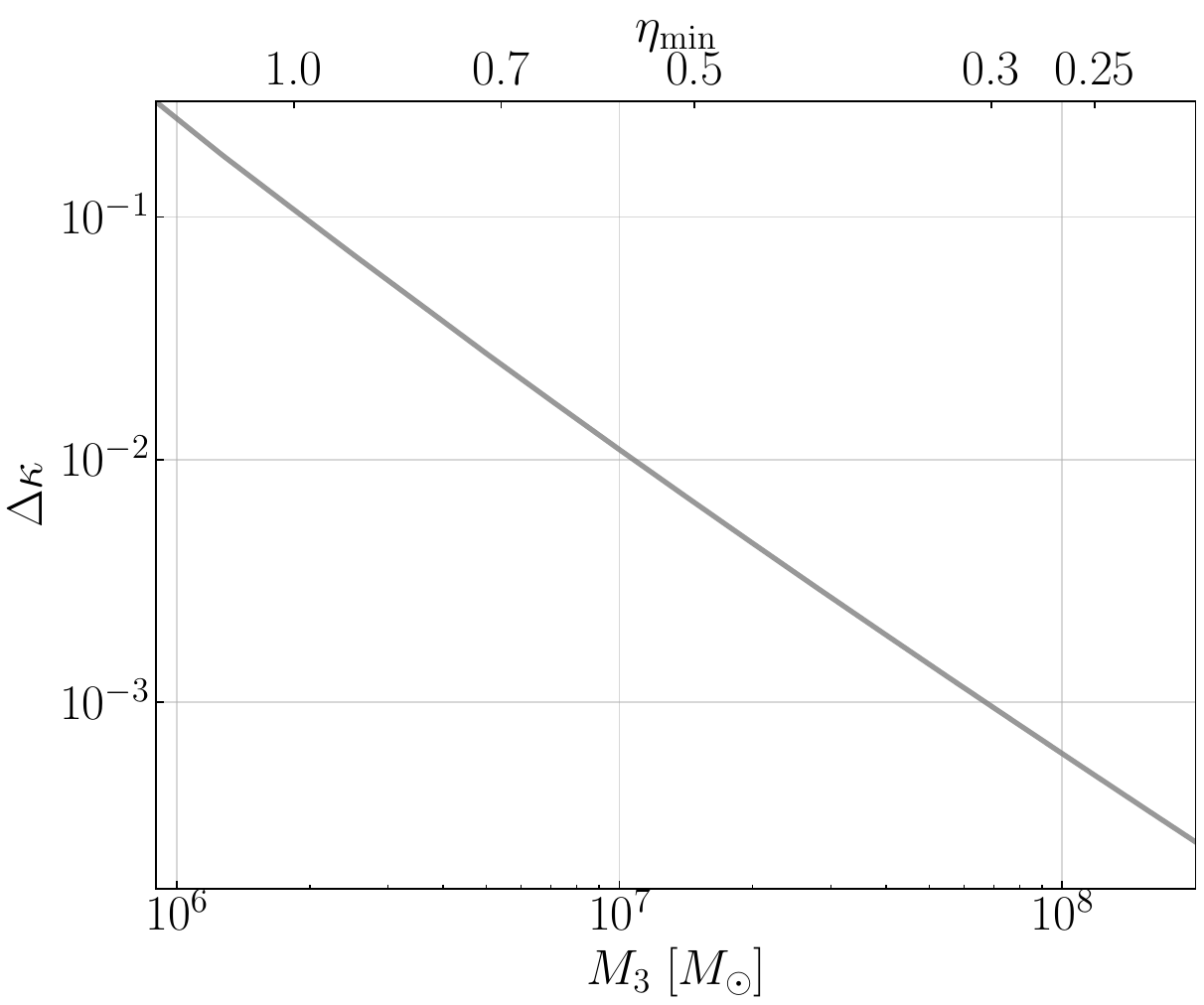}
  \caption{PE accuracy for $\kappa$ where $\kappa = M_l/M_3$ [Eq.~(\ref{eq:kappa_def})]. The source's orientation and sky location are the same as in Fig.~\ref{fig:PE_lens_vs_no_lens}. The error in $\kappa$ decreases roughly linearly with $M_3$ and it can be constrained to a $1\%$ accuracy for $M_3\gtrsim 10^7\,M_\odot$.  } 
\label{fig:PE_kappa}
\end{figure}

In Fig.~\ref{fig:PE_kappa} we show the statistical uncertainty on $\kappa$ as a function of $M_3$. The source orientation and sky location is the same as in Fig.~\ref{fig:PE_lens_vs_no_lens}. We note the error in $\kappa$ decreases roughly as $M_3^{-4/3}$ and it can be constrained to a $1\%$ accuracy for $M_3\gtrsim 10^7\,M_\odot$. 

One might understand the scaling of $\Delta \kappa$ as the following.
How well we can measure $\Delta \kappa $ depends on how well we can measure the mass of the SMBH from the precession frequency $\Omega_{\rm dS}$. Ignoring covariance with different angles, we approximately have $\Delta \log M_3 \propto \Delta \log \Omega_{\rm dS} \propto \Omega_{\rm dS}^{-2}$~\cite{Apostolatos:94, Yu:20d}. If we hold the outer orbital frequency constant, we thus have $\Omega_{\rm dS}\propto M_3^{2/3}$. Consequently, we approximately have $\Delta \kappa \propto M_3^{-4/3}$ as shown in Fig.~\ref{fig:PE_kappa}.

Meanwhile, it is interesting to note that the uncertainty in $\kappa$ is smaller than the numerically obtained uncertainty in $M_3$ when we include only the Doppler shift and the de Sitter precession in the waveform (the dotted trace in the upper panel in Fig.~\ref{fig:PE_lens_vs_no_lens}). This is because the lensing signature still helps constraining the orientation of the outer orbit $\iota_J$ in a way that cannot be mimicked by a rescaling of the mass $M_l$ (this is why we can simultaneously determine $M_l$ and $\eta$ in the static lensing case; Ref.~\cite{Takahashi:03} and Appx.~\ref{appx:PE_static_lens}). A better constrained $\iota_J$ means a better determination of parameters such as $\overline{\theta}_J$ and $\lambda_L$ that are partially degenerate with $M_3$ in the no-lensing case. As a result, the PE accuracy on $M_3$ can still be improved even we do not directly use it to evaluate the lensing.

\section{Conclusion and Discussions}
\label{sec:conclusion}
We studied the GW waveform emitted by a stellar-mass BBH in the vicinity of an SMBH, including effects such as the Doppler phase shift due to the outer orbital motion, the de Sitter-like of $\uvect{L}_i$ around $\uvect{L}_o$, and (repeated) gravitational lensing caused by the SMBH. 

For lensing, we considered not only the standard strong lensing which happens when the source is behind the lens (Sec.~\ref{sec:std_lens}), but also retro-lensing when the source is in front of the lens (Sec.~\ref{sec:retro_lens}). 

We then examined the detectability of various lensing effects by considering the mismatches they induce (Sec.~{\ref{sec:mismatch}}). For a lens with a mass of $M_3=10^{7}\,M_\odot$ and an outer orbital period of $P_o = 0.1\,{\rm yr}$, there is a $\sim 3\%$ ($\sim 10\%$) chance for the strong lensing to be detectable by LISA~\cite{Amaro-Seoane:17} (TianGO~\cite{Kuns:19}), and this probability increases with increasing $M_3$ if $P_o$ is held constant. For massive lens with $M_3\gtrsim 10^8\,M_\odot$ and compact outer orbits with $P_o\lesssim 0.1\,{\rm yr}$, there is also a small probability for the retro-lensing to be detectable for TianGO. The retro-lensing calculated with a classical cross-section [Eq.~(\ref{eq:cross_sect_cl})] is typically accurate enough for a source at $\simeq 1\,{\rm Gpc}$  given the sensitivity of TianGO. Effects of wave interference and polarization in the glory scattering [Eq.~(\ref{eq:cross_sect_qu})] might be measurable if the source is at a closer distance ($\lesssim 300\,{\rm Mpc}$) or the detector is more sensitive (e.g., DECIGO~\cite{Kawamura:20} and/or the Big Bang Observer~\cite{Harry:06}).  
On the other hand, the geometrical limit of the strong lensing [Eq.~(\ref{eq:F_std_lens_geo})] typically provides a good approximation to the full expression, Eq.~(\ref{eq:F_std_lens_full}). This is because we have $w=8\pi f M_3\gg 1$, and wave effects would show up only if we have $\eta < 1/w \ll 1$. Such an almost exact alignment is unlikely and therefore is not considered as the main case in our current study (but see, e.g., Ref.~\cite{Zhang:18}, for discussions on lensing when an exact alignment happens).

Because including the lensing does not introduce any new free parameters than what have already been used to incorporate the outer orbital dynamics, it greatly reduces the PE uncertainties, especially the statistical errors of the SMBH properties (Sec.~\ref{sec:PE_enhancement}). In fact, the error in $M_3$ can be better than in the case of static lensing, because the varying outer orbit breaks the degeneracy between $M_3$ and the source's angular position on the sky $\eta$. Furthermore, for strongly lensed BBHs, the de Sitter precession can also be detected at a longer period. Lastly, we indicated in Sec.~\ref{sec:PE_consistency} that since the mass of the SMBH can be separately inferred from the outer orbital dynamics and from the lensing effects, comparing the two inferences can thus serve as a way to test the consistency of the theoretical modeling behind each effect, and eventually, a way to test the general theory of relativity. 

We note that as our main goal here is to consider the detectability of various lensing effects and estimate their impacts on the PE, we only adopted the lowest-order approximation for each effect in our waveform construction. More careful treatments are needed by future studies if we want to build waveforms that are accurate enough to serve as detection templates. 

For example, we assumed that both the inner and the outer orbits are circular in our study for simplicity. However, eccentricities in both orbits may be expected especially if the inner binary is formed via dynamical channels (see, e.g., Refs.~\cite{OLeary:09, Antonini:12, Antonini:16, VanLandingham:16, Petrovich:17, Leigh:18, Chen:18, Fragione:19, Han:19, Samsing:20}). The qualitative effects of eccentricity have been argued in Ref.~\cite{Yu:20d}. An eccentric outer orbit might potentially enhance the PE accuracy as it reduces the period of precession. On the other hand, an eccentric inner orbit decreases the inner binary's merger time, giving the precession less time to accumulate its effect. Furthermore, if the inner binary's merger time is shorter than the outer orbital period, it would decrease the probability for lensing to happen~\cite{DOrazio:20}.   It would be crucial to properly incorporate them in future waveform studies to quantitatively understand the role of eccentricity.

The spin of the SMBH is also a critical component to be incorporated in future studies. Throughout the analysis, we have assumed that the SMBH is a non-spinning Schwarzschild BH for simplicity, whereas astrophsical SMBHs may have significant spin~\cite{Reynolds:13}. While the Lense-Thirring precession has a longer period than the de Sitter precession for sources we consider here~\cite{Yu:20d}, it has nonetheless been shown to have potentially significant role in modulating the orientation of the inner BBH~\cite{Liu:19, Fang:19, Liu:21}. Besides affecting the orbital dynamics, the spin may also modify the lensing signatures~\cite{Bozza:03, Bozza:05, Bozza:06} Therefore, similar to testing the consistency in the SMBH's mass (Sec.~\ref{sec:PE_consistency}), one may further check the consistency in the spin of the SMBH by comparing its value inferred from the Lense-Thirring effect in the orbital dynamics and that from gravitational lensing. This may serve as yet another way of testing general relativity. 

Moreover, Ref.~\cite{Torres:20} recently suggested that fast Doppler motion can further cause aberration in GW rays and is another ingredient to be added in the future. For inner binaries that are even closer to the SMBH than what we considered here, the quasinormal modes of the SMBH might be further excited~\cite{Cardoso:21}. If the inner binary is observed at a lower frequency with a GW decay timescale much longer than the duration of the observation, gaseous effect might also play a role~\cite{Chen:20, Toubiana:21} together with Lidov-Kozai oscillations~\cite{Hoang:19, Deme:20}. 


The astrophysical formation of such a hierarchical triple system is another topic that requires further dedicated studies. Because of complicated environmental effects in galactic nuclei, there are a few potential limiting requirements the triple system needs to satisfy. For example, the inner binary needs to merge efficiently before the outer orbit decays due to GW radiation (black traces in Figs.~\ref{fig:par_space_mu_m_0p1} and \ref{fig:par_space_mu_rel_cl_0p1}). The inner binary also needs to be able to survive evaporation due to dynamical
interactions with environmental stars on a timescale of typically a few Myr~\cite{Antonini:12, Yu:20a}. These conditions could be satisfied by both the gaseous channel~\cite{Bartos:17} and the dynamical channels~\cite{Samsing:20}, though the inner binary produced by some dynamical channels may have too high an eccentricity that it will merge before it orbits the SMBH by a complete cycle, disfavoring the detectability of both lensing and precession. 
In fact, we note the discussion on the astrophysical population in Ref.~\cite{DOrazio:20} applies here as well. This suggests that migration traps in AGN disks~\cite{Bellovary:16} would be particularly promising places to produce sources of interest to our study here (as also indicated in Fig.~\ref{fig:par_space_mu_m_0p1}). Another possibility is the tidal capture of a binary by the SMBH as suggested in Ref.~\cite{Chen:18}. 
More careful examination of these channels and other candidates as well as the distributions of the inner and outer orbital parameters they can produce will be of great value.

\section*{ACKNOWLEDGMENTS}
We thank Kumar Shwetketu Virbhadra, Xilong Fan, and the anonymous referee for useful comments to this work.
H.Y.\ acknowledges the support of the Sherman Fairchild
Foundation. 
B.S.\ acknowledges support by the National Science Foundation Graduate Research Fellowship under Grant No.~DGE‐1745301. Research of Y.W.\ and Y.C.\ are supported by the Simons Foundation  (Award Number 568762), the Brinson Foundation, and the National Science Foundation (Grants 
PHY--2011968, PHY--2011961 and PHY--1836809).

\appendix
\section{SPA for waveforms with fast modulation}
\label{appx:spa_fast_mod}
Suppose we can write the time-domain waveform as 
\begin{equation}
    h(t) = \Lambda(t) h_C(t) \equiv \Lambda(t) A(t) \cos\Phi(t),
\end{equation}
where $A(t)$ and $\Phi(t)$ are the amplitude and phase of the carrier waveform, and $\Lambda(t)$ is an external modulation factor induced by e.g., the precession of the source/detector plane and/or the time-variation of the lensing configuration. Here we assume that the external modulation $\Lambda(t)$ may have a fast temporal variation rate compared to the variation rate of intrinsic amplitude of the carrier, $|d\ln \Lambda(t)/dt| > |d\ln A(t)/dt|$, [but $\Lambda(t)$ still varies on a timescale longer than the typical SPA duration $t_{\rm SPA}$ which we define below]. We can thus improve the accuracy of the waveform by including more expansion terms than the lowest-order SPA does.

Specifically, we have (for $f\geq 0$)
\begin{align}
    \tilde{h}(f) &= \int h(t') e^{2\pi i f t'} dt', \\ \nonumber 
    &\simeq\frac{1}{2}\int \Lambda(t') A(t') e^{-i[\Phi(t') - 2\pi f t']} dt', \\ \nonumber  
    &\simeq\frac{1}{2} \int \left\{ \Lambda A + \left[\frac{d\Lambda}{dt} A + \Lambda\frac{dA}{dt}\right](t'-t)\right. \nonumber \\
    & \quad +\left.\left[\frac{1}{2}\frac{d^2\Lambda}{dt^2}A + \frac{d\Lambda }{dt}\frac{dA}{dt} + \frac{1}{2}\Lambda\frac{d^2A}{dt^2}\right](t'-t)^2 \right\} \nonumber \\
    & \times \exp\left\{-i\left[\Phi - 2\pi f t + \pi \frac{df}{dt}(t'-t)^2  \right]\right\} dt',
    \label{eq:h_f_spa_expansion}
\end{align}
where in the second line, we have dropped the fast oscillating term and in the third line, we have expanded all the time-dependent quantities around a time $t$ when $\left(d\Phi/dt\right)(t)=2\pi f$. For conciseness, when a time-dependent quantity is evaluated at $t$, we dropped its argument by writing $\Lambda\equiv \Lambda(t)$ and similarly for other quantities.  

If we ignore all the time derivatives on the amplitude terms as $|d\ln A /dt|<|d\ln \Lambda /dt|\ll f$ and $|d^2 \ln A/dt^2|<|d^2 \ln \Lambda(t)/dt^2|<f^2$, then we arrive at the standard (lowest-order) SPA approximation ($f\geq 0$)
\begin{align}
    \tilde{h}^{(0)}(f) &= \frac{1}{2}\frac{\Lambda A}{\sqrt{df/dt}} \exp\left[i(2\pi f t - \Phi - \pi/4)\right], \nonumber \\
    &\equiv\frac{1}{2}\Lambda \tilde{h}_C(f).
\end{align}
Following the covention in Ref.~\cite{Apostolatos:94}, we have defined the terms excluding $\Lambda /2 $ as $\tilde{h}_C(f)$. 

The lowest-order SPA is an excellent approximation if the amplitudes stay as constants over the duration when the wave oscillates at frequency $f$. The characteristic duration is given by 
\begin{align}
    t_{\rm SPA} &= \int \exp\left[-\pi \frac{df}{dt} (t-t')^2 \right] dt',\nonumber \\
    &=\sqrt{\frac{1}{df/dt}} = \sqrt{\frac{1}{3}P_i t_{\rm gw}}, 
\end{align}
where $P_i$ is the inner orbital period and $t_{\rm gw} = a_i/|da_i/dt|$ is the instantaneous GW decay timescale. In other words, the wave stays at frequency $f$ for a time given by the duration of the geometrical mean of the orbital period and the GW decay timescale. For inner binary's at $P_{i}\sim 100\,s$ and $t_{\rm gw}\sim 3\,{\rm yr}$, we have $t_{\rm SPA} \sim 0.6\,{\rm day}$ and it decreases as $f^{-11/6}$ as the inner binary evolves to higher frequencies. 

When the amplitude may change by a non-negligible amount during $t_{\rm SPA}$, we can improve the accuracy of the SPA waveform by including derivatives on the amplitude variations. Note the terms $\propto (t'-t)$ in Eq.~(\ref{eq:h_f_spa_expansion}) vanishes because the integrant is odd around $t$. Including the terms $\propto (t'-t)^2$, we have 
\begin{align}
    &\tilde{h}^{(1)}(f) = \frac{1}{2}\Lambda \tilde{h}_C(f) \nonumber \\
    &+ \frac{1}{4\pi i} \frac{\tilde{h}_C(f)}{df/dt}
    \left[
    \frac{1}{2}\frac{d^2\Lambda}{dt^2} + \frac{1}{A}\frac{d\Lambda}{dt}\frac{dA}{dt} + \frac{1}{2}\frac{\Lambda}{A}\frac{d^2A}{dt^2}
    \right].
\end{align}
If we further note 
\begin{equation}
    \frac{\tilde{h}_C}{Adf/dt}\frac{dA}{dt} = \left(\frac{d\tilde{h}_C}{df} - 2\pi i t \tilde{h}_C\right),
\end{equation}
our expression reduces to eq. (38) in Ref.~\cite{Apostolatos:94} when we ignore the terms involving second order time derivatives. Nonetheless, the $d^2\Lambda(t)/dt^2$ and the $d^2 A/dt^2$ terms also come at the $(t'-t)^2$ order and are thus left in the expression. The $d^2\Lambda(t)/dt^2$ can be particularly important because $\Lambda(t)$ can be fast varying. Indeed, Ref.~\cite{DOrazio:20} argued that duration of a strong-lensing event is approximately $0.5 \mathcal{P}_l P_o$, where $\mathcal{P}_l\sim 0.1$ is the geometrical probability for the strong-lensing to happen and $P_o\sim 0.1\,{\rm yr}$ is the period of the outer orbit.  

\section{Geometrical derivation of the lensing equation}
\label{appx:geo_derivation}

We derive here the geometrical relations shown in Eqs.~(\ref{eq:geo_relation_retro_1}) and (\ref{eq:geo_relation_retro_2}) for the retro lensing scenario (lower part of Fig.~\ref{fig:lens_geometry}). 

First, note that all the angles around $O$ sum to $2\pi$, we immediately arrive at the Ohanian lens equation
\begin{equation}
    \Delta \phi - \pi = \alpha - \theta - \theta_{S}. 
\end{equation}

We further have
\begin{align}
    D_{OS}\tan\beta &= D_{LS} \tan\left(2\pi-\Delta \phi\right),\nonumber \\
    &=D_{LS} \left\{\tan\left[(\pi-\alpha) +\theta +\theta_S\right]\right\}\nonumber \\
    &\simeq D_{LS}\left[-\tan(\alpha) + \tan \theta + \tan \theta_S\right].
\end{align}

Note further that the impact parameter can be written as 
\begin{equation}
    b = D_{OS}\theta = D_{LS}\theta_S. 
\end{equation}

Therefore, in our case we have $\theta \ll \theta_S {\simeq} D_{OS}/D_{LS}\theta (\ll 1)$. We thus arrive at Eqs.~(\ref{eq:geo_relation_retro_1}) and (\ref{eq:geo_relation_retro_2}) presented in the main text. 

\section{Understanding the improvement in $\Delta \log M_3$ due to repeated lensing}
\label{appx:PE_static_lens}

In this Appendix we explain why our PE uncertainty in $\Delta \log M_3$ (top panel in Fig.~\ref{fig:PE_lens_vs_no_lens}) is better than the results obtained in eq. (32) in Ref.~\cite{Takahashi:03}. 

\begin{figure}[bt]
  \centering
  \includegraphics[width=0.9\columnwidth]{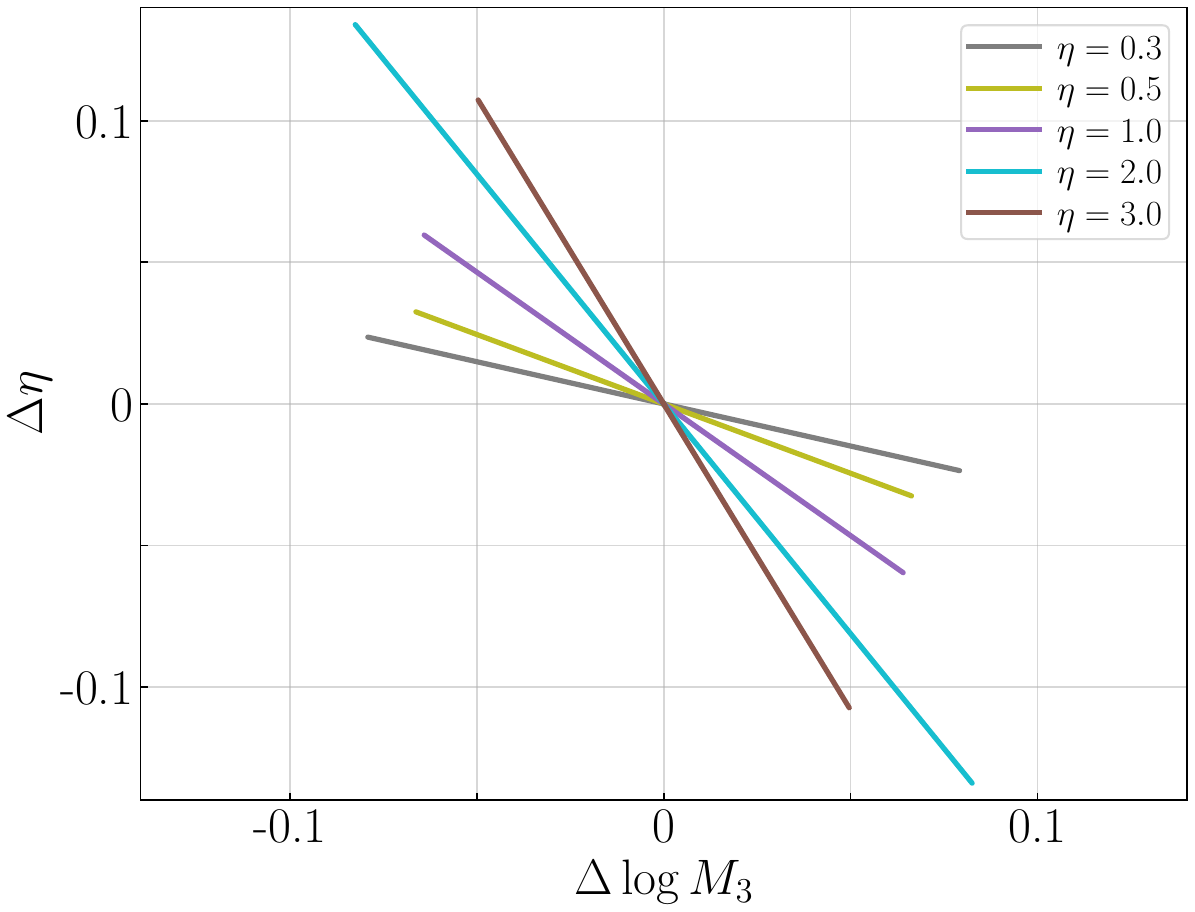}
  \caption{Error ellipses for the simple static lensing (i.e., $\eta=\beta/\theta_{\rm Ein}={\rm constant}$ for the entire waveform). At each given $\eta$, the PE error $\Delta \log M_3$ is highly correlated with $\Delta \eta$, and thus the PE error obtained by inverting the Fisher matrix is much greater than the inverse of the diagonal elements. On the other hand, the orientation of the error ellipse  varies as $\eta$ varies [Eq.~(\ref{eq:static_lens_err_ellipse})]. Therefore, in the case of repeated lensing where $\eta=\eta(t)$ due to the outer orbit's motion, different values of $\eta$ are sampled and thus breaks the degeneracy between $\log M_3$ and $\eta$, allowing a PE much better than the static lensing case [cf. eq. (32) in Ref.~\cite{Takahashi:03}.] }
\label{fig:static_lens_err_ellipse}
\end{figure}

The reason is illustrated in Fig.~\ref{fig:static_lens_err_ellipse} where we show the error ellipses between $\log M_3$ and $\eta$ for the static lensing case at different values of $\eta$. For each given $\eta$, we note $\log M_3$ and $\eta$ are highly correlated. This is because the best constrained quantity is the time delay between the primary and the secondary images, $\Delta t_l\sim 8M_3\eta$ for $\eta\ll 2$ [or $2M_3\eta^2$ for $\eta\gg 2$; Eq.~(\ref{eq:std_lens_t_delay})]. Indeed, we see that in the plot the error ellipses roughly corresponds to lines defined by 
\begin{align}
    \Delta \eta &\simeq -\frac{\Delta t_l}{\partial \left(\Delta t_l\right)/\partial \eta} \Delta \log M_3,\nonumber \\
    &\simeq 
    \begin{cases}
        -\eta \Delta \log M_3, \quad\text{ for $\eta \ll 2$,} \\
        -\frac{\eta}{2} \Delta \log M_3, \quad \text{ for $\eta \gg 2$.}
    \end{cases}
    \label{eq:static_lens_err_ellipse}
\end{align}

On the other hand, we note that the orientation of the error ellipse changes as $\eta$ changes (for a fixed $M_3$). Thus if the information at two different values of $\eta$ can be combined, the joint uncertainty will be greatly reduced. This is exactly the situation in the repeated lensing scenario. As the inner binary orbits around the SMBH, we sample the lensing signatures at different values of $\eta$, allowing the mass of the SMBH to be determined much better than the static lensing case.

\bibliography{ref.bib}

\begin{thebibliography}{83}%
\makeatletter
\providecommand \@ifxundefined [1]{%
 \@ifx{#1\undefined}
}%
\providecommand \@ifnum [1]{%
 \ifnum #1\expandafter \@firstoftwo
 \else \expandafter \@secondoftwo
 \fi
}%
\providecommand \@ifx [1]{%
 \ifx #1\expandafter \@firstoftwo
 \else \expandafter \@secondoftwo
 \fi
}%
\providecommand \natexlab [1]{#1}%
\providecommand \enquote  [1]{``#1''}%
\providecommand \bibnamefont  [1]{#1}%
\providecommand \bibfnamefont [1]{#1}%
\providecommand \citenamefont [1]{#1}%
\providecommand \href@noop [0]{\@secondoftwo}%
\providecommand \href [0]{\begingroup \@sanitize@url \@href}%
\providecommand \@href[1]{\@@startlink{#1}\@@href}%
\providecommand \@@href[1]{\endgroup#1\@@endlink}%
\providecommand \@sanitize@url [0]{\catcode `\\12\catcode `\$12\catcode
  `\&12\catcode `\#12\catcode `\^12\catcode `\_12\catcode `\%12\relax}%
\providecommand \@@startlink[1]{}%
\providecommand \@@endlink[0]{}%
\providecommand \url  [0]{\begingroup\@sanitize@url \@url }%
\providecommand \@url [1]{\endgroup\@href {#1}{\urlprefix }}%
\providecommand \urlprefix  [0]{URL }%
\providecommand \Eprint [0]{\href }%
\providecommand \doibase [0]{https://doi.org/}%
\providecommand \selectlanguage [0]{\@gobble}%
\providecommand \bibinfo  [0]{\@secondoftwo}%
\providecommand \bibfield  [0]{\@secondoftwo}%
\providecommand \translation [1]{[#1]}%
\providecommand \BibitemOpen [0]{}%
\providecommand \bibitemStop [0]{}%
\providecommand \bibitemNoStop [0]{.\EOS\space}%
\providecommand \EOS [0]{\spacefactor3000\relax}%
\providecommand \BibitemShut  [1]{\csname bibitem#1\endcsname}%
\let\auto@bib@innerbib\@empty
\bibitem [{\citenamefont {{LIGO Scientific Collaboration}}\ and\ \citenamefont
  {{Virgo Collaboration}}(2016)}]{LSC:16}%
  \BibitemOpen
  \bibfield  {author} {\bibinfo {author} {\bibnamefont {{LIGO Scientific
  Collaboration}}}\ and\ \bibinfo {author} {\bibnamefont {{Virgo
  Collaboration}}},\ }\bibfield  {title} {\bibinfo {title} {{Observation of
  Gravitational Waves from a Binary Black Hole Merger}},\ }\href
  {https://doi.org/10.1103/PhysRevLett.116.061102} {\bibfield  {journal}
  {\bibinfo  {journal} {\prl}\ }\textbf {\bibinfo {volume} {116}},\ \bibinfo
  {eid} {061102} (\bibinfo {year} {2016})},\ \Eprint
  {https://arxiv.org/abs/1602.03837} {arXiv:1602.03837 [gr-qc]} \BibitemShut
  {NoStop}%
\bibitem [{\citenamefont {{LIGO Scientific Collaboration}}(2015)}]{LSC:15}%
  \BibitemOpen
  \bibfield  {author} {\bibinfo {author} {\bibnamefont {{LIGO Scientific
  Collaboration}}},\ }\bibfield  {title} {\bibinfo {title} {{Advanced LIGO}},\
  }\href {https://doi.org/10.1088/0264-9381/32/7/074001} {\bibfield  {journal}
  {\bibinfo  {journal} {Classical and Quantum Gravity}\ }\textbf {\bibinfo
  {volume} {32}},\ \bibinfo {eid} {074001} (\bibinfo {year} {2015})},\ \Eprint
  {https://arxiv.org/abs/1411.4547} {arXiv:1411.4547 [gr-qc]} \BibitemShut
  {NoStop}%
\bibitem [{\citenamefont {{Acernese}}\ \emph {et~al.}(2015)\citenamefont
  {{Acernese}}, \citenamefont {{Agathos}}, \citenamefont {{Agatsuma}},
  \citenamefont {{Aisa}}, \citenamefont {{Allemandou}}, \citenamefont
  {{Allocca}}, \citenamefont {{Amarni}}, \citenamefont {{Astone}},\ and\
  \citenamefont {et~al.}}]{Acernese:15}%
  \BibitemOpen
  \bibfield  {author} {\bibinfo {author} {\bibfnamefont {F.}~\bibnamefont
  {{Acernese}}}, \bibinfo {author} {\bibfnamefont {M.}~\bibnamefont
  {{Agathos}}}, \bibinfo {author} {\bibfnamefont {K.}~\bibnamefont
  {{Agatsuma}}}, \bibinfo {author} {\bibfnamefont {D.}~\bibnamefont {{Aisa}}},
  \bibinfo {author} {\bibfnamefont {N.}~\bibnamefont {{Allemandou}}}, \bibinfo
  {author} {\bibfnamefont {A.}~\bibnamefont {{Allocca}}}, \bibinfo {author}
  {\bibfnamefont {J.}~\bibnamefont {{Amarni}}}, \bibinfo {author}
  {\bibfnamefont {P.}~\bibnamefont {{Astone}}},\ and\ \bibinfo {author}
  {\bibnamefont {et~al.}},\ }\bibfield  {title} {\bibinfo {title} {{Advanced
  Virgo: a second-generation interferometric gravitational wave detector}},\
  }\href {https://doi.org/10.1088/0264-9381/32/2/024001} {\bibfield  {journal}
  {\bibinfo  {journal} {Classical and Quantum Gravity}\ }\textbf {\bibinfo
  {volume} {32}},\ \bibinfo {eid} {024001} (\bibinfo {year} {2015})},\ \Eprint
  {https://arxiv.org/abs/1408.3978} {arXiv:1408.3978 [gr-qc]} \BibitemShut
  {NoStop}%
\bibitem [{\citenamefont {{Kagra Collaboration}}\ and\ \citenamefont
  {et~al.}(2019)}]{kagra:19}%
  \BibitemOpen
  \bibfield  {author} {\bibinfo {author} {\bibnamefont {{Kagra
  Collaboration}}}\ and\ \bibinfo {author} {\bibnamefont {et~al.}},\ }\bibfield
   {title} {\bibinfo {title} {{KAGRA: 2.5 generation interferometric
  gravitational wave detector}},\ }\href
  {https://doi.org/10.1038/s41550-018-0658-y} {\bibfield  {journal} {\bibinfo
  {journal} {Nature Astronomy}\ }\textbf {\bibinfo {volume} {3}},\ \bibinfo
  {pages} {35} (\bibinfo {year} {2019})},\ \Eprint
  {https://arxiv.org/abs/1811.08079} {arXiv:1811.08079 [gr-qc]} \BibitemShut
  {NoStop}%
\bibitem [{\citenamefont {{LIGO Scientific Collaboration}}\ and\ \citenamefont
  {{Virgo Collaboration}}(2019)}]{LSC:19}%
  \BibitemOpen
  \bibfield  {author} {\bibinfo {author} {\bibnamefont {{LIGO Scientific
  Collaboration}}}\ and\ \bibinfo {author} {\bibnamefont {{Virgo
  Collaboration}}},\ }\bibfield  {title} {\bibinfo {title} {{GWTC-1: A
  Gravitational-Wave Transient Catalog of Compact Binary Mergers Observed by
  LIGO and Virgo during the First and Second Observing Runs}},\ }\href
  {https://doi.org/10.1103/PhysRevX.9.031040} {\bibfield  {journal} {\bibinfo
  {journal} {Physical Review X}\ }\textbf {\bibinfo {volume} {9}},\ \bibinfo
  {eid} {031040} (\bibinfo {year} {2019})},\ \Eprint
  {https://arxiv.org/abs/1811.12907} {arXiv:1811.12907 [astro-ph.HE]}
  \BibitemShut {NoStop}%
\bibitem [{\citenamefont {{LIGO Scientific Collaboration}}\ and\ \citenamefont
  {{Virgo Collaboration}}(2021)}]{LSC:21}%
  \BibitemOpen
  \bibfield  {author} {\bibinfo {author} {\bibnamefont {{LIGO Scientific
  Collaboration}}}\ and\ \bibinfo {author} {\bibnamefont {{Virgo
  Collaboration}}},\ }\bibfield  {title} {\bibinfo {title} {{GWTC-2: Compact
  Binary Coalescences Observed by LIGO and Virgo during the First Half of the
  Third Observing Run}},\ }\href {https://doi.org/10.1103/PhysRevX.11.021053}
  {\bibfield  {journal} {\bibinfo  {journal} {Physical Review X}\ }\textbf
  {\bibinfo {volume} {11}},\ \bibinfo {eid} {021053} (\bibinfo {year}
  {2021})},\ \Eprint {https://arxiv.org/abs/2010.14527} {arXiv:2010.14527
  [gr-qc]} \BibitemShut {NoStop}%
\bibitem [{\citenamefont {{Amaro-Seoane}}\ \emph {et~al.}(2017)\citenamefont
  {{Amaro-Seoane}}, \citenamefont {{Audley}}, \citenamefont {{Babak}},
  \citenamefont {{Baker}}, \citenamefont {{Barausse}}, \citenamefont
  {{Bender}}, \citenamefont {{Berti}}, \citenamefont {{Binetruy}},\ and\
  \citenamefont {et~al.}}]{Amaro-Seoane:17}%
  \BibitemOpen
  \bibfield  {author} {\bibinfo {author} {\bibfnamefont {P.}~\bibnamefont
  {{Amaro-Seoane}}}, \bibinfo {author} {\bibfnamefont {H.}~\bibnamefont
  {{Audley}}}, \bibinfo {author} {\bibfnamefont {S.}~\bibnamefont {{Babak}}},
  \bibinfo {author} {\bibfnamefont {J.}~\bibnamefont {{Baker}}}, \bibinfo
  {author} {\bibfnamefont {E.}~\bibnamefont {{Barausse}}}, \bibinfo {author}
  {\bibfnamefont {P.}~\bibnamefont {{Bender}}}, \bibinfo {author}
  {\bibfnamefont {E.}~\bibnamefont {{Berti}}}, \bibinfo {author} {\bibfnamefont
  {P.}~\bibnamefont {{Binetruy}}},\ and\ \bibinfo {author} {\bibnamefont
  {et~al.}},\ }\bibfield  {title} {\bibinfo {title} {{Laser Interferometer
  Space Antenna}},\ }\href@noop {} {\bibfield  {journal} {\bibinfo  {journal}
  {arXiv e-prints}\ ,\ \bibinfo {eid} {arXiv:1702.00786}} (\bibinfo {year}
  {2017})},\ \Eprint {https://arxiv.org/abs/1702.00786} {arXiv:1702.00786
  [astro-ph.IM]} \BibitemShut {NoStop}%
\bibitem [{\citenamefont {{Luo}}\ \emph {et~al.}(2016)\citenamefont {{Luo}},
  \citenamefont {{Chen}}, \citenamefont {{Duan}}, \citenamefont {{Gong}},
  \citenamefont {{Hu}}, \citenamefont {{Ji}}, \citenamefont {{Liu}},
  \citenamefont {{Mei}}, \citenamefont {{Milyukov}}, \citenamefont {{Sazhin}},
  \citenamefont {{Shao}}, \citenamefont {{Toth}}, \citenamefont {{Tu}},
  \citenamefont {{Wang}}, \citenamefont {{Wang}}, \citenamefont {{Yeh}},
  \citenamefont {{Zhan}}, \citenamefont {{Zhang}}, \citenamefont {{Zharov}},\
  and\ \citenamefont {{Zhou}}}]{Luo:16}%
  \BibitemOpen
  \bibfield  {author} {\bibinfo {author} {\bibfnamefont {J.}~\bibnamefont
  {{Luo}}}, \bibinfo {author} {\bibfnamefont {L.-S.}\ \bibnamefont {{Chen}}},
  \bibinfo {author} {\bibfnamefont {H.-Z.}\ \bibnamefont {{Duan}}}, \bibinfo
  {author} {\bibfnamefont {Y.-G.}\ \bibnamefont {{Gong}}}, \bibinfo {author}
  {\bibfnamefont {S.}~\bibnamefont {{Hu}}}, \bibinfo {author} {\bibfnamefont
  {J.}~\bibnamefont {{Ji}}}, \bibinfo {author} {\bibfnamefont {Q.}~\bibnamefont
  {{Liu}}}, \bibinfo {author} {\bibfnamefont {J.}~\bibnamefont {{Mei}}},
  \bibinfo {author} {\bibfnamefont {V.}~\bibnamefont {{Milyukov}}}, \bibinfo
  {author} {\bibfnamefont {M.}~\bibnamefont {{Sazhin}}}, \bibinfo {author}
  {\bibfnamefont {C.-G.}\ \bibnamefont {{Shao}}}, \bibinfo {author}
  {\bibfnamefont {V.~T.}\ \bibnamefont {{Toth}}}, \bibinfo {author}
  {\bibfnamefont {H.-B.}\ \bibnamefont {{Tu}}}, \bibinfo {author}
  {\bibfnamefont {Y.}~\bibnamefont {{Wang}}}, \bibinfo {author} {\bibfnamefont
  {Y.}~\bibnamefont {{Wang}}}, \bibinfo {author} {\bibfnamefont {H.-C.}\
  \bibnamefont {{Yeh}}}, \bibinfo {author} {\bibfnamefont {M.-S.}\ \bibnamefont
  {{Zhan}}}, \bibinfo {author} {\bibfnamefont {Y.}~\bibnamefont {{Zhang}}},
  \bibinfo {author} {\bibfnamefont {V.}~\bibnamefont {{Zharov}}},\ and\
  \bibinfo {author} {\bibfnamefont {Z.-B.}\ \bibnamefont {{Zhou}}},\ }\bibfield
   {title} {\bibinfo {title} {{TianQin: a space-borne gravitational wave
  detector}},\ }\href {https://doi.org/10.1088/0264-9381/33/3/035010}
  {\bibfield  {journal} {\bibinfo  {journal} {Classical and Quantum Gravity}\
  }\textbf {\bibinfo {volume} {33}},\ \bibinfo {eid} {035010} (\bibinfo {year}
  {2016})},\ \Eprint {https://arxiv.org/abs/1512.02076} {arXiv:1512.02076
  [astro-ph.IM]} \BibitemShut {NoStop}%
\bibitem [{\citenamefont {Hu}\ and\ \citenamefont {Wu}(2017)}]{Hu:17}%
  \BibitemOpen
  \bibfield  {author} {\bibinfo {author} {\bibfnamefont {W.-R.}\ \bibnamefont
  {Hu}}\ and\ \bibinfo {author} {\bibfnamefont {Y.-L.}\ \bibnamefont {Wu}},\
  }\bibfield  {title} {\bibinfo {title} {{The Taiji Program in Space for
  gravitational wave physics and the nature of gravity}},\ }\href
  {https://doi.org/10.1093/nsr/nwx116} {\bibfield  {journal} {\bibinfo
  {journal} {National Science Review}\ }\textbf {\bibinfo {volume} {4}},\
  \bibinfo {pages} {685} (\bibinfo {year} {2017})},\ \Eprint
  {https://arxiv.org/abs/https://academic.oup.com/nsr/article-pdf/4/5/685/31566708/nwx116.pdf}
  {https://academic.oup.com/nsr/article-pdf/4/5/685/31566708/nwx116.pdf}
  \BibitemShut {NoStop}%
\bibitem [{\citenamefont {{Nakamura}}\ \emph {et~al.}(2016)\citenamefont
  {{Nakamura}}, \citenamefont {{Ando}}, \citenamefont {{Kinugawa}},
  \citenamefont {{Nakano}}, \citenamefont {{Eda}}, \citenamefont {{Sato}},
  \citenamefont {{Musha}}, \citenamefont {{Akutsu}}, \citenamefont {{Tanaka}},
  \citenamefont {{Seto}}, \citenamefont {{Kanda}},\ and\ \citenamefont
  {{Itoh}}}]{Nakamura:16}%
  \BibitemOpen
  \bibfield  {author} {\bibinfo {author} {\bibfnamefont {T.}~\bibnamefont
  {{Nakamura}}}, \bibinfo {author} {\bibfnamefont {M.}~\bibnamefont {{Ando}}},
  \bibinfo {author} {\bibfnamefont {T.}~\bibnamefont {{Kinugawa}}}, \bibinfo
  {author} {\bibfnamefont {H.}~\bibnamefont {{Nakano}}}, \bibinfo {author}
  {\bibfnamefont {K.}~\bibnamefont {{Eda}}}, \bibinfo {author} {\bibfnamefont
  {S.}~\bibnamefont {{Sato}}}, \bibinfo {author} {\bibfnamefont
  {M.}~\bibnamefont {{Musha}}}, \bibinfo {author} {\bibfnamefont
  {T.}~\bibnamefont {{Akutsu}}}, \bibinfo {author} {\bibfnamefont
  {T.}~\bibnamefont {{Tanaka}}}, \bibinfo {author} {\bibfnamefont
  {N.}~\bibnamefont {{Seto}}}, \bibinfo {author} {\bibfnamefont
  {N.}~\bibnamefont {{Kanda}}},\ and\ \bibinfo {author} {\bibfnamefont
  {Y.}~\bibnamefont {{Itoh}}},\ }\bibfield  {title} {\bibinfo {title}
  {{Pre-DECIGO can get the smoking gun to decide the astrophysical or
  cosmological origin of GW150914-like binary black holes}},\ }\href
  {https://doi.org/10.1093/ptep/ptw127} {\bibfield  {journal} {\bibinfo
  {journal} {Progress of Theoretical and Experimental Physics}\ }\textbf
  {\bibinfo {volume} {2016}},\ \bibinfo {eid} {093E01} (\bibinfo {year}
  {2016})},\ \Eprint {https://arxiv.org/abs/1607.00897} {arXiv:1607.00897
  [astro-ph.HE]} \BibitemShut {NoStop}%
\bibitem [{\citenamefont {{Kawamura}}\ \emph {et~al.}(2020)\citenamefont
  {{Kawamura}}, \citenamefont {{Ando}}, \citenamefont {{Seto}}, \citenamefont
  {{Sato}}, \citenamefont {{Musha}}, \citenamefont {{Kawano}}, \citenamefont
  {{Yokoyama}}, \citenamefont {{Tanaka}},\ and\ \citenamefont
  {et~al.}}]{Kawamura:20}%
  \BibitemOpen
  \bibfield  {author} {\bibinfo {author} {\bibfnamefont {S.}~\bibnamefont
  {{Kawamura}}}, \bibinfo {author} {\bibfnamefont {M.}~\bibnamefont {{Ando}}},
  \bibinfo {author} {\bibfnamefont {N.}~\bibnamefont {{Seto}}}, \bibinfo
  {author} {\bibfnamefont {S.}~\bibnamefont {{Sato}}}, \bibinfo {author}
  {\bibfnamefont {M.}~\bibnamefont {{Musha}}}, \bibinfo {author} {\bibfnamefont
  {I.}~\bibnamefont {{Kawano}}}, \bibinfo {author} {\bibfnamefont
  {J.}~\bibnamefont {{Yokoyama}}}, \bibinfo {author} {\bibfnamefont
  {T.}~\bibnamefont {{Tanaka}}},\ and\ \bibinfo {author} {\bibnamefont
  {et~al.}},\ }\bibfield  {title} {\bibinfo {title} {{Current status of space
  gravitational wave antenna DECIGO and B-DECIGO}},\ }\href@noop {} {\bibfield
  {journal} {\bibinfo  {journal} {arXiv e-prints}\ ,\ \bibinfo {eid}
  {arXiv:2006.13545}} (\bibinfo {year} {2020})},\ \Eprint
  {https://arxiv.org/abs/2006.13545} {arXiv:2006.13545 [gr-qc]} \BibitemShut
  {NoStop}%
\bibitem [{\citenamefont {{Arca Sedda}}\ \emph {et~al.}(2019)\citenamefont
  {{Arca Sedda}}, \citenamefont {{Berry}}, \citenamefont {{Jani}},
  \citenamefont {{Amaro-Seoane}}, \citenamefont {{Auclair}}, \citenamefont
  {{Baird}}, \citenamefont {{Baker}}, \citenamefont {{Berti}},\ and\
  \citenamefont {et~al.}}]{Sedda:19}%
  \BibitemOpen
  \bibfield  {author} {\bibinfo {author} {\bibfnamefont {M.}~\bibnamefont
  {{Arca Sedda}}}, \bibinfo {author} {\bibfnamefont {C.~P.~L.}\ \bibnamefont
  {{Berry}}}, \bibinfo {author} {\bibfnamefont {K.}~\bibnamefont {{Jani}}},
  \bibinfo {author} {\bibfnamefont {P.}~\bibnamefont {{Amaro-Seoane}}},
  \bibinfo {author} {\bibfnamefont {P.}~\bibnamefont {{Auclair}}}, \bibinfo
  {author} {\bibfnamefont {J.}~\bibnamefont {{Baird}}}, \bibinfo {author}
  {\bibfnamefont {T.}~\bibnamefont {{Baker}}}, \bibinfo {author} {\bibfnamefont
  {E.}~\bibnamefont {{Berti}}},\ and\ \bibinfo {author} {\bibnamefont
  {et~al.}},\ }\bibfield  {title} {\bibinfo {title} {{The Missing Link in
  Gravitational-Wave Astronomy: Discoveries waiting in the decihertz range}},\
  }\href@noop {} {\bibfield  {journal} {\bibinfo  {journal} {arXiv e-prints}\
  ,\ \bibinfo {eid} {arXiv:1908.11375}} (\bibinfo {year} {2019})},\ \Eprint
  {https://arxiv.org/abs/1908.11375} {arXiv:1908.11375 [gr-qc]} \BibitemShut
  {NoStop}%
\bibitem [{\citenamefont {Kuns}\ \emph {et~al.}(2020)\citenamefont {Kuns},
  \citenamefont {Yu}, \citenamefont {Chen},\ and\ \citenamefont
  {Adhikari}}]{Kuns:19}%
  \BibitemOpen
  \bibfield  {author} {\bibinfo {author} {\bibfnamefont {K.~A.}\ \bibnamefont
  {Kuns}}, \bibinfo {author} {\bibfnamefont {H.}~\bibnamefont {Yu}}, \bibinfo
  {author} {\bibfnamefont {Y.}~\bibnamefont {Chen}},\ and\ \bibinfo {author}
  {\bibfnamefont {R.~X.}\ \bibnamefont {Adhikari}},\ }\bibfield  {title}
  {\bibinfo {title} {Astrophysics and cosmology with a decihertz
  gravitational-wave detector: Tiango},\ }\href
  {https://doi.org/10.1103/PhysRevD.102.043001} {\bibfield  {journal} {\bibinfo
   {journal} {Phys. Rev. D}\ }\textbf {\bibinfo {volume} {102}},\ \bibinfo
  {pages} {043001} (\bibinfo {year} {2020})}\BibitemShut {NoStop}%
\bibitem [{\citenamefont {{McKernan}}\ \emph {et~al.}(2012)\citenamefont
  {{McKernan}}, \citenamefont {{Ford}}, \citenamefont {{Lyra}},\ and\
  \citenamefont {{Perets}}}]{McKernan:12}%
  \BibitemOpen
  \bibfield  {author} {\bibinfo {author} {\bibfnamefont {B.}~\bibnamefont
  {{McKernan}}}, \bibinfo {author} {\bibfnamefont {K.~E.~S.}\ \bibnamefont
  {{Ford}}}, \bibinfo {author} {\bibfnamefont {W.}~\bibnamefont {{Lyra}}},\
  and\ \bibinfo {author} {\bibfnamefont {H.~B.}\ \bibnamefont {{Perets}}},\
  }\bibfield  {title} {\bibinfo {title} {{Intermediate mass black holes in AGN
  discs - I. Production and growth}},\ }\href
  {https://doi.org/10.1111/j.1365-2966.2012.21486.x} {\bibfield  {journal}
  {\bibinfo  {journal} {\mnras}\ }\textbf {\bibinfo {volume} {425}},\ \bibinfo
  {pages} {460} (\bibinfo {year} {2012})},\ \Eprint
  {https://arxiv.org/abs/1206.2309} {arXiv:1206.2309 [astro-ph.GA]}
  \BibitemShut {NoStop}%
\bibitem [{\citenamefont {{Bartos}}\ \emph {et~al.}(2017)\citenamefont
  {{Bartos}}, \citenamefont {{Kocsis}}, \citenamefont {{Haiman}},\ and\
  \citenamefont {{M{\'a}rka}}}]{Bartos:17}%
  \BibitemOpen
  \bibfield  {author} {\bibinfo {author} {\bibfnamefont {I.}~\bibnamefont
  {{Bartos}}}, \bibinfo {author} {\bibfnamefont {B.}~\bibnamefont {{Kocsis}}},
  \bibinfo {author} {\bibfnamefont {Z.}~\bibnamefont {{Haiman}}},\ and\
  \bibinfo {author} {\bibfnamefont {S.}~\bibnamefont {{M{\'a}rka}}},\
  }\bibfield  {title} {\bibinfo {title} {{Rapid and Bright Stellar-mass Binary
  Black Hole Mergers in Active Galactic Nuclei}},\ }\href
  {https://doi.org/10.3847/1538-4357/835/2/165} {\bibfield  {journal} {\bibinfo
   {journal} {\apj}\ }\textbf {\bibinfo {volume} {835}},\ \bibinfo {eid} {165}
  (\bibinfo {year} {2017})},\ \Eprint {https://arxiv.org/abs/1602.03831}
  {arXiv:1602.03831 [astro-ph.HE]} \BibitemShut {NoStop}%
\bibitem [{\citenamefont {{Stone}}\ \emph {et~al.}(2017)\citenamefont
  {{Stone}}, \citenamefont {{Metzger}},\ and\ \citenamefont
  {{Haiman}}}]{Stone:17}%
  \BibitemOpen
  \bibfield  {author} {\bibinfo {author} {\bibfnamefont {N.~C.}\ \bibnamefont
  {{Stone}}}, \bibinfo {author} {\bibfnamefont {B.~D.}\ \bibnamefont
  {{Metzger}}},\ and\ \bibinfo {author} {\bibfnamefont {Z.}~\bibnamefont
  {{Haiman}}},\ }\bibfield  {title} {\bibinfo {title} {{Assisted inspirals of
  stellar mass black holes embedded in AGN discs: solving the `final au
  problem'}},\ }\href {https://doi.org/10.1093/mnras/stw2260} {\bibfield
  {journal} {\bibinfo  {journal} {\mnras}\ }\textbf {\bibinfo {volume} {464}},\
  \bibinfo {pages} {946} (\bibinfo {year} {2017})},\ \Eprint
  {https://arxiv.org/abs/1602.04226} {arXiv:1602.04226 [astro-ph.GA]}
  \BibitemShut {NoStop}%
\bibitem [{\citenamefont {{McKernan}}\ \emph {et~al.}(2018)\citenamefont
  {{McKernan}}, \citenamefont {{Ford}}, \citenamefont {{Bellovary}},
  \citenamefont {{Leigh}}, \citenamefont {{Haiman}}, \citenamefont {{Kocsis}},
  \citenamefont {{Lyra}}, \citenamefont {{Mac Low}}, \citenamefont {{Metzger}},
  \citenamefont {{O'Dowd}}, \citenamefont {{Endlich}},\ and\ \citenamefont
  {{Rosen}}}]{McKernan:18}%
  \BibitemOpen
  \bibfield  {author} {\bibinfo {author} {\bibfnamefont {B.}~\bibnamefont
  {{McKernan}}}, \bibinfo {author} {\bibfnamefont {K.~E.~S.}\ \bibnamefont
  {{Ford}}}, \bibinfo {author} {\bibfnamefont {J.}~\bibnamefont {{Bellovary}}},
  \bibinfo {author} {\bibfnamefont {N.~W.~C.}\ \bibnamefont {{Leigh}}},
  \bibinfo {author} {\bibfnamefont {Z.}~\bibnamefont {{Haiman}}}, \bibinfo
  {author} {\bibfnamefont {B.}~\bibnamefont {{Kocsis}}}, \bibinfo {author}
  {\bibfnamefont {W.}~\bibnamefont {{Lyra}}}, \bibinfo {author} {\bibfnamefont
  {M.~M.}\ \bibnamefont {{Mac Low}}}, \bibinfo {author} {\bibfnamefont
  {B.}~\bibnamefont {{Metzger}}}, \bibinfo {author} {\bibfnamefont
  {M.}~\bibnamefont {{O'Dowd}}}, \bibinfo {author} {\bibfnamefont
  {S.}~\bibnamefont {{Endlich}}},\ and\ \bibinfo {author} {\bibfnamefont
  {D.~J.}\ \bibnamefont {{Rosen}}},\ }\bibfield  {title} {\bibinfo {title}
  {{Constraining Stellar-mass Black Hole Mergers in AGN Disks Detectable with
  LIGO}},\ }\href {https://doi.org/10.3847/1538-4357/aadae5} {\bibfield
  {journal} {\bibinfo  {journal} {\apj}\ }\textbf {\bibinfo {volume} {866}},\
  \bibinfo {eid} {66} (\bibinfo {year} {2018})},\ \Eprint
  {https://arxiv.org/abs/1702.07818} {arXiv:1702.07818 [astro-ph.HE]}
  \BibitemShut {NoStop}%
\bibitem [{\citenamefont {{Tagawa}}\ \emph {et~al.}(2019)\citenamefont
  {{Tagawa}}, \citenamefont {{Haiman}},\ and\ \citenamefont
  {{Kocsis}}}]{Tagawa:19}%
  \BibitemOpen
  \bibfield  {author} {\bibinfo {author} {\bibfnamefont {H.}~\bibnamefont
  {{Tagawa}}}, \bibinfo {author} {\bibfnamefont {Z.}~\bibnamefont {{Haiman}}},\
  and\ \bibinfo {author} {\bibfnamefont {B.}~\bibnamefont {{Kocsis}}},\
  }\bibfield  {title} {\bibinfo {title} {{Formation and Evolution of Compact
  Object Binaries in AGN Disks}},\ }\href@noop {} {\bibfield  {journal}
  {\bibinfo  {journal} {arXiv e-prints}\ ,\ \bibinfo {eid} {arXiv:1912.08218}}
  (\bibinfo {year} {2019})},\ \Eprint {https://arxiv.org/abs/1912.08218}
  {arXiv:1912.08218 [astro-ph.GA]} \BibitemShut {NoStop}%
\bibitem [{\citenamefont {{Yang}}\ \emph {et~al.}(2019)\citenamefont {{Yang}},
  \citenamefont {{Bartos}}, \citenamefont {{Gayathri}}, \citenamefont {{Ford}},
  \citenamefont {{Haiman}}, \citenamefont {{Klimenko}}, \citenamefont
  {{Kocsis}}, \citenamefont {{M{\'a}rka}}, \citenamefont {{M{\'a}rka}},
  \citenamefont {{McKernan}},\ and\ \citenamefont {{O'Shaughnessy}}}]{Yang:19}%
  \BibitemOpen
  \bibfield  {author} {\bibinfo {author} {\bibfnamefont {Y.}~\bibnamefont
  {{Yang}}}, \bibinfo {author} {\bibfnamefont {I.}~\bibnamefont {{Bartos}}},
  \bibinfo {author} {\bibfnamefont {V.}~\bibnamefont {{Gayathri}}}, \bibinfo
  {author} {\bibfnamefont {K.~E.~S.}\ \bibnamefont {{Ford}}}, \bibinfo {author}
  {\bibfnamefont {Z.}~\bibnamefont {{Haiman}}}, \bibinfo {author}
  {\bibfnamefont {S.}~\bibnamefont {{Klimenko}}}, \bibinfo {author}
  {\bibfnamefont {B.}~\bibnamefont {{Kocsis}}}, \bibinfo {author}
  {\bibfnamefont {S.}~\bibnamefont {{M{\'a}rka}}}, \bibinfo {author}
  {\bibfnamefont {Z.}~\bibnamefont {{M{\'a}rka}}}, \bibinfo {author}
  {\bibfnamefont {B.}~\bibnamefont {{McKernan}}},\ and\ \bibinfo {author}
  {\bibfnamefont {R.}~\bibnamefont {{O'Shaughnessy}}},\ }\bibfield  {title}
  {\bibinfo {title} {{Hierarchical Black Hole Mergers in Active Galactic
  Nuclei}},\ }\href {https://doi.org/10.1103/PhysRevLett.123.181101} {\bibfield
   {journal} {\bibinfo  {journal} {\prl}\ }\textbf {\bibinfo {volume} {123}},\
  \bibinfo {eid} {181101} (\bibinfo {year} {2019})},\ \Eprint
  {https://arxiv.org/abs/1906.09281} {arXiv:1906.09281 [astro-ph.HE]}
  \BibitemShut {NoStop}%
\bibitem [{\citenamefont {{Bellovary}}\ \emph {et~al.}(2016)\citenamefont
  {{Bellovary}}, \citenamefont {{Mac Low}}, \citenamefont {{McKernan}},\ and\
  \citenamefont {{Ford}}}]{Bellovary:16}%
  \BibitemOpen
  \bibfield  {author} {\bibinfo {author} {\bibfnamefont {J.~M.}\ \bibnamefont
  {{Bellovary}}}, \bibinfo {author} {\bibfnamefont {M.-M.}\ \bibnamefont {{Mac
  Low}}}, \bibinfo {author} {\bibfnamefont {B.}~\bibnamefont {{McKernan}}},\
  and\ \bibinfo {author} {\bibfnamefont {K.~E.~S.}\ \bibnamefont {{Ford}}},\
  }\bibfield  {title} {\bibinfo {title} {{Migration Traps in Disks around
  Supermassive Black Holes}},\ }\href
  {https://doi.org/10.3847/2041-8205/819/2/L17} {\bibfield  {journal} {\bibinfo
   {journal} {\apjl}\ }\textbf {\bibinfo {volume} {819}},\ \bibinfo {eid} {L17}
  (\bibinfo {year} {2016})},\ \Eprint {https://arxiv.org/abs/1511.00005}
  {arXiv:1511.00005 [astro-ph.GA]} \BibitemShut {NoStop}%
\bibitem [{\citenamefont {{Secunda}}\ \emph {et~al.}(2019)\citenamefont
  {{Secunda}}, \citenamefont {{Bellovary}}, \citenamefont {{Mac Low}},
  \citenamefont {{Ford}}, \citenamefont {{McKernan}}, \citenamefont {{Leigh}},
  \citenamefont {{Lyra}},\ and\ \citenamefont {{S{\'a}ndor}}}]{Secunda:19}%
  \BibitemOpen
  \bibfield  {author} {\bibinfo {author} {\bibfnamefont {A.}~\bibnamefont
  {{Secunda}}}, \bibinfo {author} {\bibfnamefont {J.}~\bibnamefont
  {{Bellovary}}}, \bibinfo {author} {\bibfnamefont {M.-M.}\ \bibnamefont {{Mac
  Low}}}, \bibinfo {author} {\bibfnamefont {K.~E.~S.}\ \bibnamefont {{Ford}}},
  \bibinfo {author} {\bibfnamefont {B.}~\bibnamefont {{McKernan}}}, \bibinfo
  {author} {\bibfnamefont {N.~W.~C.}\ \bibnamefont {{Leigh}}}, \bibinfo
  {author} {\bibfnamefont {W.}~\bibnamefont {{Lyra}}},\ and\ \bibinfo {author}
  {\bibfnamefont {Z.}~\bibnamefont {{S{\'a}ndor}}},\ }\bibfield  {title}
  {\bibinfo {title} {{Orbital Migration of Interacting Stellar Mass Black Holes
  in Disks around Supermassive Black Holes}},\ }\href
  {https://doi.org/10.3847/1538-4357/ab20ca} {\bibfield  {journal} {\bibinfo
  {journal} {\apj}\ }\textbf {\bibinfo {volume} {878}},\ \bibinfo {eid} {85}
  (\bibinfo {year} {2019})},\ \Eprint {https://arxiv.org/abs/1807.02859}
  {arXiv:1807.02859 [astro-ph.HE]} \BibitemShut {NoStop}%
\bibitem [{\citenamefont {{O'Leary}}\ \emph {et~al.}(2009)\citenamefont
  {{O'Leary}}, \citenamefont {{Kocsis}},\ and\ \citenamefont
  {{Loeb}}}]{OLeary:09}%
  \BibitemOpen
  \bibfield  {author} {\bibinfo {author} {\bibfnamefont {R.~M.}\ \bibnamefont
  {{O'Leary}}}, \bibinfo {author} {\bibfnamefont {B.}~\bibnamefont
  {{Kocsis}}},\ and\ \bibinfo {author} {\bibfnamefont {A.}~\bibnamefont
  {{Loeb}}},\ }\bibfield  {title} {\bibinfo {title} {{Gravitational waves from
  scattering of stellar-mass black holes in galactic nuclei}},\ }\href
  {https://doi.org/10.1111/j.1365-2966.2009.14653.x} {\bibfield  {journal}
  {\bibinfo  {journal} {\mnras}\ }\textbf {\bibinfo {volume} {395}},\ \bibinfo
  {pages} {2127} (\bibinfo {year} {2009})},\ \Eprint
  {https://arxiv.org/abs/0807.2638} {arXiv:0807.2638 [astro-ph]} \BibitemShut
  {NoStop}%
\bibitem [{\citenamefont {{Antonini}}\ and\ \citenamefont
  {{Perets}}(2012)}]{Antonini:12}%
  \BibitemOpen
  \bibfield  {author} {\bibinfo {author} {\bibfnamefont {F.}~\bibnamefont
  {{Antonini}}}\ and\ \bibinfo {author} {\bibfnamefont {H.~B.}\ \bibnamefont
  {{Perets}}},\ }\bibfield  {title} {\bibinfo {title} {{Secular Evolution of
  Compact Binaries near Massive Black Holes: Gravitational Wave Sources and
  Other Exotica}},\ }\href {https://doi.org/10.1088/0004-637X/757/1/27}
  {\bibfield  {journal} {\bibinfo  {journal} {\apj}\ }\textbf {\bibinfo
  {volume} {757}},\ \bibinfo {eid} {27} (\bibinfo {year} {2012})},\ \Eprint
  {https://arxiv.org/abs/1203.2938} {arXiv:1203.2938 [astro-ph.GA]}
  \BibitemShut {NoStop}%
\bibitem [{\citenamefont {{Antonini}}\ and\ \citenamefont
  {{Rasio}}(2016)}]{Antonini:16}%
  \BibitemOpen
  \bibfield  {author} {\bibinfo {author} {\bibfnamefont {F.}~\bibnamefont
  {{Antonini}}}\ and\ \bibinfo {author} {\bibfnamefont {F.~A.}\ \bibnamefont
  {{Rasio}}},\ }\bibfield  {title} {\bibinfo {title} {{Merging Black Hole
  Binaries in Galactic Nuclei: Implications for Advanced-LIGO Detections}},\
  }\href {https://doi.org/10.3847/0004-637X/831/2/187} {\bibfield  {journal}
  {\bibinfo  {journal} {\apj}\ }\textbf {\bibinfo {volume} {831}},\ \bibinfo
  {eid} {187} (\bibinfo {year} {2016})},\ \Eprint
  {https://arxiv.org/abs/1606.04889} {arXiv:1606.04889 [astro-ph.HE]}
  \BibitemShut {NoStop}%
\bibitem [{\citenamefont {{VanLandingham}}\ \emph {et~al.}(2016)\citenamefont
  {{VanLandingham}}, \citenamefont {{Miller}}, \citenamefont {{Hamilton}},\
  and\ \citenamefont {{Richardson}}}]{VanLandingham:16}%
  \BibitemOpen
  \bibfield  {author} {\bibinfo {author} {\bibfnamefont {J.~H.}\ \bibnamefont
  {{VanLandingham}}}, \bibinfo {author} {\bibfnamefont {M.~C.}\ \bibnamefont
  {{Miller}}}, \bibinfo {author} {\bibfnamefont {D.~P.}\ \bibnamefont
  {{Hamilton}}},\ and\ \bibinfo {author} {\bibfnamefont {D.~C.}\ \bibnamefont
  {{Richardson}}},\ }\bibfield  {title} {\bibinfo {title} {{The Role of the
  Kozai--Lidov Mechanism in Black Hole Binary Mergers in Galactic Centers}},\
  }\href {https://doi.org/10.3847/0004-637X/828/2/77} {\bibfield  {journal}
  {\bibinfo  {journal} {\apj}\ }\textbf {\bibinfo {volume} {828}},\ \bibinfo
  {eid} {77} (\bibinfo {year} {2016})},\ \Eprint
  {https://arxiv.org/abs/1604.04948} {arXiv:1604.04948 [astro-ph.HE]}
  \BibitemShut {NoStop}%
\bibitem [{\citenamefont {{Petrovich}}\ and\ \citenamefont
  {{Antonini}}(2017)}]{Petrovich:17}%
  \BibitemOpen
  \bibfield  {author} {\bibinfo {author} {\bibfnamefont {C.}~\bibnamefont
  {{Petrovich}}}\ and\ \bibinfo {author} {\bibfnamefont {F.}~\bibnamefont
  {{Antonini}}},\ }\bibfield  {title} {\bibinfo {title} {{Greatly Enhanced
  Merger Rates of Compact-object Binaries in Non-spherical Nuclear Star
  Clusters}},\ }\href {https://doi.org/10.3847/1538-4357/aa8628} {\bibfield
  {journal} {\bibinfo  {journal} {\apj}\ }\textbf {\bibinfo {volume} {846}},\
  \bibinfo {eid} {146} (\bibinfo {year} {2017})},\ \Eprint
  {https://arxiv.org/abs/1705.05848} {arXiv:1705.05848 [astro-ph.HE]}
  \BibitemShut {NoStop}%
\bibitem [{\citenamefont {{Leigh}}\ \emph {et~al.}(2018)\citenamefont
  {{Leigh}}, \citenamefont {{Geller}}, \citenamefont {{McKernan}},
  \citenamefont {{Ford}}, \citenamefont {{Mac Low}}, \citenamefont
  {{Bellovary}}, \citenamefont {{Haiman}}, \citenamefont {{Lyra}},
  \citenamefont {{Samsing}}, \citenamefont {{O'Dowd}}, \citenamefont
  {{Kocsis}},\ and\ \citenamefont {{Endlich}}}]{Leigh:18}%
  \BibitemOpen
  \bibfield  {author} {\bibinfo {author} {\bibfnamefont {N.~W.~C.}\
  \bibnamefont {{Leigh}}}, \bibinfo {author} {\bibfnamefont {A.~M.}\
  \bibnamefont {{Geller}}}, \bibinfo {author} {\bibfnamefont {B.}~\bibnamefont
  {{McKernan}}}, \bibinfo {author} {\bibfnamefont {K.~E.~S.}\ \bibnamefont
  {{Ford}}}, \bibinfo {author} {\bibfnamefont {M.~M.}\ \bibnamefont {{Mac
  Low}}}, \bibinfo {author} {\bibfnamefont {J.}~\bibnamefont {{Bellovary}}},
  \bibinfo {author} {\bibfnamefont {Z.}~\bibnamefont {{Haiman}}}, \bibinfo
  {author} {\bibfnamefont {W.}~\bibnamefont {{Lyra}}}, \bibinfo {author}
  {\bibfnamefont {J.}~\bibnamefont {{Samsing}}}, \bibinfo {author}
  {\bibfnamefont {M.}~\bibnamefont {{O'Dowd}}}, \bibinfo {author}
  {\bibfnamefont {B.}~\bibnamefont {{Kocsis}}},\ and\ \bibinfo {author}
  {\bibfnamefont {S.}~\bibnamefont {{Endlich}}},\ }\bibfield  {title} {\bibinfo
  {title} {{On the rate of black hole binary mergers in galactic nuclei due to
  dynamical hardening}},\ }\href {https://doi.org/10.1093/mnras/stx3134}
  {\bibfield  {journal} {\bibinfo  {journal} {\mnras}\ }\textbf {\bibinfo
  {volume} {474}},\ \bibinfo {pages} {5672} (\bibinfo {year} {2018})},\ \Eprint
  {https://arxiv.org/abs/1711.10494} {arXiv:1711.10494 [astro-ph.GA]}
  \BibitemShut {NoStop}%
\bibitem [{\citenamefont {{Chen}}\ and\ \citenamefont {{Han}}(2018)}]{Chen:18}%
  \BibitemOpen
  \bibfield  {author} {\bibinfo {author} {\bibfnamefont {X.}~\bibnamefont
  {{Chen}}}\ and\ \bibinfo {author} {\bibfnamefont {W.-B.}\ \bibnamefont
  {{Han}}},\ }\bibfield  {title} {\bibinfo {title} {{Extreme-mass-ratio
  inspirals produced by tidal capture of binary black holes}},\ }\href
  {https://doi.org/10.1038/s42005-018-0053-0} {\bibfield  {journal} {\bibinfo
  {journal} {Communications Physics}\ }\textbf {\bibinfo {volume} {1}},\
  \bibinfo {eid} {53} (\bibinfo {year} {2018})},\ \Eprint
  {https://arxiv.org/abs/1801.05780} {arXiv:1801.05780 [astro-ph.HE]}
  \BibitemShut {NoStop}%
\bibitem [{\citenamefont {{Fragione}}\ \emph {et~al.}(2019)\citenamefont
  {{Fragione}}, \citenamefont {{Leigh}},\ and\ \citenamefont
  {{Perna}}}]{Fragione:19}%
  \BibitemOpen
  \bibfield  {author} {\bibinfo {author} {\bibfnamefont {G.}~\bibnamefont
  {{Fragione}}}, \bibinfo {author} {\bibfnamefont {N.~W.~C.}\ \bibnamefont
  {{Leigh}}},\ and\ \bibinfo {author} {\bibfnamefont {R.}~\bibnamefont
  {{Perna}}},\ }\bibfield  {title} {\bibinfo {title} {{Black hole and neutron
  star mergers in galactic nuclei: the role of triples}},\ }\href
  {https://doi.org/10.1093/mnras/stz1803} {\bibfield  {journal} {\bibinfo
  {journal} {\mnras}\ }\textbf {\bibinfo {volume} {488}},\ \bibinfo {pages}
  {2825} (\bibinfo {year} {2019})},\ \Eprint {https://arxiv.org/abs/1903.09160}
  {arXiv:1903.09160 [astro-ph.GA]} \BibitemShut {NoStop}%
\bibitem [{\citenamefont {{Han}}\ and\ \citenamefont {{Chen}}(2019)}]{Han:19}%
  \BibitemOpen
  \bibfield  {author} {\bibinfo {author} {\bibfnamefont {W.-B.}\ \bibnamefont
  {{Han}}}\ and\ \bibinfo {author} {\bibfnamefont {X.}~\bibnamefont {{Chen}}},\
  }\bibfield  {title} {\bibinfo {title} {{Testing general relativity using
  binary extreme-mass-ratio inspirals}},\ }\href
  {https://doi.org/10.1093/mnrasl/slz021} {\bibfield  {journal} {\bibinfo
  {journal} {\mnras}\ }\textbf {\bibinfo {volume} {485}},\ \bibinfo {pages}
  {L29} (\bibinfo {year} {2019})},\ \Eprint {https://arxiv.org/abs/1801.07060}
  {arXiv:1801.07060 [gr-qc]} \BibitemShut {NoStop}%
\bibitem [{\citenamefont {{Samsing}}\ \emph {et~al.}(2020)\citenamefont
  {{Samsing}}, \citenamefont {{Bartos}}, \citenamefont {{D'Orazio}},
  \citenamefont {{Haiman}}, \citenamefont {{Kocsis}}, \citenamefont {{Leigh}},
  \citenamefont {{Liu}}, \citenamefont {{Pessah}},\ and\ \citenamefont
  {{Tagawa}}}]{Samsing:20}%
  \BibitemOpen
  \bibfield  {author} {\bibinfo {author} {\bibfnamefont {J.}~\bibnamefont
  {{Samsing}}}, \bibinfo {author} {\bibfnamefont {I.}~\bibnamefont {{Bartos}}},
  \bibinfo {author} {\bibfnamefont {D.~J.}\ \bibnamefont {{D'Orazio}}},
  \bibinfo {author} {\bibfnamefont {Z.}~\bibnamefont {{Haiman}}}, \bibinfo
  {author} {\bibfnamefont {B.}~\bibnamefont {{Kocsis}}}, \bibinfo {author}
  {\bibfnamefont {N.~W.~C.}\ \bibnamefont {{Leigh}}}, \bibinfo {author}
  {\bibfnamefont {B.}~\bibnamefont {{Liu}}}, \bibinfo {author} {\bibfnamefont
  {M.~E.}\ \bibnamefont {{Pessah}}},\ and\ \bibinfo {author} {\bibfnamefont
  {H.}~\bibnamefont {{Tagawa}}},\ }\bibfield  {title} {\bibinfo {title}
  {{Active Galactic Nuclei as Factories for Eccentric Black Hole Mergers}},\
  }\href@noop {} {\bibfield  {journal} {\bibinfo  {journal} {arXiv e-prints}\
  ,\ \bibinfo {eid} {arXiv:2010.09765}} (\bibinfo {year} {2020})},\ \Eprint
  {https://arxiv.org/abs/2010.09765} {arXiv:2010.09765 [astro-ph.HE]}
  \BibitemShut {NoStop}%
\bibitem [{\citenamefont {{Chen}}\ \emph {et~al.}(2020)\citenamefont {{Chen}},
  \citenamefont {{Xuan}},\ and\ \citenamefont {{Peng}}}]{Chen:20}%
  \BibitemOpen
  \bibfield  {author} {\bibinfo {author} {\bibfnamefont {X.}~\bibnamefont
  {{Chen}}}, \bibinfo {author} {\bibfnamefont {Z.-Y.}\ \bibnamefont {{Xuan}}},\
  and\ \bibinfo {author} {\bibfnamefont {P.}~\bibnamefont {{Peng}}},\
  }\bibfield  {title} {\bibinfo {title} {{Fake Massive Black Holes in the
  Milli-Hertz Gravitational-wave Band}},\ }\href
  {https://doi.org/10.3847/1538-4357/ab919f} {\bibfield  {journal} {\bibinfo
  {journal} {\apj}\ }\textbf {\bibinfo {volume} {896}},\ \bibinfo {eid} {171}
  (\bibinfo {year} {2020})},\ \Eprint {https://arxiv.org/abs/2003.08639}
  {arXiv:2003.08639 [astro-ph.HE]} \BibitemShut {NoStop}%
\bibitem [{\citenamefont {{Chen}}(2020)}]{Chen:20b}%
  \BibitemOpen
  \bibfield  {author} {\bibinfo {author} {\bibfnamefont {X.}~\bibnamefont
  {{Chen}}},\ }\bibfield  {title} {\bibinfo {title} {{Distortion of
  Gravitational-wave Signals by Astrophysical Environments}},\ }\href@noop {}
  {\bibfield  {journal} {\bibinfo  {journal} {arXiv e-prints}\ ,\ \bibinfo
  {eid} {arXiv:2009.07626}} (\bibinfo {year} {2020})},\ \Eprint
  {https://arxiv.org/abs/2009.07626} {arXiv:2009.07626 [astro-ph.HE]}
  \BibitemShut {NoStop}%
\bibitem [{\citenamefont {{Caputo}}\ \emph {et~al.}(2020)\citenamefont
  {{Caputo}}, \citenamefont {{Sberna}}, \citenamefont {{Toubiana}},
  \citenamefont {{Babak}}, \citenamefont {{Barausse}}, \citenamefont
  {{Marsat}},\ and\ \citenamefont {{Pani}}}]{Caputo:20}%
  \BibitemOpen
  \bibfield  {author} {\bibinfo {author} {\bibfnamefont {A.}~\bibnamefont
  {{Caputo}}}, \bibinfo {author} {\bibfnamefont {L.}~\bibnamefont {{Sberna}}},
  \bibinfo {author} {\bibfnamefont {A.}~\bibnamefont {{Toubiana}}}, \bibinfo
  {author} {\bibfnamefont {S.}~\bibnamefont {{Babak}}}, \bibinfo {author}
  {\bibfnamefont {E.}~\bibnamefont {{Barausse}}}, \bibinfo {author}
  {\bibfnamefont {S.}~\bibnamefont {{Marsat}}},\ and\ \bibinfo {author}
  {\bibfnamefont {P.}~\bibnamefont {{Pani}}},\ }\bibfield  {title} {\bibinfo
  {title} {{Gravitational-wave Detection and Parameter Estimation for Accreting
  Black-hole Binaries and Their Electromagnetic Counterpart}},\ }\href
  {https://doi.org/10.3847/1538-4357/ab7b66} {\bibfield  {journal} {\bibinfo
  {journal} {\apj}\ }\textbf {\bibinfo {volume} {892}},\ \bibinfo {eid} {90}
  (\bibinfo {year} {2020})},\ \Eprint {https://arxiv.org/abs/2001.03620}
  {arXiv:2001.03620 [astro-ph.HE]} \BibitemShut {NoStop}%
\bibitem [{\citenamefont {{Toubiana}}\ \emph {et~al.}(2021)\citenamefont
  {{Toubiana}}, \citenamefont {{Sberna}}, \citenamefont {{Caputo}},
  \citenamefont {{Cusin}}, \citenamefont {{Marsat}}, \citenamefont {{Jani}},
  \citenamefont {{Babak}}, \citenamefont {{Barausse}}, \citenamefont
  {{Caprini}}, \citenamefont {{Pani}}, \citenamefont {{Sesana}},\ and\
  \citenamefont {{Tamanini}}}]{Toubiana:21}%
  \BibitemOpen
  \bibfield  {author} {\bibinfo {author} {\bibfnamefont {A.}~\bibnamefont
  {{Toubiana}}}, \bibinfo {author} {\bibfnamefont {L.}~\bibnamefont
  {{Sberna}}}, \bibinfo {author} {\bibfnamefont {A.}~\bibnamefont {{Caputo}}},
  \bibinfo {author} {\bibfnamefont {G.}~\bibnamefont {{Cusin}}}, \bibinfo
  {author} {\bibfnamefont {S.}~\bibnamefont {{Marsat}}}, \bibinfo {author}
  {\bibfnamefont {K.}~\bibnamefont {{Jani}}}, \bibinfo {author} {\bibfnamefont
  {S.}~\bibnamefont {{Babak}}}, \bibinfo {author} {\bibfnamefont
  {E.}~\bibnamefont {{Barausse}}}, \bibinfo {author} {\bibfnamefont
  {C.}~\bibnamefont {{Caprini}}}, \bibinfo {author} {\bibfnamefont
  {P.}~\bibnamefont {{Pani}}}, \bibinfo {author} {\bibfnamefont
  {A.}~\bibnamefont {{Sesana}}},\ and\ \bibinfo {author} {\bibfnamefont
  {N.}~\bibnamefont {{Tamanini}}},\ }\bibfield  {title} {\bibinfo {title}
  {{Detectable Environmental Effects in GW190521-like Black-Hole Binaries with
  LISA}},\ }\href {https://doi.org/10.1103/PhysRevLett.126.101105} {\bibfield
  {journal} {\bibinfo  {journal} {\prl}\ }\textbf {\bibinfo {volume} {126}},\
  \bibinfo {eid} {101105} (\bibinfo {year} {2021})},\ \Eprint
  {https://arxiv.org/abs/2010.06056} {arXiv:2010.06056 [astro-ph.HE]}
  \BibitemShut {NoStop}%
\bibitem [{\citenamefont {{Inayoshi}}\ \emph {et~al.}(2017)\citenamefont
  {{Inayoshi}}, \citenamefont {{Tamanini}}, \citenamefont {{Caprini}},\ and\
  \citenamefont {{Haiman}}}]{Inayoshi:17}%
  \BibitemOpen
  \bibfield  {author} {\bibinfo {author} {\bibfnamefont {K.}~\bibnamefont
  {{Inayoshi}}}, \bibinfo {author} {\bibfnamefont {N.}~\bibnamefont
  {{Tamanini}}}, \bibinfo {author} {\bibfnamefont {C.}~\bibnamefont
  {{Caprini}}},\ and\ \bibinfo {author} {\bibfnamefont {Z.}~\bibnamefont
  {{Haiman}}},\ }\bibfield  {title} {\bibinfo {title} {{Probing stellar binary
  black hole formation in galactic nuclei via the imprint of their center of
  mass acceleration on their gravitational wave signal}},\ }\href
  {https://doi.org/10.1103/PhysRevD.96.063014} {\bibfield  {journal} {\bibinfo
  {journal} {\prd}\ }\textbf {\bibinfo {volume} {96}},\ \bibinfo {eid} {063014}
  (\bibinfo {year} {2017})},\ \Eprint {https://arxiv.org/abs/1702.06529}
  {arXiv:1702.06529 [astro-ph.HE]} \BibitemShut {NoStop}%
\bibitem [{\citenamefont {{Randall}}\ and\ \citenamefont
  {{Xianyu}}(2019)}]{Randall:19}%
  \BibitemOpen
  \bibfield  {author} {\bibinfo {author} {\bibfnamefont {L.}~\bibnamefont
  {{Randall}}}\ and\ \bibinfo {author} {\bibfnamefont {Z.-Z.}\ \bibnamefont
  {{Xianyu}}},\ }\bibfield  {title} {\bibinfo {title} {{A Direct Probe of Mass
  Density near Inspiraling Binary Black Holes}},\ }\href
  {https://doi.org/10.3847/1538-4357/ab20c6} {\bibfield  {journal} {\bibinfo
  {journal} {\apj}\ }\textbf {\bibinfo {volume} {878}},\ \bibinfo {eid} {75}
  (\bibinfo {year} {2019})},\ \Eprint {https://arxiv.org/abs/1805.05335}
  {arXiv:1805.05335 [gr-qc]} \BibitemShut {NoStop}%
\bibitem [{\citenamefont {{Hoang}}\ \emph {et~al.}(2019)\citenamefont
  {{Hoang}}, \citenamefont {{Naoz}}, \citenamefont {{Kocsis}}, \citenamefont
  {{Farr}},\ and\ \citenamefont {{McIver}}}]{Hoang:19}%
  \BibitemOpen
  \bibfield  {author} {\bibinfo {author} {\bibfnamefont {B.-M.}\ \bibnamefont
  {{Hoang}}}, \bibinfo {author} {\bibfnamefont {S.}~\bibnamefont {{Naoz}}},
  \bibinfo {author} {\bibfnamefont {B.}~\bibnamefont {{Kocsis}}}, \bibinfo
  {author} {\bibfnamefont {W.~M.}\ \bibnamefont {{Farr}}},\ and\ \bibinfo
  {author} {\bibfnamefont {J.}~\bibnamefont {{McIver}}},\ }\bibfield  {title}
  {\bibinfo {title} {{Detecting Supermassive Black Hole-induced Binary
  Eccentricity Oscillations with LISA}},\ }\href
  {https://doi.org/10.3847/2041-8213/ab14f7} {\bibfield  {journal} {\bibinfo
  {journal} {\apjl}\ }\textbf {\bibinfo {volume} {875}},\ \bibinfo {eid} {L31}
  (\bibinfo {year} {2019})},\ \Eprint {https://arxiv.org/abs/1903.00134}
  {arXiv:1903.00134 [astro-ph.HE]} \BibitemShut {NoStop}%
\bibitem [{\citenamefont {{Deme}}\ \emph {et~al.}(2020)\citenamefont {{Deme}},
  \citenamefont {{Hoang}}, \citenamefont {{Naoz}},\ and\ \citenamefont
  {{Kocsis}}}]{Deme:20}%
  \BibitemOpen
  \bibfield  {author} {\bibinfo {author} {\bibfnamefont {B.}~\bibnamefont
  {{Deme}}}, \bibinfo {author} {\bibfnamefont {B.-M.}\ \bibnamefont {{Hoang}}},
  \bibinfo {author} {\bibfnamefont {S.}~\bibnamefont {{Naoz}}},\ and\ \bibinfo
  {author} {\bibfnamefont {B.}~\bibnamefont {{Kocsis}}},\ }\bibfield  {title}
  {\bibinfo {title} {{Detecting Kozai-Lidov Imprints on the Gravitational Waves
  of Intermediate-mass Black Holes in Galactic Nuclei}},\ }\href
  {https://doi.org/10.3847/1538-4357/abafa3} {\bibfield  {journal} {\bibinfo
  {journal} {\apj}\ }\textbf {\bibinfo {volume} {901}},\ \bibinfo {eid} {125}
  (\bibinfo {year} {2020})},\ \Eprint {https://arxiv.org/abs/2005.03677}
  {arXiv:2005.03677 [astro-ph.HE]} \BibitemShut {NoStop}%
\bibitem [{\citenamefont {{Chandramouli}}\ and\ \citenamefont
  {{Yunes}}(2021)}]{Chandramouli:21}%
  \BibitemOpen
  \bibfield  {author} {\bibinfo {author} {\bibfnamefont {R.~S.}\ \bibnamefont
  {{Chandramouli}}}\ and\ \bibinfo {author} {\bibfnamefont {N.}~\bibnamefont
  {{Yunes}}},\ }\bibfield  {title} {\bibinfo {title} {{The Trouble with
  Triples: Ready-to-use Analytic Model for Gravitational Waves from a
  Hierarchical Triple with Kozai-Lidov Oscillations}},\ }\href@noop {}
  {\bibfield  {journal} {\bibinfo  {journal} {arXiv e-prints}\ ,\ \bibinfo
  {eid} {arXiv:2107.00741}} (\bibinfo {year} {2021})},\ \Eprint
  {https://arxiv.org/abs/2107.00741} {arXiv:2107.00741 [gr-qc]} \BibitemShut
  {NoStop}%
\bibitem [{\citenamefont {{Yu}}\ and\ \citenamefont {{Chen}}(2021)}]{Yu:20d}%
  \BibitemOpen
  \bibfield  {author} {\bibinfo {author} {\bibfnamefont {H.}~\bibnamefont
  {{Yu}}}\ and\ \bibinfo {author} {\bibfnamefont {Y.}~\bibnamefont {{Chen}}},\
  }\bibfield  {title} {\bibinfo {title} {{Direct Determination of Supermassive
  Black Hole Properties with Gravitational-Wave Radiation from Surrounding
  Stellar-Mass Black Hole Binaries}},\ }\href
  {https://doi.org/10.1103/PhysRevLett.126.021101} {\bibfield  {journal}
  {\bibinfo  {journal} {\prl}\ }\textbf {\bibinfo {volume} {126}},\ \bibinfo
  {eid} {021101} (\bibinfo {year} {2021})},\ \Eprint
  {https://arxiv.org/abs/2009.02579} {arXiv:2009.02579 [gr-qc]} \BibitemShut
  {NoStop}%
\bibitem [{\citenamefont {{Will}}(2018)}]{Will:18}%
  \BibitemOpen
  \bibfield  {author} {\bibinfo {author} {\bibfnamefont {C.~M.}\ \bibnamefont
  {{Will}}},\ }\bibfield  {title} {\bibinfo {title} {{New General Relativistic
  Contribution to Mercury's Perihelion Advance}},\ }\href
  {https://doi.org/10.1103/PhysRevLett.120.191101} {\bibfield  {journal}
  {\bibinfo  {journal} {\prl}\ }\textbf {\bibinfo {volume} {120}},\ \bibinfo
  {eid} {191101} (\bibinfo {year} {2018})},\ \Eprint
  {https://arxiv.org/abs/1802.05304} {arXiv:1802.05304 [gr-qc]} \BibitemShut
  {NoStop}%
\bibitem [{\citenamefont {{Liu}}\ \emph {et~al.}(2019)\citenamefont {{Liu}},
  \citenamefont {{Lai}},\ and\ \citenamefont {{Wang}}}]{Liu:19}%
  \BibitemOpen
  \bibfield  {author} {\bibinfo {author} {\bibfnamefont {B.}~\bibnamefont
  {{Liu}}}, \bibinfo {author} {\bibfnamefont {D.}~\bibnamefont {{Lai}}},\ and\
  \bibinfo {author} {\bibfnamefont {Y.-H.}\ \bibnamefont {{Wang}}},\ }\bibfield
   {title} {\bibinfo {title} {{Binary Mergers near a Supermassive Black Hole:
  Relativistic Effects in Triples}},\ }\href
  {https://doi.org/10.3847/2041-8213/ab40c0} {\bibfield  {journal} {\bibinfo
  {journal} {\apjl}\ }\textbf {\bibinfo {volume} {883}},\ \bibinfo {eid} {L7}
  (\bibinfo {year} {2019})},\ \Eprint {https://arxiv.org/abs/1906.07726}
  {arXiv:1906.07726 [astro-ph.HE]} \BibitemShut {NoStop}%
\bibitem [{\citenamefont {{Kuntz}}\ \emph {et~al.}(2021)\citenamefont
  {{Kuntz}}, \citenamefont {{Serra}},\ and\ \citenamefont
  {{Trincherini}}}]{Kuntz:21}%
  \BibitemOpen
  \bibfield  {author} {\bibinfo {author} {\bibfnamefont {A.}~\bibnamefont
  {{Kuntz}}}, \bibinfo {author} {\bibfnamefont {F.}~\bibnamefont {{Serra}}},\
  and\ \bibinfo {author} {\bibfnamefont {E.}~\bibnamefont {{Trincherini}}},\
  }\bibfield  {title} {\bibinfo {title} {{Effective two-body approach to the
  hierarchical three-body problem}},\ }\href
  {https://doi.org/10.1103/PhysRevD.104.024016} {\bibfield  {journal} {\bibinfo
   {journal} {\prd}\ }\textbf {\bibinfo {volume} {104}},\ \bibinfo {eid}
  {024016} (\bibinfo {year} {2021})},\ \Eprint
  {https://arxiv.org/abs/2104.13387} {arXiv:2104.13387 [hep-th]} \BibitemShut
  {NoStop}%
\bibitem [{\citenamefont {{Peterson}}(2014)}]{Peterson:14}%
  \BibitemOpen
  \bibfield  {author} {\bibinfo {author} {\bibfnamefont {B.~M.}\ \bibnamefont
  {{Peterson}}},\ }\bibfield  {title} {\bibinfo {title} {{Measuring the Masses
  of Supermassive Black Holes}},\ }\href
  {https://doi.org/10.1007/s11214-013-9987-4} {\bibfield  {journal} {\bibinfo
  {journal} {\ssr}\ }\textbf {\bibinfo {volume} {183}},\ \bibinfo {pages} {253}
  (\bibinfo {year} {2014})}\BibitemShut {NoStop}%
\bibitem [{\citenamefont {{D'Orazio}}\ and\ \citenamefont
  {{Loeb}}(2020)}]{DOrazio:20}%
  \BibitemOpen
  \bibfield  {author} {\bibinfo {author} {\bibfnamefont {D.~J.}\ \bibnamefont
  {{D'Orazio}}}\ and\ \bibinfo {author} {\bibfnamefont {A.}~\bibnamefont
  {{Loeb}}},\ }\bibfield  {title} {\bibinfo {title} {{Repeated gravitational
  lensing of gravitational waves in hierarchical black hole triples}},\ }\href
  {https://doi.org/10.1103/PhysRevD.101.083031} {\bibfield  {journal} {\bibinfo
   {journal} {\prd}\ }\textbf {\bibinfo {volume} {101}},\ \bibinfo {eid}
  {083031} (\bibinfo {year} {2020})},\ \Eprint
  {https://arxiv.org/abs/1910.02966} {arXiv:1910.02966 [astro-ph.HE]}
  \BibitemShut {NoStop}%
\bibitem [{\citenamefont {{Virbhadra}}\ and\ \citenamefont
  {{Ellis}}(2000)}]{Virbhadra:00}%
  \BibitemOpen
  \bibfield  {author} {\bibinfo {author} {\bibfnamefont {K.~S.}\ \bibnamefont
  {{Virbhadra}}}\ and\ \bibinfo {author} {\bibfnamefont {G.~F.~R.}\
  \bibnamefont {{Ellis}}},\ }\bibfield  {title} {\bibinfo {title}
  {{Schwarzschild black hole lensing}},\ }\href
  {https://doi.org/10.1103/PhysRevD.62.084003} {\bibfield  {journal} {\bibinfo
  {journal} {\prd}\ }\textbf {\bibinfo {volume} {62}},\ \bibinfo {eid} {084003}
  (\bibinfo {year} {2000})},\ \Eprint {https://arxiv.org/abs/astro-ph/9904193}
  {arXiv:astro-ph/9904193 [astro-ph]} \BibitemShut {NoStop}%
\bibitem [{\citenamefont {{Bozza}}(2002)}]{Bozza:02}%
  \BibitemOpen
  \bibfield  {author} {\bibinfo {author} {\bibfnamefont {V.}~\bibnamefont
  {{Bozza}}},\ }\bibfield  {title} {\bibinfo {title} {{Gravitational lensing in
  the strong field limit}},\ }\href
  {https://doi.org/10.1103/PhysRevD.66.103001} {\bibfield  {journal} {\bibinfo
  {journal} {\prd}\ }\textbf {\bibinfo {volume} {66}},\ \bibinfo {eid} {103001}
  (\bibinfo {year} {2002})},\ \Eprint {https://arxiv.org/abs/gr-qc/0208075}
  {arXiv:gr-qc/0208075 [gr-qc]} \BibitemShut {NoStop}%
\bibitem [{\citenamefont {{Bozza}}\ and\ \citenamefont
  {{Scarpetta}}(2007)}]{Bozza:07}%
  \BibitemOpen
  \bibfield  {author} {\bibinfo {author} {\bibfnamefont {V.}~\bibnamefont
  {{Bozza}}}\ and\ \bibinfo {author} {\bibfnamefont {G.}~\bibnamefont
  {{Scarpetta}}},\ }\bibfield  {title} {\bibinfo {title} {{Strong deflection
  limit of black hole gravitational lensing with arbitrary source distances}},\
  }\href {https://doi.org/10.1103/PhysRevD.76.083008} {\bibfield  {journal}
  {\bibinfo  {journal} {\prd}\ }\textbf {\bibinfo {volume} {76}},\ \bibinfo
  {eid} {083008} (\bibinfo {year} {2007})},\ \Eprint
  {https://arxiv.org/abs/0705.0246} {arXiv:0705.0246 [gr-qc]} \BibitemShut
  {NoStop}%
\bibitem [{\citenamefont {{Eiroa}}\ and\ \citenamefont
  {{Sendra}}(2011)}]{Eiroa:11}%
  \BibitemOpen
  \bibfield  {author} {\bibinfo {author} {\bibfnamefont {E.~F.}\ \bibnamefont
  {{Eiroa}}}\ and\ \bibinfo {author} {\bibfnamefont {C.~M.}\ \bibnamefont
  {{Sendra}}},\ }\bibfield  {title} {\bibinfo {title} {{Gravitational lensing
  by a regular black hole}},\ }\href
  {https://doi.org/10.1088/0264-9381/28/8/085008} {\bibfield  {journal}
  {\bibinfo  {journal} {Classical and Quantum Gravity}\ }\textbf {\bibinfo
  {volume} {28}},\ \bibinfo {eid} {085008} (\bibinfo {year} {2011})},\ \Eprint
  {https://arxiv.org/abs/1011.2455} {arXiv:1011.2455 [gr-qc]} \BibitemShut
  {NoStop}%
\bibitem [{\citenamefont {{Virbhadra}}(2009)}]{Virbhadra:09}%
  \BibitemOpen
  \bibfield  {author} {\bibinfo {author} {\bibfnamefont {K.~S.}\ \bibnamefont
  {{Virbhadra}}},\ }\bibfield  {title} {\bibinfo {title} {{Relativistic images
  of Schwarzschild black hole lensing}},\ }\href
  {https://doi.org/10.1103/PhysRevD.79.083004} {\bibfield  {journal} {\bibinfo
  {journal} {\prd}\ }\textbf {\bibinfo {volume} {79}},\ \bibinfo {eid} {083004}
  (\bibinfo {year} {2009})},\ \Eprint {https://arxiv.org/abs/0810.2109}
  {arXiv:0810.2109 [gr-qc]} \BibitemShut {NoStop}%
\bibitem [{\citenamefont {{Luminet}}(1979)}]{Luminet:79}%
  \BibitemOpen
  \bibfield  {author} {\bibinfo {author} {\bibfnamefont {J.~P.}\ \bibnamefont
  {{Luminet}}},\ }\bibfield  {title} {\bibinfo {title} {{Image of a spherical
  black hole with thin accretion disk.}},\ }\href@noop {} {\bibfield  {journal}
  {\bibinfo  {journal} {\aap}\ }\textbf {\bibinfo {volume} {75}},\ \bibinfo
  {pages} {228} (\bibinfo {year} {1979})}\BibitemShut {NoStop}%
\bibitem [{\citenamefont {{Holz}}\ and\ \citenamefont
  {{Wheeler}}(2002)}]{Holz:02}%
  \BibitemOpen
  \bibfield  {author} {\bibinfo {author} {\bibfnamefont {D.~E.}\ \bibnamefont
  {{Holz}}}\ and\ \bibinfo {author} {\bibfnamefont {J.~A.}\ \bibnamefont
  {{Wheeler}}},\ }\bibfield  {title} {\bibinfo {title} {{Retro-MACHOs:
  {\ensuremath{\pi}} in the Sky?}},\ }\href {https://doi.org/10.1086/342463}
  {\bibfield  {journal} {\bibinfo  {journal} {\apj}\ }\textbf {\bibinfo
  {volume} {578}},\ \bibinfo {pages} {330} (\bibinfo {year} {2002})},\ \Eprint
  {https://arxiv.org/abs/astro-ph/0209039} {arXiv:astro-ph/0209039 [astro-ph]}
  \BibitemShut {NoStop}%
\bibitem [{\citenamefont {{Eiroa}}\ and\ \citenamefont
  {{Torres}}(2004)}]{Eiroa:04}%
  \BibitemOpen
  \bibfield  {author} {\bibinfo {author} {\bibfnamefont {E.~F.}\ \bibnamefont
  {{Eiroa}}}\ and\ \bibinfo {author} {\bibfnamefont {D.~F.}\ \bibnamefont
  {{Torres}}},\ }\bibfield  {title} {\bibinfo {title} {{Strong field limit
  analysis of gravitational retrolensing}},\ }\href
  {https://doi.org/10.1103/PhysRevD.69.063004} {\bibfield  {journal} {\bibinfo
  {journal} {\prd}\ }\textbf {\bibinfo {volume} {69}},\ \bibinfo {eid} {063004}
  (\bibinfo {year} {2004})},\ \Eprint {https://arxiv.org/abs/gr-qc/0311013}
  {arXiv:gr-qc/0311013 [gr-qc]} \BibitemShut {NoStop}%
\bibitem [{\citenamefont {{Bozza}}(2010)}]{Bozza:10}%
  \BibitemOpen
  \bibfield  {author} {\bibinfo {author} {\bibfnamefont {V.}~\bibnamefont
  {{Bozza}}},\ }\bibfield  {title} {\bibinfo {title} {{Gravitational lensing by
  black holes}},\ }\href {https://doi.org/10.1007/s10714-010-0988-2} {\bibfield
   {journal} {\bibinfo  {journal} {General Relativity and Gravitation}\
  }\textbf {\bibinfo {volume} {42}},\ \bibinfo {pages} {2269} (\bibinfo {year}
  {2010})},\ \Eprint {https://arxiv.org/abs/0911.2187} {arXiv:0911.2187
  [gr-qc]} \BibitemShut {NoStop}%
\bibitem [{\citenamefont {{Eiroa}}(2012)}]{Eiroa:12}%
  \BibitemOpen
  \bibfield  {author} {\bibinfo {author} {\bibfnamefont {E.~F.}\ \bibnamefont
  {{Eiroa}}},\ }\bibfield  {title} {\bibinfo {title} {{Strong deflection
  gravitational lensing}},\ }\href@noop {} {\bibfield  {journal} {\bibinfo
  {journal} {arXiv e-prints}\ ,\ \bibinfo {eid} {arXiv:1212.4535}} (\bibinfo
  {year} {2012})},\ \Eprint {https://arxiv.org/abs/1212.4535} {arXiv:1212.4535
  [gr-qc]} \BibitemShut {NoStop}%
\bibitem [{\citenamefont {{Takahashi}}\ and\ \citenamefont
  {{Nakamura}}(2003)}]{Takahashi:03}%
  \BibitemOpen
  \bibfield  {author} {\bibinfo {author} {\bibfnamefont {R.}~\bibnamefont
  {{Takahashi}}}\ and\ \bibinfo {author} {\bibfnamefont {T.}~\bibnamefont
  {{Nakamura}}},\ }\bibfield  {title} {\bibinfo {title} {{Wave Effects in the
  Gravitational Lensing of Gravitational Waves from Chirping Binaries}},\
  }\href {https://doi.org/10.1086/377430} {\bibfield  {journal} {\bibinfo
  {journal} {\apj}\ }\textbf {\bibinfo {volume} {595}},\ \bibinfo {pages}
  {1039} (\bibinfo {year} {2003})},\ \Eprint
  {https://arxiv.org/abs/astro-ph/0305055} {arXiv:astro-ph/0305055 [astro-ph]}
  \BibitemShut {NoStop}%
\bibitem [{\citenamefont {{Apostolatos}}\ \emph {et~al.}(1994)\citenamefont
  {{Apostolatos}}, \citenamefont {{Cutler}}, \citenamefont {{Sussman}},\ and\
  \citenamefont {{Thorne}}}]{Apostolatos:94}%
  \BibitemOpen
  \bibfield  {author} {\bibinfo {author} {\bibfnamefont {T.~A.}\ \bibnamefont
  {{Apostolatos}}}, \bibinfo {author} {\bibfnamefont {C.}~\bibnamefont
  {{Cutler}}}, \bibinfo {author} {\bibfnamefont {G.~J.}\ \bibnamefont
  {{Sussman}}},\ and\ \bibinfo {author} {\bibfnamefont {K.~S.}\ \bibnamefont
  {{Thorne}}},\ }\bibfield  {title} {\bibinfo {title} {{Spin-induced orbital
  precession and its modulation of the gravitational waveforms from merging
  binaries}},\ }\href {https://doi.org/10.1103/PhysRevD.49.6274} {\bibfield
  {journal} {\bibinfo  {journal} {\prd}\ }\textbf {\bibinfo {volume} {49}},\
  \bibinfo {pages} {6274} (\bibinfo {year} {1994})}\BibitemShut {NoStop}%
\bibitem [{\citenamefont {{Cutler}}(1998)}]{Cutler:98}%
  \BibitemOpen
  \bibfield  {author} {\bibinfo {author} {\bibfnamefont {C.}~\bibnamefont
  {{Cutler}}},\ }\bibfield  {title} {\bibinfo {title} {{Angular resolution of
  the LISA gravitational wave detector}},\ }\href
  {https://doi.org/10.1103/PhysRevD.57.7089} {\bibfield  {journal} {\bibinfo
  {journal} {\prd}\ }\textbf {\bibinfo {volume} {57}},\ \bibinfo {pages} {7089}
  (\bibinfo {year} {1998})},\ \Eprint {https://arxiv.org/abs/gr-qc/9703068}
  {arXiv:gr-qc/9703068 [gr-qc]} \BibitemShut {NoStop}%
\bibitem [{\citenamefont {{Dhurandhar}}\ \emph {et~al.}(2005)\citenamefont
  {{Dhurandhar}}, \citenamefont {{Nayak}}, \citenamefont {{Koshti}},\ and\
  \citenamefont {{Vinet}}}]{Dhurandhar:05}%
  \BibitemOpen
  \bibfield  {author} {\bibinfo {author} {\bibfnamefont {S.~V.}\ \bibnamefont
  {{Dhurandhar}}}, \bibinfo {author} {\bibfnamefont {K.~R.}\ \bibnamefont
  {{Nayak}}}, \bibinfo {author} {\bibfnamefont {S.}~\bibnamefont {{Koshti}}},\
  and\ \bibinfo {author} {\bibfnamefont {J.~Y.}\ \bibnamefont {{Vinet}}},\
  }\bibfield  {title} {\bibinfo {title} {{Fundamentals of the LISA stable
  flight formation}},\ }\href {https://doi.org/10.1088/0264-9381/22/3/002}
  {\bibfield  {journal} {\bibinfo  {journal} {Classical and Quantum Gravity}\
  }\textbf {\bibinfo {volume} {22}},\ \bibinfo {pages} {481} (\bibinfo {year}
  {2005})},\ \Eprint {https://arxiv.org/abs/gr-qc/0410093} {arXiv:gr-qc/0410093
  [astro-ph]} \BibitemShut {NoStop}%
\bibitem [{\citenamefont {{D'Orazio}}\ and\ \citenamefont {{Di
  Stefano}}(2018)}]{DOrazio:18}%
  \BibitemOpen
  \bibfield  {author} {\bibinfo {author} {\bibfnamefont {D.~J.}\ \bibnamefont
  {{D'Orazio}}}\ and\ \bibinfo {author} {\bibfnamefont {R.}~\bibnamefont {{Di
  Stefano}}},\ }\bibfield  {title} {\bibinfo {title} {{Periodic self-lensing
  from accreting massive black hole binaries}},\ }\href
  {https://doi.org/10.1093/mnras/stx2936} {\bibfield  {journal} {\bibinfo
  {journal} {\mnras}\ }\textbf {\bibinfo {volume} {474}},\ \bibinfo {pages}
  {2975} (\bibinfo {year} {2018})},\ \Eprint {https://arxiv.org/abs/1707.02335}
  {arXiv:1707.02335 [astro-ph.HE]} \BibitemShut {NoStop}%
\bibitem [{\citenamefont {{Nakamura}}\ and\ \citenamefont
  {{Deguchi}}(1999)}]{Nakamura:99}%
  \BibitemOpen
  \bibfield  {author} {\bibinfo {author} {\bibfnamefont {T.~T.}\ \bibnamefont
  {{Nakamura}}}\ and\ \bibinfo {author} {\bibfnamefont {S.}~\bibnamefont
  {{Deguchi}}},\ }\bibfield  {title} {\bibinfo {title} {{Wave Optics in
  Gravitational Lensing}},\ }\href {https://doi.org/10.1143/PTPS.133.137}
  {\bibfield  {journal} {\bibinfo  {journal} {Progress of Theoretical Physics
  Supplement}\ }\textbf {\bibinfo {volume} {133}},\ \bibinfo {pages} {137}
  (\bibinfo {year} {1999})}\BibitemShut {NoStop}%
\bibitem [{\citenamefont {{Matzner}}\ \emph {et~al.}(1985)\citenamefont
  {{Matzner}}, \citenamefont {{Dewitte-Morette}}, \citenamefont {{Nelson}},\
  and\ \citenamefont {{Zhang}}}]{Matzner:85}%
  \BibitemOpen
  \bibfield  {author} {\bibinfo {author} {\bibfnamefont {R.~A.}\ \bibnamefont
  {{Matzner}}}, \bibinfo {author} {\bibfnamefont {C.}~\bibnamefont
  {{Dewitte-Morette}}}, \bibinfo {author} {\bibfnamefont {B.}~\bibnamefont
  {{Nelson}}},\ and\ \bibinfo {author} {\bibfnamefont {T.-R.}\ \bibnamefont
  {{Zhang}}},\ }\bibfield  {title} {\bibinfo {title} {{Glory scattering by
  black holes}},\ }\href {https://doi.org/10.1103/PhysRevD.31.1869} {\bibfield
  {journal} {\bibinfo  {journal} {\prd}\ }\textbf {\bibinfo {volume} {31}},\
  \bibinfo {pages} {1869} (\bibinfo {year} {1985})}\BibitemShut {NoStop}%
\bibitem [{\citenamefont {{Bozza}}\ and\ \citenamefont
  {{Mancini}}(2004)}]{Bozza:04}%
  \BibitemOpen
  \bibfield  {author} {\bibinfo {author} {\bibfnamefont {V.}~\bibnamefont
  {{Bozza}}}\ and\ \bibinfo {author} {\bibfnamefont {L.}~\bibnamefont
  {{Mancini}}},\ }\bibfield  {title} {\bibinfo {title} {{Time Delay in Black
  Hole Gravitational Lensing as a Distance Estimator}},\ }\href
  {https://doi.org/10.1023/B:GERG.0000010486.58026.4f} {\bibfield  {journal}
  {\bibinfo  {journal} {General Relativity and Gravitation}\ }\textbf {\bibinfo
  {volume} {36}},\ \bibinfo {pages} {435} (\bibinfo {year} {2004})},\ \Eprint
  {https://arxiv.org/abs/gr-qc/0305007} {arXiv:gr-qc/0305007 [gr-qc]}
  \BibitemShut {NoStop}%
\bibitem [{\citenamefont {{Robson}}\ \emph {et~al.}(2019)\citenamefont
  {{Robson}}, \citenamefont {{Cornish}},\ and\ \citenamefont
  {{Liu}}}]{Robson:19}%
  \BibitemOpen
  \bibfield  {author} {\bibinfo {author} {\bibfnamefont {T.}~\bibnamefont
  {{Robson}}}, \bibinfo {author} {\bibfnamefont {N.~J.}\ \bibnamefont
  {{Cornish}}},\ and\ \bibinfo {author} {\bibfnamefont {C.}~\bibnamefont
  {{Liu}}},\ }\bibfield  {title} {\bibinfo {title} {{The construction and use
  of LISA sensitivity curves}},\ }\href
  {https://doi.org/10.1088/1361-6382/ab1101} {\bibfield  {journal} {\bibinfo
  {journal} {Classical and Quantum Gravity}\ }\textbf {\bibinfo {volume}
  {36}},\ \bibinfo {eid} {105011} (\bibinfo {year} {2019})},\ \Eprint
  {https://arxiv.org/abs/1803.01944} {arXiv:1803.01944 [astro-ph.HE]}
  \BibitemShut {NoStop}%
\bibitem [{\citenamefont {{Claudel}}\ \emph {et~al.}(2001)\citenamefont
  {{Claudel}}, \citenamefont {{Virbhadra}},\ and\ \citenamefont
  {{Ellis}}}]{Claudel:01}%
  \BibitemOpen
  \bibfield  {author} {\bibinfo {author} {\bibfnamefont {C.-M.}\ \bibnamefont
  {{Claudel}}}, \bibinfo {author} {\bibfnamefont {K.~S.}\ \bibnamefont
  {{Virbhadra}}},\ and\ \bibinfo {author} {\bibfnamefont {G.~F.~R.}\
  \bibnamefont {{Ellis}}},\ }\bibfield  {title} {\bibinfo {title} {{The
  geometry of photon surfaces}},\ }\href {https://doi.org/10.1063/1.1308507}
  {\bibfield  {journal} {\bibinfo  {journal} {Journal of Mathematical Physics}\
  }\textbf {\bibinfo {volume} {42}},\ \bibinfo {pages} {818} (\bibinfo {year}
  {2001})},\ \Eprint {https://arxiv.org/abs/gr-qc/0005050} {arXiv:gr-qc/0005050
  [gr-qc]} \BibitemShut {NoStop}%
\bibitem [{Note1()}]{Note1}%
  \BibitemOpen
  \bibinfo {note} {While using $(M_3, a_o)$ is conceptually simple, we
  nonetheless use $(M_3, \omega _o)$ when calculating the Fisher matrices in
  Sec.~\ref {sec:PE} as it is more numerically accurate.}\BibitemShut {Stop}%
\bibitem [{Note2()}]{Note2}%
  \BibitemOpen
  \bibinfo {note} {This expression assumes a circular outer orbit and $M_3\gg
  M_{1,(2)}$. When the outer orbit is elliptical, the eccentricity can be
  constrained from the Doppler shift~\cite {Yu:20d}. In this case, the
  instantaneous precession rate (see, e.g., Ref.~\cite {Barker:75}) should be
  used.}\BibitemShut {Stop}%
\bibitem [{\citenamefont {{Lindblom}}\ \emph {et~al.}(2008)\citenamefont
  {{Lindblom}}, \citenamefont {{Owen}},\ and\ \citenamefont
  {{Brown}}}]{Lindblom:08}%
  \BibitemOpen
  \bibfield  {author} {\bibinfo {author} {\bibfnamefont {L.}~\bibnamefont
  {{Lindblom}}}, \bibinfo {author} {\bibfnamefont {B.~J.}\ \bibnamefont
  {{Owen}}},\ and\ \bibinfo {author} {\bibfnamefont {D.~A.}\ \bibnamefont
  {{Brown}}},\ }\bibfield  {title} {\bibinfo {title} {{Model waveform accuracy
  standards for gravitational wave data analysis}},\ }\href
  {https://doi.org/10.1103/PhysRevD.78.124020} {\bibfield  {journal} {\bibinfo
  {journal} {\prd}\ }\textbf {\bibinfo {volume} {78}},\ \bibinfo {eid} {124020}
  (\bibinfo {year} {2008})},\ \Eprint {https://arxiv.org/abs/0809.3844}
  {arXiv:0809.3844 [gr-qc]} \BibitemShut {NoStop}%
\bibitem [{\citenamefont {{Fang}}\ \emph {et~al.}(2019)\citenamefont {{Fang}},
  \citenamefont {{Chen}},\ and\ \citenamefont {{Huang}}}]{Fang:19}%
  \BibitemOpen
  \bibfield  {author} {\bibinfo {author} {\bibfnamefont {Y.}~\bibnamefont
  {{Fang}}}, \bibinfo {author} {\bibfnamefont {X.}~\bibnamefont {{Chen}}},\
  and\ \bibinfo {author} {\bibfnamefont {Q.-G.}\ \bibnamefont {{Huang}}},\
  }\bibfield  {title} {\bibinfo {title} {{Impact of a Spinning Supermassive
  Black Hole on the Orbit and Gravitational Waves of a Nearby Compact
  Binary}},\ }\href {https://doi.org/10.3847/1538-4357/ab510e} {\bibfield
  {journal} {\bibinfo  {journal} {\apj}\ }\textbf {\bibinfo {volume} {887}},\
  \bibinfo {eid} {210} (\bibinfo {year} {2019})},\ \Eprint
  {https://arxiv.org/abs/1908.01443} {arXiv:1908.01443 [astro-ph.HE]}
  \BibitemShut {NoStop}%
\bibitem [{\citenamefont {{Harry}}\ \emph {et~al.}(2006)\citenamefont
  {{Harry}}, \citenamefont {{Fritschel}}, \citenamefont {{Shaddock}},
  \citenamefont {{Folkner}},\ and\ \citenamefont {{Phinney}}}]{Harry:06}%
  \BibitemOpen
  \bibfield  {author} {\bibinfo {author} {\bibfnamefont {G.~M.}\ \bibnamefont
  {{Harry}}}, \bibinfo {author} {\bibfnamefont {P.}~\bibnamefont
  {{Fritschel}}}, \bibinfo {author} {\bibfnamefont {D.~A.}\ \bibnamefont
  {{Shaddock}}}, \bibinfo {author} {\bibfnamefont {W.}~\bibnamefont
  {{Folkner}}},\ and\ \bibinfo {author} {\bibfnamefont {E.~S.}\ \bibnamefont
  {{Phinney}}},\ }\bibfield  {title} {\bibinfo {title} {{Laser interferometry
  for the Big Bang Observer}},\ }\href
  {https://doi.org/10.1088/0264-9381/23/15/008} {\bibfield  {journal} {\bibinfo
   {journal} {Classical and Quantum Gravity}\ }\textbf {\bibinfo {volume}
  {23}},\ \bibinfo {pages} {4887} (\bibinfo {year} {2006})}\BibitemShut
  {NoStop}%
\bibitem [{\citenamefont {{Will}}(2003)}]{Will:03}%
  \BibitemOpen
  \bibfield  {author} {\bibinfo {author} {\bibfnamefont {C.~M.}\ \bibnamefont
  {{Will}}},\ }\bibfield  {title} {\bibinfo {title} {{Propagation Speed of
  Gravity and the Relativistic Time Delay}},\ }\href
  {https://doi.org/10.1086/375164} {\bibfield  {journal} {\bibinfo  {journal}
  {\apj}\ }\textbf {\bibinfo {volume} {590}},\ \bibinfo {pages} {683} (\bibinfo
  {year} {2003})},\ \Eprint {https://arxiv.org/abs/astro-ph/0301145}
  {arXiv:astro-ph/0301145 [astro-ph]} \BibitemShut {NoStop}%
\bibitem [{\citenamefont {{Will}}(2014)}]{Will:14}%
  \BibitemOpen
  \bibfield  {author} {\bibinfo {author} {\bibfnamefont {C.~M.}\ \bibnamefont
  {{Will}}},\ }\bibfield  {title} {\bibinfo {title} {{The Confrontation between
  General Relativity and Experiment}},\ }\href
  {https://doi.org/10.12942/lrr-2014-4} {\bibfield  {journal} {\bibinfo
  {journal} {Living Reviews in Relativity}\ }\textbf {\bibinfo {volume} {17}},\
  \bibinfo {eid} {4} (\bibinfo {year} {2014})},\ \Eprint
  {https://arxiv.org/abs/1403.7377} {arXiv:1403.7377 [gr-qc]} \BibitemShut
  {NoStop}%
\bibitem [{\citenamefont {{Hongsheng}}\ and\ \citenamefont
  {{Xilong}}(2018)}]{Zhang:18}%
  \BibitemOpen
  \bibfield  {author} {\bibinfo {author} {\bibfnamefont {Z.}~\bibnamefont
  {{Hongsheng}}}\ and\ \bibinfo {author} {\bibfnamefont {F.}~\bibnamefont
  {{Xilong}}},\ }\bibfield  {title} {\bibinfo {title} {{Poisson-Arago spot for
  gravitational waves}},\ }\href@noop {} {\bibfield  {journal} {\bibinfo
  {journal} {arXiv e-prints}\ ,\ \bibinfo {eid} {arXiv:1809.06511}} (\bibinfo
  {year} {2018})},\ \Eprint {https://arxiv.org/abs/1809.06511}
  {arXiv:1809.06511 [gr-qc]} \BibitemShut {NoStop}%
\bibitem [{\citenamefont {{Reynolds}}(2013)}]{Reynolds:13}%
  \BibitemOpen
  \bibfield  {author} {\bibinfo {author} {\bibfnamefont {C.~S.}\ \bibnamefont
  {{Reynolds}}},\ }\bibfield  {title} {\bibinfo {title} {{The spin of
  supermassive black holes}},\ }\href
  {https://doi.org/10.1088/0264-9381/30/24/244004} {\bibfield  {journal}
  {\bibinfo  {journal} {Classical and Quantum Gravity}\ }\textbf {\bibinfo
  {volume} {30}},\ \bibinfo {eid} {244004} (\bibinfo {year} {2013})},\ \Eprint
  {https://arxiv.org/abs/1307.3246} {arXiv:1307.3246 [astro-ph.HE]}
  \BibitemShut {NoStop}%
\bibitem [{\citenamefont {{Liu}}\ and\ \citenamefont {{Lai}}(2021)}]{Liu:21}%
  \BibitemOpen
  \bibfield  {author} {\bibinfo {author} {\bibfnamefont {B.}~\bibnamefont
  {{Liu}}}\ and\ \bibinfo {author} {\bibfnamefont {D.}~\bibnamefont {{Lai}}},\
  }\bibfield  {title} {\bibinfo {title} {{Probing the Spins of Supermassive
  Black Holes with Gravitational Waves from Surrounding Compact Binaries}},\
  }\href@noop {} {\bibfield  {journal} {\bibinfo  {journal} {arXiv e-prints}\
  ,\ \bibinfo {eid} {arXiv:2105.02230}} (\bibinfo {year} {2021})},\ \Eprint
  {https://arxiv.org/abs/2105.02230} {arXiv:2105.02230 [astro-ph.HE]}
  \BibitemShut {NoStop}%
\bibitem [{\citenamefont {{Bozza}}(2003)}]{Bozza:03}%
  \BibitemOpen
  \bibfield  {author} {\bibinfo {author} {\bibfnamefont {V.}~\bibnamefont
  {{Bozza}}},\ }\bibfield  {title} {\bibinfo {title} {{Quasiequatorial
  gravitational lensing by spinning black holes in the strong field limit}},\
  }\href {https://doi.org/10.1103/PhysRevD.67.103006} {\bibfield  {journal}
  {\bibinfo  {journal} {\prd}\ }\textbf {\bibinfo {volume} {67}},\ \bibinfo
  {eid} {103006} (\bibinfo {year} {2003})},\ \Eprint
  {https://arxiv.org/abs/gr-qc/0210109} {arXiv:gr-qc/0210109 [gr-qc]}
  \BibitemShut {NoStop}%
\bibitem [{\citenamefont {{Bozza}}\ \emph {et~al.}(2005)\citenamefont
  {{Bozza}}, \citenamefont {{de Luca}}, \citenamefont {{Scarpetta}},\ and\
  \citenamefont {{Sereno}}}]{Bozza:05}%
  \BibitemOpen
  \bibfield  {author} {\bibinfo {author} {\bibfnamefont {V.}~\bibnamefont
  {{Bozza}}}, \bibinfo {author} {\bibfnamefont {F.}~\bibnamefont {{de Luca}}},
  \bibinfo {author} {\bibfnamefont {G.}~\bibnamefont {{Scarpetta}}},\ and\
  \bibinfo {author} {\bibfnamefont {M.}~\bibnamefont {{Sereno}}},\ }\bibfield
  {title} {\bibinfo {title} {{Analytic Kerr black hole lensing for equatorial
  observers in the strong deflection limit}},\ }\href
  {https://doi.org/10.1103/PhysRevD.72.083003} {\bibfield  {journal} {\bibinfo
  {journal} {\prd}\ }\textbf {\bibinfo {volume} {72}},\ \bibinfo {eid} {083003}
  (\bibinfo {year} {2005})},\ \Eprint {https://arxiv.org/abs/gr-qc/0507137}
  {arXiv:gr-qc/0507137 [gr-qc]} \BibitemShut {NoStop}%
\bibitem [{\citenamefont {{Bozza}}\ \emph {et~al.}(2006)\citenamefont
  {{Bozza}}, \citenamefont {{de Luca}},\ and\ \citenamefont
  {{Scarpetta}}}]{Bozza:06}%
  \BibitemOpen
  \bibfield  {author} {\bibinfo {author} {\bibfnamefont {V.}~\bibnamefont
  {{Bozza}}}, \bibinfo {author} {\bibfnamefont {F.}~\bibnamefont {{de Luca}}},\
  and\ \bibinfo {author} {\bibfnamefont {G.}~\bibnamefont {{Scarpetta}}},\
  }\bibfield  {title} {\bibinfo {title} {{Kerr black hole lensing for generic
  observers in the strong deflection limit}},\ }\href
  {https://doi.org/10.1103/PhysRevD.74.063001} {\bibfield  {journal} {\bibinfo
  {journal} {\prd}\ }\textbf {\bibinfo {volume} {74}},\ \bibinfo {eid} {063001}
  (\bibinfo {year} {2006})},\ \Eprint {https://arxiv.org/abs/gr-qc/0604093}
  {arXiv:gr-qc/0604093 [gr-qc]} \BibitemShut {NoStop}%
\bibitem [{\citenamefont {{Torres-Orjuela}}\ \emph {et~al.}(2020)\citenamefont
  {{Torres-Orjuela}}, \citenamefont {{Chen}},\ and\ \citenamefont
  {{Amaro-Seoane}}}]{Torres:20}%
  \BibitemOpen
  \bibfield  {author} {\bibinfo {author} {\bibfnamefont {A.}~\bibnamefont
  {{Torres-Orjuela}}}, \bibinfo {author} {\bibfnamefont {X.}~\bibnamefont
  {{Chen}}},\ and\ \bibinfo {author} {\bibfnamefont {P.}~\bibnamefont
  {{Amaro-Seoane}}},\ }\bibfield  {title} {\bibinfo {title} {{Excitation of
  gravitational wave modes by a center-of-mass velocity of the source}},\
  }\href@noop {} {\bibfield  {journal} {\bibinfo  {journal} {arXiv e-prints}\
  ,\ \bibinfo {eid} {arXiv:2010.15856}} (\bibinfo {year} {2020})},\ \Eprint
  {https://arxiv.org/abs/2010.15856} {arXiv:2010.15856 [astro-ph.CO]}
  \BibitemShut {NoStop}%
\bibitem [{\citenamefont {{Cardoso}}\ \emph {et~al.}(2021)\citenamefont
  {{Cardoso}}, \citenamefont {{Duque}},\ and\ \citenamefont
  {{Khanna}}}]{Cardoso:21}%
  \BibitemOpen
  \bibfield  {author} {\bibinfo {author} {\bibfnamefont {V.}~\bibnamefont
  {{Cardoso}}}, \bibinfo {author} {\bibfnamefont {F.}~\bibnamefont {{Duque}}},\
  and\ \bibinfo {author} {\bibfnamefont {G.}~\bibnamefont {{Khanna}}},\
  }\bibfield  {title} {\bibinfo {title} {{Gravitational tuning forks and
  hierarchical triple systems}},\ }\href@noop {} {\bibfield  {journal}
  {\bibinfo  {journal} {arXiv e-prints}\ ,\ \bibinfo {eid} {arXiv:2101.01186}}
  (\bibinfo {year} {2021})},\ \Eprint {https://arxiv.org/abs/2101.01186}
  {arXiv:2101.01186 [gr-qc]} \BibitemShut {NoStop}%
\bibitem [{\citenamefont {{Yu}}\ \emph {et~al.}(2020)\citenamefont {{Yu}},
  \citenamefont {{Ma}}, \citenamefont {{Giesler}},\ and\ \citenamefont
  {{Chen}}}]{Yu:20a}%
  \BibitemOpen
  \bibfield  {author} {\bibinfo {author} {\bibfnamefont {H.}~\bibnamefont
  {{Yu}}}, \bibinfo {author} {\bibfnamefont {S.}~\bibnamefont {{Ma}}}, \bibinfo
  {author} {\bibfnamefont {M.}~\bibnamefont {{Giesler}}},\ and\ \bibinfo
  {author} {\bibfnamefont {Y.}~\bibnamefont {{Chen}}},\ }\bibfield  {title}
  {\bibinfo {title} {{Spin and eccentricity evolution in triple systems: From
  the Lidov-Kozai interaction to the final merger of the inner binary}},\
  }\href {https://doi.org/10.1103/PhysRevD.102.123009} {\bibfield  {journal}
  {\bibinfo  {journal} {\prd}\ }\textbf {\bibinfo {volume} {102}},\ \bibinfo
  {eid} {123009} (\bibinfo {year} {2020})},\ \Eprint
  {https://arxiv.org/abs/2007.12978} {arXiv:2007.12978 [gr-qc]} \BibitemShut
  {NoStop}%
\bibitem [{\citenamefont {{Barker}}\ and\ \citenamefont
  {{O'Connell}}(1975)}]{Barker:75}%
  \BibitemOpen
  \bibfield  {author} {\bibinfo {author} {\bibfnamefont {B.~M.}\ \bibnamefont
  {{Barker}}}\ and\ \bibinfo {author} {\bibfnamefont {R.~F.}\ \bibnamefont
  {{O'Connell}}},\ }\bibfield  {title} {\bibinfo {title} {{Gravitational
  two-body problem with arbitrary masses, spins, and quadrupole moments}},\
  }\href {https://doi.org/10.1103/PhysRevD.12.329} {\bibfield  {journal}
  {\bibinfo  {journal} {\prd}\ }\textbf {\bibinfo {volume} {12}},\ \bibinfo
  {pages} {329} (\bibinfo {year} {1975})}\BibitemShut {NoStop}%
\end{thebibliography}%

\end{document}